\documentclass[a4paper,11pt]{article}
\usepackage{caption}
\usepackage{jcappub}
\usepackage{lineno}
\usepackage{fontawesome}
\usepackage{orcidlink}
\usepackage{todonotes}
\usepackage{subcaption}
\usepackage{enumitem}
\usepackage[normalem]{ulem}
\usepackage{xcolor}

\definecolor{r}{HTML}{EE6677}
\definecolor{b}{HTML}{4477AA}
\definecolor{g}{HTML}{228833}
\definecolor{y}{HTML}{CCBB44}
\definecolor{k}{HTML}{BBBBBB}
\definecolor{p}{HTML}{AA3377}
\definecolor{c}{HTML}{66CCEE}

\arxivnumber{2510.13805}
\title{\texttt{SBi3PCF:} Simulation-based inference with the integrated 3PCF}

\author[a,b,c]{David Gebauer \orcidlink{0009-0002-2889-3704},}
\author[d,e,b,c]{Anik Halder \orcidlink{0000-0002-0352-9351},}
\author[b,c]{Stella Seitz,}
\author[f,g]{Dhayaa Anbajagane}
\affiliation[a]{Fakultät für Physik, Universität Bielefeld, Postfach 100131, 33501 Bielefeld, Germany}
\affiliation[b]{Universitäts-Sternwarte, Fakultät für Physik, Ludwig-Maximilians-Universität München,\\Scheinerstraße 1, 81679 München, Germany}
\affiliation[c]{Max Planck Institute for Extraterrestrial Physics, Giessenbachstr. 1, 85748 Garching, Germany}
\affiliation[d]{Institute of Astronomy and Kavli Institute for Cosmology, University of Cambridge, Madingley Road, Cambridge CB3 0HA, UK}
\affiliation[e]{Jesus College, Jesus Lane, Cambridge, CB5 8BL, UK}
\affiliation[f]{Department of Astronomy and Astrophysics, University of Chicago, Chicago, 60637, USA}
\affiliation[g]{Kavli Institute for Cosmological Physics, University of Chicago, Chicago, IL 60637, USA}

\emailAdd{david.gebauer@uni-bielefeld.de, ah2425@cam.ac.uk}

\abstract{
We present \texttt{SBi3PCF}, a simulation-based inference (SBI) framework for analysing a higher-order weak lensing statistic, the integrated 3-point correlation function (i3PCF). Our approach forward-models the cosmic shear field using the \texttt{CosmoGridV1} suite of N-body simulations, including a comprehensive set of systematic effects such as intrinsic alignment, baryonic feedback, photometric redshift uncertainty, shear calibration bias, and shape noise. Using this, we have produced a set of DES Y3-like synthetic measurements for 2-point shear correlation functions $\xi_{\pm}$ (2PCFs) and i3PCFs $\zeta_{\pm}$ across 6 cosmological and 11 systematic parameters. Having validated these measurements against theoretical predictions and thoroughly examined for potential systematic biases, we have found that the impact of source galaxy clustering and reduced shear on the i3PCF is negligible for Stage-III surveys. Furthermore, we have tested the Gaussianity assumption for the likelihood of our data vector and found that while the sampling distribution of the 2PCF can be well approximated by a Gaussian function, the likelihood of the combined 2PCF + i3PCF data vector including filter sizes of $90'$ and larger can deviate from this assumption. Our SBI pipeline employs masked autoregressive flows to perform neural likelihood estimation and is validated to give statistically accurate posterior estimates. On mock data, we find that including the i3PCF yields a substantial $63.8\%$ median improvement in the figure of merit for $\Omega_m - \sigma_8 - w_0$. These findings are consistent with previous works on the i3PCF and demonstrate that our SBI framework can achieve the accuracy and realism needed to analyse the i3PCF in wide-area weak lensing surveys.}

\begin{document}

\maketitle
\flushbottom

\section{Introduction}

Cosmic shear describes the coherent distortions imprinted in the observed shapes of background galaxies along the line-of-sight by the weak gravitational lensing (WL) effect from the foreground matter distribution \cite{BartelmannSchneider2001, Schneider2006, Kilbinger2015}. Hence, by analysing statistical correlations in this cosmic shear field, we can probe the properties of the matter density field and measure key cosmological parameters governing the theories of large-scale cosmic structure (LSS) formation as well as details of the Universe's expansion history.

Current Stage-III WL surveys, such as the Dark Energy Survey (DES) \cite{DES_overview, DES_Y3_data, Gatti2021}, the Kilo-Degree Survey (KiDS) \cite{Kids_overview, Wright2024}, and the Hyper Suprime-Cam (HSC) survey \cite{HSC_overview}, have already placed powerful constraints on cosmological parameters, primarily by measuring two-point correlation functions (2PCFs) of the shear field \cite{DESY3_results, Kids_results, HSC_results, Anbajagane2025}. While the 2PCF probes the information contained in the second moment of the density field, it is insensitive to the higher-order moments of the inherently non-Gaussian late-time matter density distribution \cite{Bernardeau2002}.

To unlock this extra information, we must turn to higher-order statistics (HOS) \cite{Bernardeau2002}. The logical next step is the three-point cosmic shear correlation function (3PCF), which measures shape correlations between triplets of galaxies. However, a full 3PCF analysis is challenging for multiple reasons. Not only is it computationally demanding to evaluate correlations all possible triangle configurations \cite{Jarvis2019}, but it is also much more difficult to model the bispectrum compared to power spectrum \cite{Sugiyama2024}. Another practical challenge due to having many possible triangle configurations is the increased size of the data vector (DV), necessitating higher numbers of simulations required for covariance estimation for a traditional Gaussian-likelihood based MCMC analysis. Work on HOS is currently ongoing using many different approaches. Some have analytic frameworks for modelling, such as aperture mass statistics \cite{Gomes2025, Porth2024, Gatti2022, Gomes2025_des_a, Sugiyama2025, Gomes2025_des_b} or the weak lensing convergence 1-point PDF \cite{Uhlemann2023, Castiblanco2024, Friedrich2025, Thiele2020, Thiele2023}. Others, are fully based on simulations, such as Refs.~\cite{Gong2024, Fluri2022, Cheng2020, Cheng2025, Gatti2024b, Gatti2025, Jeffrey2021, Jeffrey2025, Novaes2025}.

A more practical and efficient alternative to cosmic shear 3PCFs is the integrated 3-point correlation function (i3PCF) \cite{Halder2021, Halder2023, Gong2023}. The i3PCF correlates the locally measured 2PCF in small patches of the sky with the 1-point WL shear aperture mass, effectively capturing information from squeezed triangle configurations of the lensing bispectrum directly from the cosmic shear data without the need for any convergence map-making. From the modelling side, over many other HOS, it has the advantage that it can be described analytically and allows for a physical interpretation using the response approach to perturbation theory \cite{Halder2022, Barreira_2017}.

Harnessing the cosmological information captured by HOS is of paramount importance as we enter the era of Stage-IV surveys with the Euclid mission \cite{Euclid_overview, Euclid_overview2, Euclid_HOWLS} and the Vera C. Rubin Observatory's Legacy Survey of Space and Time (LSST) \cite{LSST_overview}. These surveys will dramatically increase the statistical power of WL by observing billions of galaxies. However, this unprecedented precision also renders the measurements far more sensitive to systematic effects, such as baryonic feedback, source clustering, and source blending \citep{Deshpande2024}. Analytically modelling the impact of these systematics for each HOS separately will therefore be infeasible. This necessitates a shift away from traditional analytical methods towards simulation-based inference (SBI) approaches. By employing generative neural networks, such as normalizing flows (NF), SBI frameworks can utilize suites of forward-modelled simulations that incorporate realistic survey properties and systematics, to learn the likelihood function of the summary statistic, bypassing the need for analytical approximations such as the widely used Gaussian likelihood function, which HOS are not expected to follow. This approach is perfectly suited to the challenges of modern WL analyses, enabling a robust inference from complex observables like the i3PCF and other HOS \cite{Alsing2019, Wietersheim-Kramsta2025, Jeffrey2021, Jeffrey2025, Abellan2025, Mancini2024, Deistler2025, Gatti2024b, Gatti2025}.
To our knowledge, this work represents the first exploration of an SBI framework to jointly analyse the traditional real-space cosmic shear two-point correlation functions $\xi_\pm$ alongside a higher-order cosmic shear statistic in the form of the integrated 3-point correlation function. While previous SBI studies have analysed higher-order statistics from weak lensing convergence maps \cite{Jeffrey2025, Gatti2025}, and traditional likelihood-based analyses have combined real-space 2PCFs with compressed three-point shear statistics \cite{Burger2024, Sugiyama2025}, our approach bridges these methodologies. Working with the cosmic shear field directly bypasses any need for the reconstruction of a convergence map (even in the forward modelling process), which is not at all straightforward in the presence of complicated survey geometry and masks.

In this work, we develop and validate \texttt{SBi3PCF}, a complete SBI framework for an i3PCF analysis. For this, we have constructed a forward model based on the \texttt{CosmoGridV1} N-body simulations \cite{Kacprzak2022}, capable of generating realistic cosmic shear mocks that include survey masks, shape noise, and a range of crucial systematic effects including baryonic feedback effects using the baryonification model included at the map level \cite{Schneider2015, Schneider2019, Giri2021, Fluri2022}.\footnote{For a discussion on the performance of baryonification compared to full hydrodynamical simulations see Refs.~\cite{Schneider2019, Giri2021, Bigwood2024}. Other approaches to include baryonic effects at the map level can be seen in Refs.~\cite{Anbajagane2024, Zhou2025}.} We also publicly release \href{https://github.com/D-Gebauer/CosmoFuse}{\texttt{CosmoFuse} \faGithub}, a Python package for rapidly measuring the i3PCF from tens of thousands of pixelated shear maps on GPUs, making the generation of large training sets for SBI with the i3PCF feasible. Leveraging this, we have built an inference pipeline for both cosmic shear 2PCFs alone and a combined analysis of 2PCFs and i3PCFs. For each of these, we have trained masked autoregressive flows \cite{Papamakarios2017} to estimate the likelihood of these statistics. Through detailed checks against analytical predictions at the summary statistic level, and detailed coverage tests and simulated analyses at the inference level, we have rigorously stress-tested our SBI forward-modelling pipeline. We show that our framework produces statistically reliable posteriors and successfully recovers the true cosmological parameters from mock data without biases. In line with previous studies using analytical approaches, we find a significant improvement in cosmological constraining power gained by adding the i3PCF to a standard 2PCF cosmic shear analysis.

We structure this paper as follows. In Section~\ref{sec:cosmic_shear_theory}, we review the theory of cosmic shear correlation functions and relevant systematic effects. Section~\ref{sec:forward_model} details our simulation-based forward model. Our SBI pipeline is described in Section~\ref{sec:sbi_pipeline}. We present our validation tests and main results in Section~\ref{sec:results}, and conclude in Section~\ref{sec:conclusions}.

\section{Cosmic shear correlation functions}\label{sec:cosmic_shear_theory}

In this section, we briefly review the theory of cosmic shear correlation functions, including the 2PCF and i3PCF, as well as relevant weak lensing systematic effects. For a comprehensive review of weak lensing we direct the interested reader to Refs.~\cite{BartelmannSchneider2001, Mandelbaum_2018}

\subsection{Gravitational lensing}

\noindent Gravitational lensing has two components, convergence $\kappa$ and shear $\vec{\gamma}$. Convergence magnifies the shape of the observed source isotropically, while shear anisotropically distorts the light bundle coming from the source in some direction. As shear is a spin-2 field, we can decompose it into a modulus and an angle $\vec{\gamma} = \gamma e^{2i\varphi} = \gamma_1 + i \gamma_2$ \cite{Seitz1994} and describe it as a complex quantity. Convergence and shear can be described as fields that are functions of angular position $\vec{\theta}$ and (comoving) distance $\chi$ describing the field at some point along the light-cone. These fields depend on the lensing potential $\Psi$ which can be expressed as a weighted line-of-sight integral of the 3-dimensional Newtonian gravitational potential $\Phi$ between the source and the observer \cite{Narayan1996}:
\begin{equation}\label{eq:lensing-pot}
    \Psi(\vec{\theta}, \chi) = \frac{2}{c^2} \int_0^{\chi} d\chi' \frac{\chi - \chi'}{\chi \chi'} \Phi(\chi' \vec{\theta}, \chi') \,
\end{equation}
where $\Phi$ is described by the Poisson equation in comoving coordinates
\begin{equation}
    \nabla^2 \Phi = \frac{3}{2} H_0^2 \Omega_m \frac{\delta}{a} \,
\end{equation}
where $\delta_m$ is the 3D matter overdensity $(\rho - \Bar{\rho})/\Bar{\rho}$, $H_0$ is the Hubble constant, $\Omega_m$ is the total matter density parameter, $a$ is the scale factor, and $c$ is the speed of light. Convergence and shear are in turn described as second-order derivatives of the lensing potential \citep{Schneider2006}:
\begin{equation}
    \kappa = \frac{1}{2} \nabla^2_\theta \Psi \, , \qquad
    \gamma_1 = \frac{1}{2}(\Psi_{11}- \Psi_{22}) \, , \qquad
    \gamma_2 = \Psi_{12} \, , 
\end{equation}
where $\Psi_{ij} = \partial\Psi / (\partial\theta_i \partial\theta_i)$. In Fourier space and working in the flat-sky approximation, $\kappa$ and $\gamma$ are related through \cite{KaiserSquires1993}
\begin{equation}\label{eq:kappa2gamma}
    \gamma(\vec{\ell}) = \frac{(\ell_x + i \ell_y)^2}{\ell^2}\kappa(\vec{\ell}) = e^{2i\phi_{\ell}} \kappa(\vec{\ell}), \qquad \qquad \ell>0
\end{equation}
with $\ell = \sqrt{\ell^2_x + \ell^2_y}$ and the polar angle $\phi_{\ell} = \arctan \left( \frac{\ell_y}{\ell_x} \right)$ with $\vec{\ell}$ the 2D Fourier wave mode. As $\kappa$ is a scalar quantity, for modelling simplicity, it is easiest to work with any WL statistic in terms of $\kappa$, and only convert to $\gamma$ in the end. The $\kappa$ field in real space is given by \cite{BartelmannSchneider2001}
\begin{equation}\label{eq:convergence-los-ss}
    \kappa(\vec{\theta}, \chi) = \frac{3 H_0^2 \Omega_m}{2 c^2} \int_0^{\chi} d\chi' \frac{\chi - \chi'}{\chi \chi'} \frac{\delta_m(\chi' \vec{\theta}, \chi')}{a(\chi')} \ .
\end{equation}
Note that we assume a spatially flat Universe throughout this paper. 
For sources following a normalised distribution $n_s^i(\chi)$ in a tomographic bin $i$, the \textit{effective} convergence in some direction \cite{BartelmannSchneider2001} reads:
\begin{equation}\label{eq:convergence-los-nz}
    \begin{split}
        \kappa^i_{\text{eff}}(\vec{\theta}) & = \int_0^{\chi_{\text{max}}} d\chi \: n_s^i(\chi) \kappa(\vec{\theta}, \chi) \\
        & = \frac{3 H_0^2 \Omega_m}{2 c^2} \int_0^{\chi_{\text{max}}} d\chi \ g^i(\chi) \chi \frac{\delta_m(\chi \vec{\theta}, \chi)}{a(\chi)},
    \end{split}
\end{equation}
with $g^i(\chi)$ given by:
\begin{equation}\label{eq:g-lensing}
    g^i(\chi) = \int_{\chi}^{\chi_{\text{max}}} d\chi' \: n_s^i(\chi') \frac{\chi' - \chi}{\chi'}.
\end{equation}

\subsection{2-point correlation functions}

The Gaussian information content\footnote{in other words, the information contained up to the second-connected moment of a field.} from a random field can be extracted using the 2-point correlation function (2PCF), which corresponds to the real-space equivalent of the power spectrum. As the observed cosmic shear is described by two components, we can define two non-vanishing correlation functions $\xi_{\pm}$ for an angular separation $\vec{\alpha}$ \cite{SchneiderLombardi2003, Jarvis2004}:
\begin{equation}\label{eq:shear_2PCF_definition}
    \begin{split}
        \xi_+(\vec{\alpha}) = \left\langle \gamma(\vec{\theta_1}) \gamma^*(\vec{\theta_2}) \right\rangle = \left\langle \gamma_{t,1} \gamma_{t,2} \right\rangle + \left\langle \gamma_{\times,1} \gamma_{\times,2} \right\rangle, \\
        \xi_-(\vec{\alpha}) = \left\langle \gamma(\vec{\theta_1}) \gamma^*(\vec{\theta_2}) e^{-4i\vartheta} \right\rangle = \left\langle \gamma_{t,1} \gamma_{t,2} \right\rangle - \left\langle \gamma_{\times,1} \gamma_{\times,2} \right\rangle,
    \end{split}
\end{equation}

\noindent where $\gamma_t$ and $\gamma_\times$ are the tangential and cross components of shear, respectively. These are obtained from projecting the global shear $\gamma_{1,2}$ onto the separation vector $\vec{\alpha}$. For more details see Appendix~\ref{sec:cosmofuse}.

In cosmic shear surveys, the 2PCFs are estimated globally from the whole footprint. Analytically, they are related to their Fourier space counterpart, the convergence power spectrum $P_\kappa^{2D}$, by a Hankel transform
\begin{equation}
    \xi_\pm(\alpha) = \int \frac{d\ell \ell}{2 \pi} P_{\kappa}^{2D}(\ell)J_{0/4}(\ell \alpha),
\end{equation}
where $J_{n}$ is the n-th order Bessel function. The convergence power spectrum is obtained by a line-of-sight integral of the 3D matter power spectrum $P^{3D}_{\delta}(k,\chi)$\footnote{note that in describing the power spectrum $P^{3D}_{\delta}(k,z)$ here and in several other places in the paper, we have expressed the time coordinate usually represented by redshift $z$, in terms of the comoving distance $\chi$, since a deterministic mapping between $z$ and $\chi$ exists.}
\begin{equation}\label{eq:convergence_power_spectrum}
	P_{\kappa}^{ij}({\ell}) = \int d\chi \frac{q^i(\chi)q^j(\chi)}{\chi^2}P^{3D}_{\delta} \left({k}=\frac{{\ell}}{\chi},\chi \right)
\end{equation}
where $q^i(\chi)$ is the lensing kernel, which using Eq.~\eqref{eq:g-lensing} can be written as
\begin{equation}
    q^i(\chi) = \frac{3H_0^2\Omega_{m}}{2c^2}\frac{\chi}{a(\chi)} g^i(\chi).
\end{equation}
\subsection{Integrated 3-point correlation functions}

Due to non-linear gravitational evolution, the cosmic matter density field is non-Gaussian distributed \citep{Bernardeau2002} with significant information contained in the higher-order moments of the field. To extract information from these higher-order moments, summary statistics other than 2PCFs are required. An obvious extension to the 2PCF would be to use the 3-point correlation function \citep{Schneider2005, Secco2022b}, or its counterpart in Fourier space, the bispectrum. However, efficiently measuring them on real data as well as modelling them analytically is challenging. In the context of our simulation-based inference (SBI) framework, a full tomographic 3PCF analysis is particularly unfeasible. Evaluating it across the required $\mathcal{O}(10^5)$ simulations is computationally prohibitive, and the resulting uncompressed data vector would be too large for stable covariance estimation prior to MOPED compression. Expanding this pipeline to a full 3PCF analysis therefore, remains a natural extension for future work.

To overcome these practical limitations while still extracting information from the third moment of the density field, we focus on the integrated bispectrum $\mathcal{B}_{\pm}(\ell)$ (or the integrated 3PCF in real space). The i3PCF offers a significant advantage in measurement ease and allows for straightforward validation against established analytical frameworks. It is an integration of the bispectrum that is mostly sensitive to squeezed configurations ($k_3 \ll k_1 \approx k_2$) \citep{Chiang2014} given by a correlation of the mean density within a sub-volume to the position-dependent power spectrum within that sub-volume. This approach has been extended to the weak lensing field in Refs.~\cite{Halder2021, Munshi_2023}. When dividing the sky into many patches with their centres at positions $\theta_c$, one can obtain the integrated bispectrum as an ensemble average of the product of the aperture mass $M_{ap}(\theta_c)$ and the position-dependent cosmic shear power spectra in those patches. 

The aperture mass is the 1-point mean shear signal and is related to the large-scale mode $k_3$ that effectively modulates the small-scale power spectra described by the two small-scale modes. Following \cite{Halder2023}, we can then express the 2D integrated bispectrum $\mathcal{B}_{\pm}(\ell)$ as:
\begin{equation}\label{eq:integrated_bispectrum}
    \begin{split}
        \mathcal{B}_{\pm}^{ijk}(\ell) = \int d\chi \frac{q^i(\chi )q^j(\chi )q^k(\chi )}{\chi^4} \int_{\vec\ell_1}\int_{\vec\ell_2} B^{3D}_{\delta} \left( \frac{\ell_1}{\chi}, \frac{\ell_2}{\chi}, \frac{|-\vec\ell_{1}-\vec\ell_{2}|}{\chi} ; \chi \right) e^{2i(\vartheta_{\vec\ell_2}\mp \vartheta_{-\vec\ell_{1}-\vec\ell_{2}})}\\ 
        \times U(\vec\ell_{1})W(\vec\ell_{2}+\vec\ell)W(-\vec\ell_{1}-\vec\ell_{2}-\vec\ell) \ ,
    \end{split}
\end{equation}
where the compensated filter $U(\ell)$ is used for measuring $M_{ap}$ while the top-hat filters $W(\ell)$ are used for measuring the position-dependent cosmic shear power spectra (which we further detail in Section~\ref{sec:measuring_corrs}). The real space counterpart of this is the integrated 3-point correlation function \citep{Halder2022} that in a tomographic setting (with redshift bins indicated as $i,j,k)$ reads:
\begin{equation}\label{eq:i3PCF}
    \zeta^{ijk}_{\pm}(\alpha) = \left\langle M^i_{ap}(\theta_c) \xi^{jk}_{\pm}(\alpha; \theta_c) \right\rangle = \frac{1}{A_{2pt}(\alpha)}\int \frac{d\ell \ell}{2\pi} \mathcal{B}^{ijk}_{\pm} (\ell) J_{0/4} (\ell \alpha),
\end{equation}
 with the projected integrated bispectrum expressed through Eq.~\eqref{eq:integrated_bispectrum}. The area normalisation factor $A_{2pt}(\alpha)$ is given by
\begin{equation}
     A_{2pt}(\alpha) = \int \frac{d \phi_{\vec\alpha}}{2\pi}\int d^2\theta \ W(-\vec\theta_c+\vec\theta)W(-\vec\theta_c + \vec\theta + \vec\alpha).
\end{equation}

\subsection{Weak lensing systematic effects}\label{sec:systematics_theory}

There are multiple systematic effects in weak lensing observations that directly impact measurements and interpretation of $\xi_\pm$ and $\zeta_\pm$. If neglected, these can lead to biases in cosmological parameter inference. In this section, we briefly describe some key weak lensing systematics and how we treat them in our work.

\subsubsection{Photometric redshift uncertainty}

With redshifts of galaxies not measured spectroscopically in wide-area weak lensing surveys, they have to be estimated from the observed galaxy photometry in few wavelength bands, which in itself is a very challenging task \cite{Newman_2022}. This leads to uncertainties in characterizing the true underlying redshift distribution of source galaxies in a given tomographic redshift bin $i$. The associated uncertainty, at first order, can be modelled as a constant shift $\Delta z^i$ in the mean of the redshift distribution $n^i_s(z)$. For this, we use separate uncorrelated shift parameters $\Delta z^i$ for each tomographic bin, as described in \cite{Secco2022a}. This transforms the redshift distribution as

\begin{equation}
    \label{eq:photometric_uncertainty_model}
    n^i_s(z) \longrightarrow n^i_s(z + \Delta z^i) \, .
\end{equation}

\subsubsection{Shear calibration bias}\label{sec:systematics_mult}

Errors in the shear estimation pipeline from observed galaxy ellipticities can lead to two types of biases in the shape measurement of source galaxies: additive and multiplicative. Following the approach of current lensing shear estimation methodologies \citep{Gatti2021, HSC_overview, Anbajagane2025catalog}, we take the additive bias to be negligible. As the correlation level, the multiplicative component is modelled as a factor $1+m_i$ multiplied directly to $\xi_\pm$ and $\zeta_\pm$ for each tomographic bin \cite{Heymans2006}: 

\begin{align}
    \label{eq:shear_2pcf_multiplicative_bias_model}
    \xi_{\pm}^{ij}(\alpha) &\longrightarrow (1+m_i)(1+m_j)\ \xi_{\pm}^{ij}(\alpha) \ , \\
    \label{eq:shear_i3pcf_multiplicative_bias_model}
    \zeta_{\pm}^{ijk}(\alpha) &\longrightarrow (1+m_i)(1+m_j)(1+m_k)\ \zeta_{\pm}^{ijk}(\alpha) \ .
\end{align}

\subsubsection{Intrinsic alignment}\label{sec:systematics_ia}

Intrinsic alignment (IA) describes the intrinsic shapes of galaxies (mainly due to local tidal forces) prior to any gravitational lensing imprint in their observed shapes. This leads to physical correlations of galaxy shapes that can mimic a gravitational lensing signal and can lead to significant biases in cosmological inference when unaccounted for. In the following, we will give a basic overview of the IA model implemented in this paper and refer the reader to Refs.~\cite{Troxel_2015, Joachimi_2015, Krause_2015} for more details. We use the so-called \textit{non-linear linear alignment} (NLA) model \citep{BridleKing2007}, in which the intrinsic alignment power spectrum of source galaxies is described by 
\begin{equation}
    P^{3D}_{\text{IA}}(k, z) = A_{\text{IA}}(z) P^{3D}_{\delta}(k,z),
\end{equation}
where $P^{3D}_{\delta}$ is the non-linear matter power spectrum. This model has been shown to accurately reflect intrinsic shape correlations of red elliptical galaxies down to separations of $6\ h^{-1} Mpc$ \citep{Singh2015}. The intrinsic alignment amplitude $A_{\text{IA}}(z)$ is given by
\begin{equation}\label{eq:NLA}
    A_{\text{IA}}(z) = A_{\text{IA},0} \frac{C_1 \rho_m}{D(z)}\left( \frac{1+z}{1+z_0} \right)^\eta,
\end{equation}
where $D(z)$ is the growth factor (normalised to unity at $z=0$) and $C_1$ is a normalisation constant first introduced in \cite{HirataSeljak2004} which is often given the value $5\times10^{14} / (H_0/100)^2 M_\odot^{-1} Mpc^3$ \citep{Joachimi2011}. In this work, we will assume no redshift dependence, so $\eta = 0$,\footnote{This prevents projection effects biasing the posterior of $A_{\text{IA},0}$ to zero.} and will use $A_{\text{IA}}$ instead of $A_{\text{IA},0}$ to describe the amplitude.

In the NLA model, the IA contribution to the 2D lensing convergence power spectrum is given by a line-of-sight projection of $P^{3D}_{\text{IA}}(k, z)$, similar to Eq.~\eqref{eq:convergence-los-nz}. Overall, the impact of IA to the total convergence power spectrum (and hence the shear 2PCF) can be expressed as a modification to the lensing kernel $q^i(\chi)$ that is sensitive to the distribution of source galaxies $n^i_{s}(z)$:
\begin{equation}\label{eq:ia-los}
    q^i(\chi) \rightarrow q^i(\chi) - A_{\text{IA}}(z(\chi)) \frac{n^i_{s}(z(\chi))}{\Bar{n^i_{s}}}\frac{dz}{d\chi}.
\end{equation}

The impact of the NLA model at the i3PCF level can also be encoded with the same transformation to the lensing kernel as presented in the above Eq.~\eqref{eq:ia-los}. We refer the reader to \cite{Gong2023} for further details. We also point the reader to recent developments towards the incorporation of higher-order intrinsic alignment effects into weak lensing simulations for Stage-IV surveys, e.g.~the tidal alignment and tidal torquing (TATT) model \cite{joachim_TATT_2025}. We note that the \texttt{CosmoGridV1} simulations do not provide the shell tidal fields and hence do not allow for the incorporation of IA effects beyond the NLA model. However, for the angular scales ($> 15'$; see Section \ref{sec:measuring_corrs}) that we are interested in and for Stage-III survey-like uncertainties of the i3PCF data vector, we deem the NLA model sufficient for our present work. 

\subsubsection{Baryonic feedback}\label{sec:baryonic_feedback}

Baryonic physics, encompassing processes such as gas cooling, star formation, and energetic feedback from supernovae (SNe) and active galactic nuclei (AGN), significantly impacts the distribution of matter in the Universe on small scales \cite{Chisari2018, Amon2022, Lehman2024}. These processes, when unaccounted for, can pose as a crucial systematic effect for weak lensing (WL) studies, particularly for higher-order statistics (HOS) that are designed to probe the non-Gaussian features of the cosmic density field, with high signal-to-noise on these smaller scales. To account for the impact that such redistribution of baryons has on the dark matter distribution via gravitational coupling, the \textit{Baryonification} model \citep{Schneider2015} has been widely adopted. In this framework, firstly, particles in a pure dark matter-only N-body simulation are displaced according to some particular baryonic processes. Specifically, the density profile of a pure dark matter only halo (prior to any baryonification) is described by two density components
\begin{equation}\label{eq:baryonification_dmo}
    \rho_{dmo}(r) = \rho_{NFW}(r) + \rho_{2h},
\end{equation}
where $\rho_{NFW}$ represents a Navarro-Frenk-White profile \citep{Navarro1997} and $\rho_{2h}$ represents the 2-halo term describing the density contributions from neighbouring halos. The \textit{baryonified dark matter} profile in turn is described as the sum of collisionless matter $\rho_{clm}(r)$, gas $\rho_{gas}(r)$, and central galaxies $\rho_{cga}(r)$:
\begin{equation}
    \rho_{dmb}(r) = \rho_{clm}(r) + \rho_{gas}(r) + \rho_{cga}(r) + \rho_{2h},
\end{equation}
where the collisionless matter component $\rho_{clm}(r)$ includes the dark matter halo described by $\rho_{NFW}$ in Eq.~\eqref{eq:baryonification_dmo}, in addition to stellar mass in satellite galaxies for the case including baryons).

In the full model of Ref.~\cite{Schneider2015}, baryonification then describes a transformation $\rho_{dmo}(r) \rightarrow \rho_{dmb}(r)$ depending on 14 parameters describing the gas profile, stars, dark matter, and background density. Out of these 14 parameters, Ref.~\cite{Giri2021} found that WL observables are mostly impacted by the parameter $M_c$ describing the mass dependence of the gas profile. This dependence and its redshift evolution can be modelled as a power law \cite{Kacprzak2022}:
\begin{equation}
    M_c = M_c^0 \left( 1 + z \right)^\nu,
\end{equation}
\noindent with $M_c^0$ and $\nu$ as variable parameters and the rest of the 14 parameters fixed to the best-fitting values from X-ray observations \cite{Schneider2019}. This method has been shown to accurately reproduce weak lensing 2PCF observations from hydrodynamical simulations provided they are flexible enough \cite{Schneider2019, Giri2021, Bigwood2024, Zhou2025}. The modelling and understanding of baryonic feedback is an area of active research, and more informed alternatives to encode the impact of feedback in dark matter simulations for future SBI lensing analyses might be required with improved understanding of feedback effects \cite{Lucie-Smith_2025}.

\subsubsection{Reduced shear}\label{sec:systematics_red}

When measuring galaxy shapes in photometric surveys we do not observe $\gamma$ directly \cite{Seitz1996}. Rather, the observed shape is sensitive to a quantity called the \textit{reduced shear} $\gamma^\text{red}$, which for a pixelated cosmic shear map is given by \cite{Schneider1995, Bernstein2002}:
\begin{equation}
    \gamma^\text{red}(\theta_{\text{pix}}) = \frac{\gamma(\theta_{\text{pix}})}{1 - \kappa(\theta_{\text{pix}})}.
\end{equation}
where $\kappa(\theta_{\text{pix}})$ is the corresponding pixelised (unobserved) convergence field. We include this test in Section~\ref{sec:including_systematics} to see the impact of reduced shear on $\xi_{\pm}$ and $\zeta_{\pm}$ at the fiducial cosmology.

\subsubsection{Source galaxy clustering}\label{sec:systematics_sc}

Source galaxy clustering (SC) refers to the clustering of background source galaxies \cite{Schneider2002}, used for estimating the WL signal, which is usually assumed to be zero. Ref.~\cite{Gatti2024} showed that SC can severely impact some higher-order statistics, such as third-order convergence moments $\langle \kappa_\text{obs}^3 \rangle$ \cite{Linke2025}, peak counts \cite{Zuercher2022}, and wavelet phase harmonics \cite{Cheng2020} that are all WL statistics measured on the reconstructed convergence field. However, second-order cosmic shear statistics, such as 2PCFs, are seen to be not affected by SC \citep{Linke2025}. 

This has not been tested for the real space cosmic shear i3PCF, and therefore warrants an investigation that we perform at the fiducial cosmology in Section~\ref{sec:including_systematics}. We also measure the impact of SC on 2PCFs as a sanity check. Here, we briefly outline how we include SC in our forward model following Ref.~\cite{Gatti2024} which we refer the reader for further details. Firstly, the cosmic shear signal without SC at the position of a given pixel $\theta_{\text{pix}}$ in a simulated WL map can be written as:
\begin{equation}\label{eq:mock_no_sc}
    \gamma(\theta_{\text{pix}}) = \frac{\sum_s \Bar{n}(s)\gamma(\theta_{\text{pix}},s)}{\sum_s \Bar{n}(s)} + \frac{\sum_g w_g e_g}{\sum_g w_g},
\end{equation}
\noindent where the WL signal (first) term is summed over all source shells\footnote{denoting slices in redshifts.} $s$ and the shape noise (second) term over all galaxies $g$ in the source galaxy catalogue that fall within the given pixel. $w_g$ and $e_g$ are the weights and ellipticities of the galaxies, and $\Bar{n}(s)$ is the spatial average of the source galaxy density. To include source clustering in simulated observations, Eq.~\eqref{eq:mock_no_sc} has to be modified as \citep{Gatti2024}:
\begin{equation}\label{eq:mock_sc}
\begin{split}
        \gamma(\theta_{\text{pix}}) &= \frac{\sum_s \Bar{n}(s) [1 + b_g \delta(\theta_{\text{pix}},s)] \gamma(\theta_{\text{pix}},s)}{\sum_s \Bar{n}(s)[1 + b_g \delta(\theta_{\text{pix}},s)]}\\ 
        &+ \left( \frac{\sum_s \Bar{n}(s) }{\sum_s \Bar{n}(s)[1 + b_g \delta(\theta_{\text{pix}},s)]} \right)^\frac{1}{2} F(\theta_{\text{pix}})\frac{\sum_g w_g e_g}{\sum_g w_g}.
\end{split}
\end{equation}
The signal (first) term now has additional factors of $[1 + b_g \delta(\theta_{\text{pix}},s)]$ which correlates the galaxy count with the underlying matter density field at $s$, indicating that the source galaxies are a biased tracer (assuming linear galaxy bias $b_g$). The shape noise (second) term now has two multiplicative factors, where the first fractional term correlates the variance of the shape noise with the underlying density field. The second factor, $F(\theta_{\text{pix}})$, compensates for the correlations already present in the observed galaxy catalogue used to generate the mock shape noise. For the 4 tomographic bins of the DES Y3 source galaxy catalogue, $F(\theta_{\text{pix}})$ can be calculated as \citep{Gatti2024}
\begin{equation}
    F(\theta_{\text{pix}}) = A \sqrt{1 - B \sigma_e^2(\theta_{\text{pix}})},
\end{equation}
\noindent with the two sets of per-bin constants $A$ = [0.97, 0.985, 0.990, 0.995] and $B$ = [0.1, 0.05, 0.035, 0.035] which have been estimated from simulations used in Ref.~\cite{Gatti2024} for a source-clustering study of several weak lensing convergence HOS in DES Y3.

\section{Forward model for cosmic shear correlation functions}\label{sec:forward_model}

\noindent This section describes the framework for our simulation-based forward model that we use in order to produce accurate predictions for 2PCF and i3PCF at different points in the cosmological and systematic parameter space of interest to us. We generate weak lensing full-sky maps based on the publicly available \texttt{CosmoGridV1} N-body simulation suite \cite{Kacprzak2022} for a set of cosmological parameters: $\Omega_m, \Omega_b, H_0, n_s, \sigma_8, w_0, M_c^0$ and $\nu$ and use the corresponding parameter ranges and priors that were used to run the simulations (more details in Section~\ref{sec:CosmoGrid}). For the other systematic parameters, such as the intrinsic alignment amplitude, the shear calibration biases and the photo-z uncertainty, we follow the DES Y3 priors \cite{DESY3_results, Secco2022a}. We have sampled 17500 $A_{\text{IA}}$ values from a Sobol sequence \cite{Sobol1967}, leading to a separate value for each full-sky map. Similarly, the photometric redshift uncertainties have one value per full-sky simulation, drawn from a per-bin Gaussian distribution. We present the full set of prior ranges in Table~\ref{tab:prior_range}.

Apart from these priors, there are further restrictions in the $\Omega_m - \sigma_8$ and $\Omega_m - w_0$ plane where some combinations of these parameters from the \texttt{CosmoGridV1} simulation set are removed (cf.~the plotted prior in Figure~\ref{fig:full_posterior} in our Appendix~\ref{sec:full_post}). For more details about the specific choice of parameter ranges for which the \texttt{CosmoGridV1} simulations were run, we refer the reader to Section~2 of Ref.~\cite{Kacprzak2022}.

\begin{table}[htbp]
    \centering
    \begin{tabular}{|ll|}
        \hline
        \textbf{Parameter} & \textbf{Prior Distribution} \\
        \hline
        $\Omega_m$ & $U[0.10, 0.50]$ \\
        $\Omega_b$ & $U[0.03, 0.06]$ \\
        $H_0$ & $U[64.0, 82.0]$ \\
        $n_s$ & $U[0.87, 1.07]$ \\
        $\sigma_8$ & $U[0.40, 1.40]$ \\
        $w_0$ & $U[-2.00, -0.33]$ \\
        \hline
        $\log M_c^0$ & $U[12.0, 15.0]$ \\
        $\nu$ & $U[-2.00, 2.00]$ \\
        \hline
        $A_\text{IA}$ & $U[-5.00, 5.00]$ \\
        \hline 
        $m_1$ & $\mathcal{N}( -0.0063, 0.0091)$ \\
        $m_2$ & $\mathcal{N}( -0.0198, 0.0078)$ \\
        $m_3$ & $\mathcal{N}( -0.0241, 0.0076)$ \\
        $m_4$ & $\mathcal{N}( -0.0369, 0.0076)$ \\
        $\Delta z_1$ & $\mathcal{N}(0.0, 0.018)$ \\
        $\Delta z_2$ & $\mathcal{N}(0.0, 0.015)$ \\
        $\Delta z_3$ & $\mathcal{N}(0.0, 0.011)$ \\
        $\Delta z_4$ & $\mathcal{N}(0.0, 0.017)$ \\
        \hline
    \end{tabular}
    \caption{The prior ranges used for inference. The priors for the cosmological, baryonification, and intrinsic alignment parameters are uniform flat priors (denoted by $U$ with the corresponding bounds). For the shear calibration and photometric redshift uncertainty parameters, we use Gaussian prior distributions (denoted as $\mathcal{N}$ with the mean and standard deviations of the distributions).}
    \label{tab:prior_range}
\end{table}

\subsection{CosmoGridV1 N-body simulations particle shells}\label{sec:CosmoGrid}

The \texttt{CosmoGridV1} \citep{Kacprzak2022, Fluri2022} suite of N-body simulations was run using \texttt{PKDGRAV3} \citep{Potter2017} for boxes of size $900 Mpc/h$ filled with $832^3$ dark matter particles with a mass between $3.5 \times 10^{10} h^{-1}M_{\odot}$ and $17.5 \times 10^{10} h^{-1}M_{\odot}$, depending on cosmology. Although this mass resolution does not fully resolve the inner substructures of small dark matter halos, it is sufficient for the statistics considered in this work. For the tomographic bins and scales analysed here, the weak lensing signal is predominantly sensitive to halos with masses $\log_{10}(M/M_\odot)>12$ \cite{Lucie-Smith_2025}, which are robustly resolved in the CosmoGridV1 suite. Unresolved substructure primarily impacts the signal on very small scales (sub-arcminute). Because we employ conservative scale cuts ($\geq15'$ for the local 2PCF), we exclude the regime where missing substructure would bias the signal. Furthermore, on those discarded small scales, other systematic uncertainties, such as baryonic feedback and observational data effects, would vastly dominate over the contributions from dark matter substructure.

All simulations include three degenerate neutrino species with $\sum m_\nu = 0.06 eV$. The boxes are evolved in 70 time steps equally spaced in proper time from $z=99$ to $z=4$ and another 70 steps from $z=4$ to $z=0$.
The grid is made up of 2500 different flat $w$CDM cosmologies varying $\Omega_m, \Omega_b, w_0, n_s, H_0, \sigma_8$, as well as two parameters describing baryonification $M_c^0, \, \nu$ (for an explanation, see Refs.~\cite{Kacprzak2022, Schneider2015}). The parameters are drawn from a Sobol sequence, which is a quasi-random distribution, close to a draw from a uniform distribution. At every cosmology, 7 simulations were run with different initial conditions. Aside from this, a set of fiducial simulations also exists, comprising 13 different cosmologies: one at the fiducial value and six where each cosmological parameter is changed by a small amount while others are held fixed. At every cosmology, these are run for 200 different initial conditions. 

The simulations are provided in the form of light-cone particle shells i.e.~full-sky maps of particle counts in a given pixel within some redshift interval. These maps are in the \texttt{HEALPIX} \citep{Gorski2005} format with a resolution of $\texttt{NSIDE} = 512$, splitting the sky into $12 \times 512^2$ pixels of equal area. To create these shells, the boxes at different redshifts are stacked and rotated, with the order of the shells permuted, to have multiple possible lines-of-sight that only cross a given structure once, thereby increasing the number of quasi-independent realisations. More details on this can be found in \cite{Kacprzak2022, Sgier2019}.

The final data product consists of 69 (68 for a few cosmologies) particle shells. The set of shells used in this work includes baryonic effects as described in Section~\ref{sec:baryonic_feedback}. The baryonification model described in Ref.~\cite{Schneider2015} only works in 3-dimensional boxes. Ref.~\cite{Kacprzak2022} presents the shell baryonification method, which can be used on the pixelised particle shells provided by CosmoGridV1. This method works with projected halo profiles and has been shown to agree with the original work in 3D boxes \citep{Fluri2022}.

\subsection{Creating weak lensing maps including systematics}

To create weak lensing maps from the provided particle shells we use the code \texttt{UFalcon2} \citep{Sgier2019, Sgier2021, Reeves2024}, which works as follows. In the first step, the particle shells are projected to convergence maps for the four DES Y3 source tomographic bins. These full-sky maps are then converted to shear maps, the DES Y3 mask is applied, and then shape noise is added. The convergence can be obtained from a line-of-sight integral of the matter overdensity (see Eq.~\eqref{eq:convergence-los-nz}). For pixelised particle shells as \texttt{CosmoGridV1} provides, this can be written as a sum over the particle shells $s$ \citep{Sgier2019}
\begin{equation}\label{eq:data-kappa-los-pixel}
    \kappa (\theta_{\text{pix}}) = \frac{3}{2} \Omega_m \sum_s W_s \frac{H_0}{c} \left[ \frac{N_{\text{pix}}}{4\pi}\frac{V_\text{sim}}{N_\text{part}^\text{sim}} \left( \frac{H_0}{c}\right)^2 \frac{n_p(\theta_\text{pix}, \Delta \chi_s)}{\mathcal{D}^2(z_s)} - \left( \frac{c}{H_0} \Delta \mathcal{D}_s \right)\right],
\end{equation}
where $n_p$ is the number of particles in a given pixel in that shell. $W_s$ is the weight for each shell, depending on the distribution of source galaxies (through the lensing kernel):
\begin{equation}\label{eq:data-los-w}
    W_s = \left( \int_{\Delta z_s} dz \frac{ (1+z) H_0 \chi(z)}{H(z)} \int_z^{z_\text{max}} dz' n(z') \frac{\chi(z') - \chi(z)}{\chi_z'} \right) / \left( \int_{\Delta z_s} \frac{dz H_0}{H(z)} \right)
\end{equation}
The resulting shear maps can be converted from the full-sky convergence maps through a Fourier transform pair using Eq.~\eqref{eq:kappa2gamma}, which is akin to Kaiser-Squires reconstruction \cite{KaiserSquires1993}.

\begin{figure}[htbp]
    \centering
    \includegraphics[width=\linewidth]{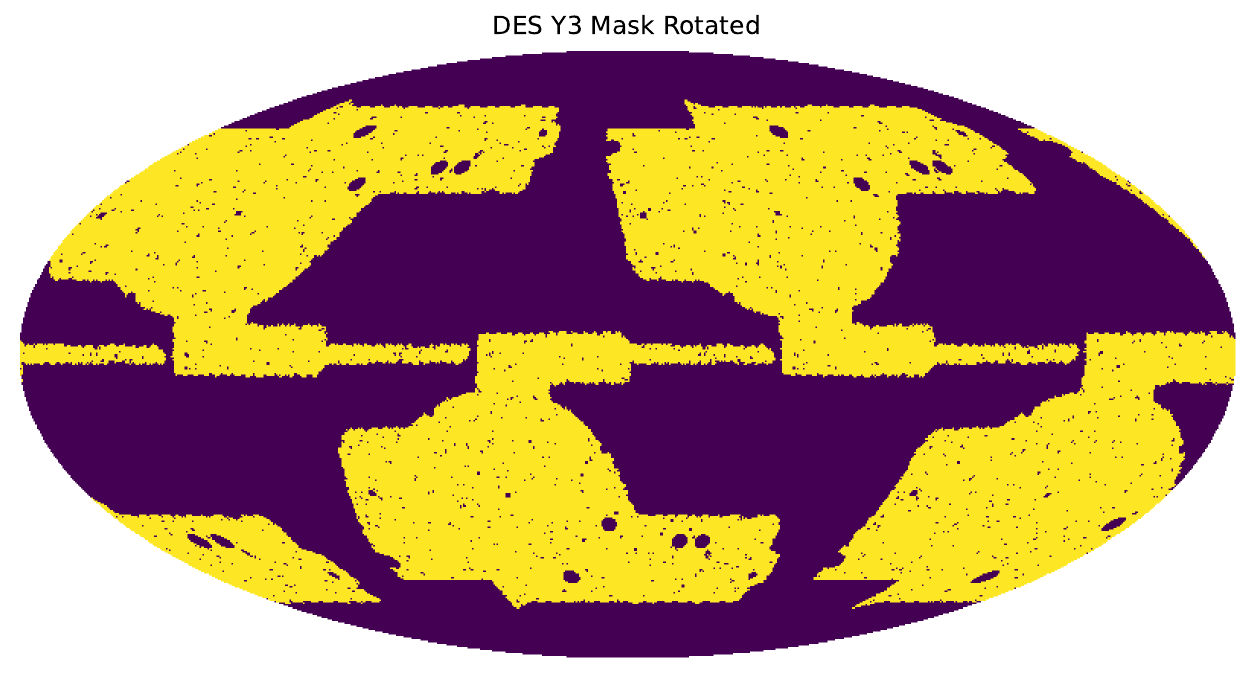}
    \caption{The DES mask rotated to four non overlapping positions in the sky.}
    \label{fig:DES_footprints}
\end{figure}

By then rotating these full-sky shear maps before applying the DES mask, up to four non-overlapping footprints can be extracted (see Figure~\ref{fig:DES_footprints}). For the simulation suite grid, this led to $2500$ cosmologies $\times 7$ realisations $\times 4$ footprints $= 70,000$ independent mocks. To estimate the covariance matrix of the data vector at the fiducial cosmology, the 200 fiducial maps are rotated to 20 random positions before the 4 independent footprints are cut out, resulting in $200$ realisations $\times 20$ random orientations $\times 4$ footprints $= 16,000$ footprints.

To each of these footprints, shape noise is then added. Noise in cosmic shear describes the effect of random intrinsic ellipticities on the measurement. To realistically model this, we take the galaxies from the observed source galaxy catalogue of DES Y3 \cite{Gatti2021} and randomly rotate their ellipticities to decorrelate them. By summing up the ellipticities of galaxies (with the corresponding weights) that fall within a given pixel, we can get per-pixel shape noise, which is then added to the simulated footprints. This leads to simulated shear maps with the same weight per pixel as in the DES Y3 observations.

We also add the systematic effects described in Section~\ref{sec:systematics_theory} to our mock footprints. For grid cosmologies, we add photometric redshift uncertainty with 1 shift parameter per tomographic bin, shear calibration bias with one parameter per bin, and intrinsic alignment (Eq.~\eqref{eq:NLA}) as per the NLA model with a free amplitude $A_{\text{IA}}$ with redshift dependence $\eta = 0$. Reduced shear and source galaxy clustering are only added to a separate set of fiducial simulations, on which we verified that these are negligible for the work under consideration (see Section~\ref{sec:including_systematics}).

\begin{figure}[htbp!]
    \centering
    \includegraphics[width=\linewidth]{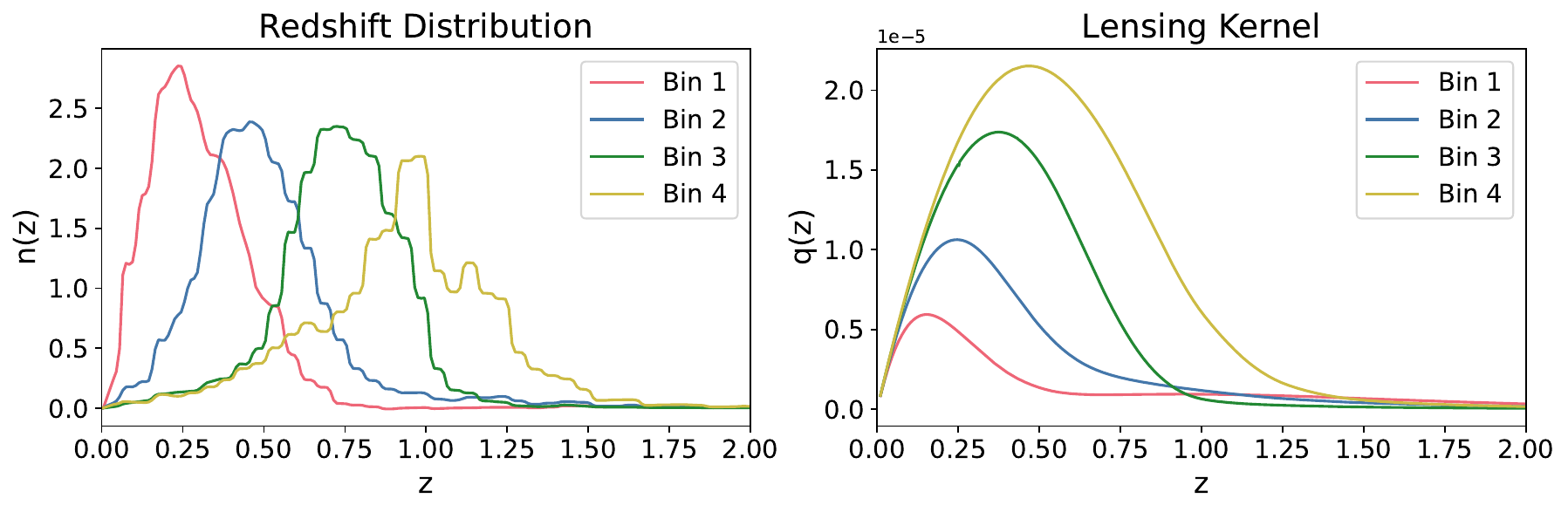}
    \caption{Left: The redshift distribution of source galaxies for the 4 DES Y3 tomographic bins. The galaxy number densities are 1.476, 1.479, 1.484 and 1.461 per square arcminute respectively in these bins. Right: The lensing kernel for each bin.}
    \label{fig:DES_nz}
\end{figure}

\subsection{Measuring the correlation functions}\label{sec:measuring_corrs}

With all the mock footprints created, we now present the methodology to estimate the 2PCF and i3PCF data vectors from each cosmic shear $\gamma$ footprint. 

To calculate shear 2PCFs $\xi_{\pm}$, it is best to decompose the spin-2 $\gamma$ field into the so-called \textit{tangential} and \textit{cross} shear components with respect to a convenient reference point, i.e.~relative to each galaxy (or pixel under consideration) every 2PCF pair at a given angular separation is considered. The \textit{tangential} component is positive, orthogonal to the line connecting the reference galaxy (pixel), while the cross \textit{component} is rotated by 45 degrees to the connecting line. Mathematically, this gives us the following transformation in terms of the $\gamma_1$ and $\gamma_2$ shear components that describes the $\gamma$ field in the original coordinate system:
\begin{equation}\label{eq:relative_shear}
    \begin{split}
        \gamma_t &= -\gamma_1 \cos( 2 \vartheta ) - \gamma_2 \sin( 2 \vartheta )\\
        \gamma_{\times} &= -\gamma_1 \sin( 2 \vartheta ) + \gamma_2 \cos( 2 \vartheta ),
    \end{split}
\end{equation}
with $\vartheta$ being the polar angle between the global $x$-axis and that of the line connecting a galaxy under consideration to the reference point. The shear 2PCFs (Eq.~\eqref{eq:shear_2PCF_definition}) are then measured as the sum over all pairs that fall within an angular separation bin $\alpha_{\min} \le \alpha \le \alpha_{\max}$:
\begin{equation}\label{eq:2PCF_measure}
    \xi_{\pm}(\alpha) = \frac{\sum_{ij} w_i w_j \left( \gamma^i_t \gamma^j_t \pm \gamma^i_\times \gamma^j_\times \right) }{\sum_{ij}w_i w_j},
\end{equation}
where $w$ are the pixel weights obtained by summing up the weights of all individual galaxies that fall within a pixel. In this work, we use 8 logarithmically spaced bins between 15 and 250 arcminutes for the global 2PCFs. The lower separation was chosen as it is twice the resolution of our shear maps at $\texttt{NSIDE} = 512$ to ensure that enough pairs fall within the smallest bin and that we do not consider pairs at separations smaller than the resolution of a pixel. This also automatically excludes the scales most affected by baryonic feedback effects. The maximal angular separation for the 2PCF was chosen to match the DES Y3 setup \cite{DESY3_results, Secco2022a}.

To obtain the i3PCFs, we measure the local 2PCFs $\xi_{\pm}(\alpha; \theta_c)$ in small patches with radius $\theta_{\text{ap}}$ with the same angular separation bins as the global 2PCFs with the larger bins truncated following $\alpha_{\max} \le 2 \theta_{\text{ap}}-5'$ to not have separations larger than the diameter of a given filter. On the other hand, we estimate the aperture mass as the weighted sum over all pixels' tangential shear relative to the patch centre $\theta_c$:
\begin{equation}\label{eq:aperture_mass}
    M_{\text{ap}}(\theta_c; \theta_\text{ap}) = \frac{\sum_i w_i \gamma_{t,i} Q(\vartheta_i; \theta_\text{ap})}{\sum_i w_i},
\end{equation}
here $\vartheta_i$ are the pixels angular separations from the patch centre and $Q$ is the compensated filter centred at location $\theta_c$ with its functional form given by \citep{kilbinger2005}:
\begin{equation}
    Q(\vartheta; \theta_\text{ap}) = \frac{\vartheta^2}{4 \pi \theta_\text{ap}^4} \exp \left( - \frac{\vartheta^2}{2 \theta_\text{ap}^2} \right).
\end{equation}
Practically, we evaluate Eq.~\eqref{eq:aperture_mass} for pixels up to $5 \theta_{\text{ap}}$ separation from the patch centre, as 
\begin{equation}
    \lim_{\vartheta \rightarrow \inf} Q(\vartheta) = 0 .
\end{equation}
The i3PCF is then estimated by taking the mean product of the local 2PCFs and aperture mass measurements over all patches while subtracting out the product of the average aperture mass and average 2PCF over the whole footprint:
\begin{equation}
    \zeta_\pm = \frac{\sum_p M_\text{ap}^p \ \xi^p_\pm}{N_\text{patches}} -  \frac{\sum_p M_\text{ap}^p}{N_\text{patches}} \frac{\sum_p \xi^p_\pm}{N_\text{patches}} ,
\end{equation}
where we sum over the patch indices $p$ for the total number of patches within a footprint ($N_\text{patches}$). These calculations are computationally intensive, especially the computation of the local 2PCFs for the large number of mocks required for SBI. To perform them efficiently, we developed the public Python package \texttt{CosmoFuse}, which leverages just-in-time (JIT) compilation and GPU parallelization to accelerate the measurement of both local 2PCFs and aperture masses. Further details on the package and its validation are provided in Appendix~\ref{sec:cosmofuse}.

\begin{figure}[htbp!]
    \centering
    \includegraphics[width=1.\linewidth]{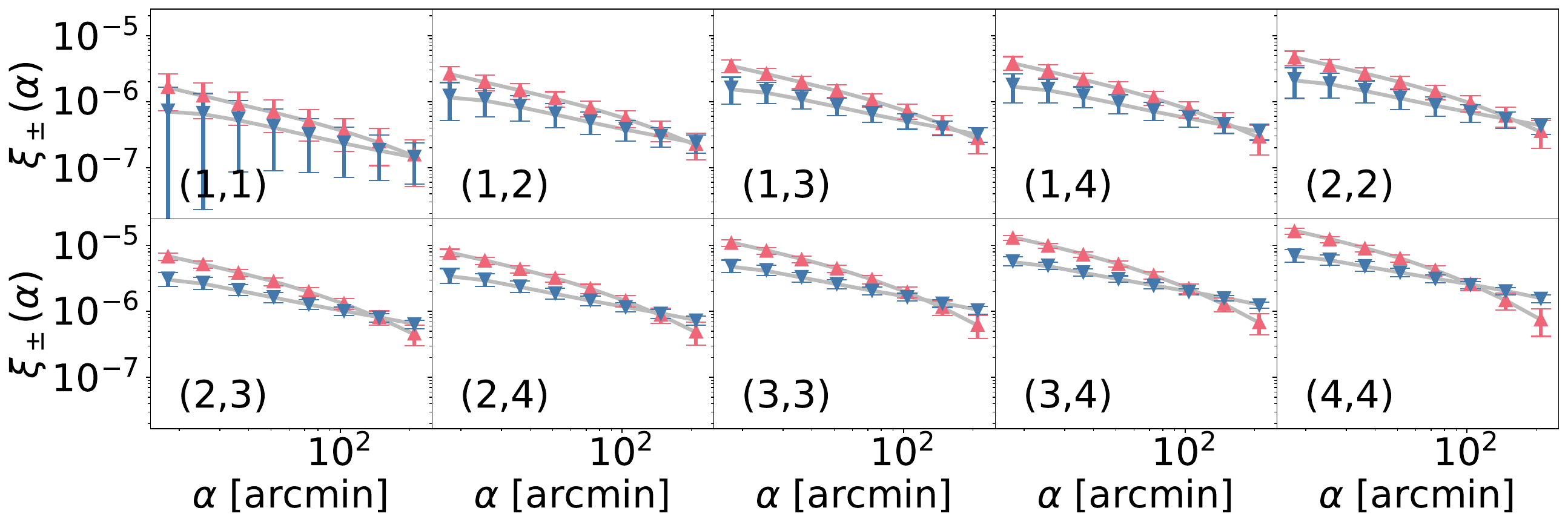}
    \caption{Comparison of the 2PCFs $\xi_+$ (\textcolor{r}{$\triangle$}) and $\xi_-$ (\textcolor{b}{$\triangledown$}) and measured in the \texttt{CosmoGridV1} noiseless simulations at fiducial cosmology against theoretical predictions (grey). The simulation measurements are without shape noise, meaning the error bars reflect cosmic variance for a DES Y3-like survey footprint.}
    \label{fig:2PCF_theory_val}
\end{figure}

\begin{figure}[htbp!]
    \centering
    \includegraphics[width=1.\linewidth]{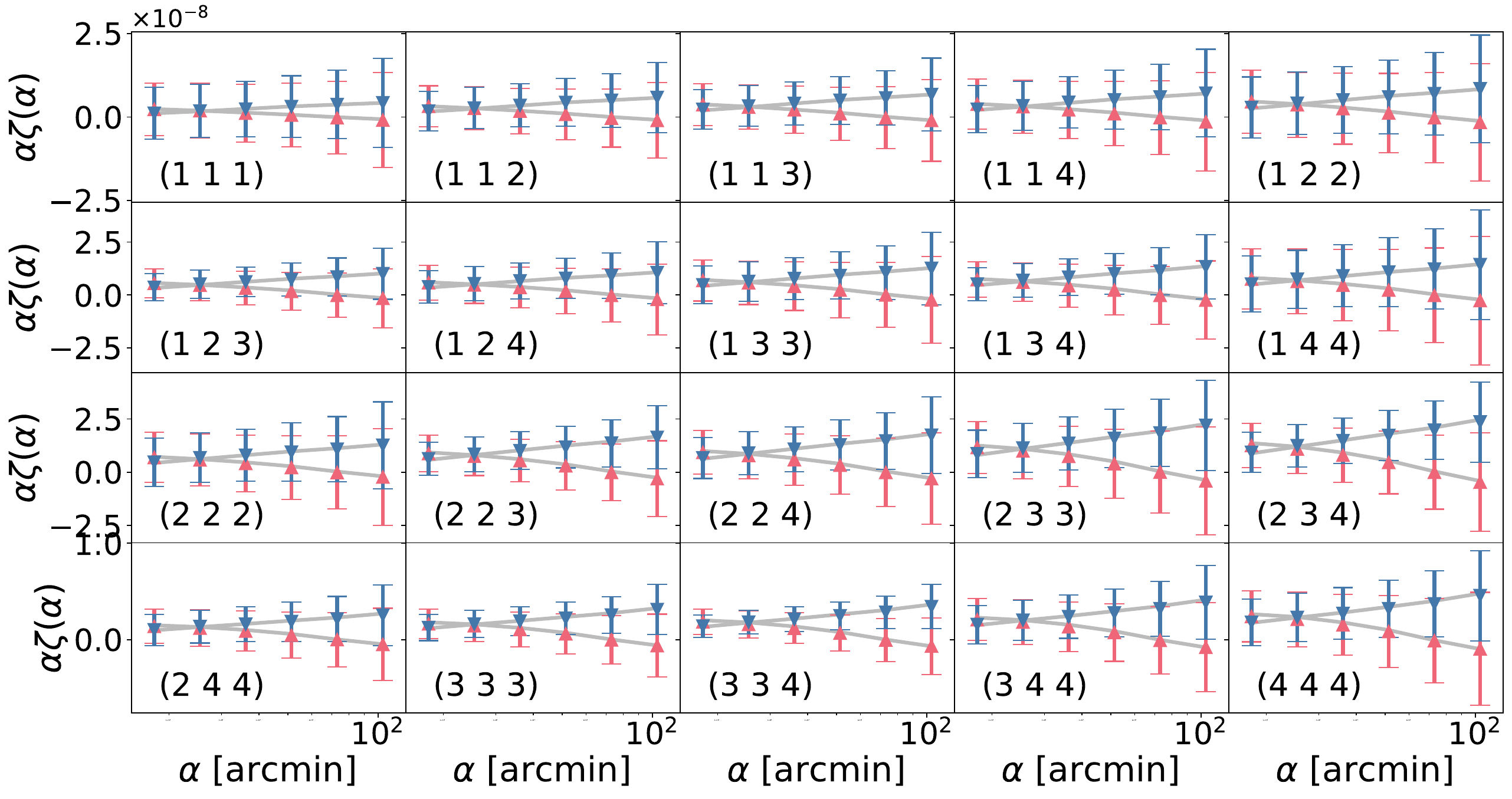}
    \caption{Same as Figure~\ref{fig:2PCF_theory_val} but for the i3PCFs $\zeta_+$ (\textcolor{orange}{$\triangle$}) and $\zeta_-$ (\textcolor{blue}{$\triangledown$}) with filter size $90'$.}
    \label{fig:i3PCF_theory_val}
\end{figure}

We have validated the forward model against theoretical predictions for the 2PCF and i3PCF. Figure~\ref{fig:2PCF_theory_val} shows these comparison for the 2PCF and Figure~\ref{fig:i3PCF_theory_val} for the i3PCF with filter size $90'$ at the fiducial cosmology. We have made the comparisons at multiple grid cosmologies too (that we do not show here) and found excellent agreement between the theoretical and simulation measurements, giving us confidence in the simulation results. The theoretical predictions were obtained using \texttt{CLASS} \cite{Lesgourgues2011, Blas2011} and \texttt{HMCode} \cite{Mead2021} following the modelling methodologies from Refs.~\cite{Halder2021, Halder2022, Gong2023}.\footnote{The code for theory computations is publicly available at \href{https://github.com/anikhalder/i3PCF}{i3PCF \faGithub}.}

\section{Simulation-based inference framework}\label{sec:sbi_pipeline}

Simulation-based inference (SBI) describes an approach in which, instead of assuming an explicit analytical form for the likelihood, it is estimated from simulated data. This is extremely powerful for inference with observables where no analytical model for the likelihood exists, such as weak lensing CDFs \cite{Anbajagane2023}, scattering transforms \cite{Gatti2024b, Gatti2025}, or field-level inference \cite{Jeffrey2021, Jeffrey2025}. To run SBI, two things are required. First, a forward model that is able to produce realistic mock observations $x$ for any parameter $\theta$ within the prior, and second, a generative model such as normalising flows (NFs) to learn this distribution of parameters $\theta$ and data $x$.

While it is possible to learn the full joint distribution $p(x,\theta)$, it is easier to train on the conditional densities $p(x|\theta)$ or $p(\theta | x)$. In this work, we use \textit{normalizing flows (NFs)} to estimate the likelihood $p(x|\theta)$, also known as \textit{neural likelihood estimation (NLE)}, which we describe in the next section. While NFs are used in this work, other machine learning architectures such as mixture density networks, can also be employed for NLE. \cite{Alsing2019}

\subsection{Neural likelihood estimation}

NFs parametrise a transformation from a base distribution $\pi(x)$ to a target distribution $p(x)$ through a series of invertible mappings. The term `flow' refers to this sequence of transformations, and `normalizing' comes from the fact that the flow transforms a complex, unknown distribution into a simple, normalized base distribution (e.g.~a standard Gaussian) for which the probability density is trivial to compute. It is made up of individual layers of artificial neurons, with each layer taking the transformed distribution from the prior layer as input and producing another distribution after performing a transformation on it, as output. For the loss function, NFs directly maximise the likelihood $ p(x|\theta)$ provided a set of training data $x_{\text{train}}$ at labelled parameters $\theta_{\text{train}}$. Practically, the task is then to learn an estimator for the likelihood function characterised by some neural network parameters $\varphi$: $\hat{p}_\varphi(x_{\text{train}}|\theta_{\text{train}})$ This is called neural likelihood estimation (NLE) and was first proposed using NFs in Ref.~\cite{Papamakarios2019}. After training the NF, one can use it as a likelihood function in sampling algorithms such as MCMC for Bayesian posterior inferences. The MCMC steps can be performed very efficiently, as evaluating even complex NFs is quick, and can be done in parallel on GPUs to facilitate multiple chains to run concurrently. This means that the most expensive part of SBI is acquiring the training dataset $\{x_{\text{train}}, \theta_{\text{train}} \}$. In Ref.~\cite{Alsing2019}, it has been discussed that the number of simulations needed for NLE can be much lower than the number of simulations used for estimating a fixed covariance in a likelihood-assumed MCMC inference, making this a viable alternative even for cases where the likelihood function is tractable. However one should note that with a low number of simulations, an effect similar to the \textit{Dodelson-Schneider factor} that impacts traditional covariance estimation, affects the learned likelihood in NLE too \cite{Homer2025}.

The computationally most expensive step of training NFs is the calculation of the Jacobians of the transformation $\varphi$. An architecture optimised to efficiently calculate these Jacobians are the so-called \textit{Masked Autoencoders for Distribution Estimation (MADEs)} \cite{Germain2015}. They are autoregressive networks, meaning that each output dimension depends only on the previous input dimension. Therefore, this transformation can be represented by a triangular matrix, making it very efficient to compute the determinant of such matrices (and hence the Jacobian). The likelihood is then given by:
\begin{equation}
\begin{aligned}
    \hat{p}_\varphi(x | \boldsymbol{\theta}) &= \mathcal{N} \left[ \mathbf{\varphi}(x, \boldsymbol{\theta}) \mid 0, \mathbf{I} \right] \times \left| \frac{\partial \mathbf{\varphi}(x, \boldsymbol{\theta})}{\partial x} \right| \\
    &= \prod_i \hat{p}_\varphi(t_i \mid t_{i-1}, \boldsymbol{\theta}),
\end{aligned}
\end{equation}
where $\varphi$ is the transformation parametrised by the MADE and $\mathcal{N}$ is the Gaussian base distribution with zero mean and identity covariance. The loss function that is minimised during training, is specified by:
\begin{equation}
    - \ln U \equiv - \sum_i^{N_{\text{sims}}} \ln \hat{p}_\varphi \left( x_i | \theta_i \right),
\end{equation}
therefore, directly maximising the likelihood. For a more detailed description of this architecture, we refer the reader to Ref.~\cite{Alsing2019}. 

A downside of MADEs is their dependence on the order of the input dimensions. This problem is solved by stacking multiple MADEs, passing the output of each to the input of the next. By shuffling the order of the inputs in between, any dependence on the order can be neglected. These constructs are then called \textit{masked autoregressive flows (MAFs)} \cite{Papamakarios2017}. They also exhibit much higher expressivity, allowing more complex distributions to be estimated. As MAFs are invertible, it is trivial to evaluate the likelihood of a given sample. In this work we use the MAFs implemented in the \texttt{sbi} Python package \cite{sbi2020}.

\subsection{Hyperparameter tuning}

\noindent To find the optimal architecture for the normalizing flows, we use the \texttt{optuna} \cite{optuna_2019} package for hyperparameter tuning. For this, we varied the number of transforms, the depth and width of the individual MADEs, the learning rate, the training batch size, and the early stopping patience. This was run for a total of 300 steps with a \texttt{TPESampler} \cite{bergstra2011} using 100 random start-up trials. We then use the model that achieved the highest likelihood on the validation set for each trial. The optimised hyperparameters are listed in Table~\ref{tab:hyperparameters}:

\begin{table}[htbp]
    \centering
    \begin{tabular}{|l||r|r|}
        \hline
        Hyperparameter & 2PCF & 2PCF + i3PCF \\
        \hline \hline
        Transforms & 10 & 8\\
        Features & 64 & 32\\
        Blocks & 3 & 2\\
        Learning Rate & $5 \times 10^{-5}$ & $1.32 \times 10^{-4}$ \\
        Batch Size & 16 &  64 \\
        Patience & 15 & 30 \\
        \hline 
    \end{tabular}
    \caption{The hyperparameters of the best performing NF for estimating the likelihood of each data vector.}
    \label{tab:hyperparameters}
\end{table}

\subsection{Inference pipeline}

We use a flat prior in the parameter space spanned by \texttt{CosmoGridV1} as described in Section~\ref{sec:forward_model}. The parameter ranges are shown in Table~\ref{tab:prior_range}. The raw, uncompressed data vector for the 2PCF-only analysis consists of 160 elements (10 tomographic bin combinations $\times$ 2 shear components of $\xi_{\pm}$ $\times$ 8 angular bins). The joint 2PCF + i3PCF data vector, including all 5 i3PCF filter sizes, has a total length of 1420. To reduce the dimensionality of our data vectors we employ the \texttt{MOPED} compression technique (\textit{Massively Optimised Parameter Estimation and Data Compression}) \cite{Heavens2000}. It linearly transforms the data vector in a way that maximises the Fisher information of the data vector at a fiducial point in parameter space. This transformation can be written as 
\begin{equation}
    y_\alpha = b^T_\alpha x,
\end{equation}
where $y_\alpha$ is the compressed vector with respect to a parameter $\alpha$, $x$ is the original data vector and $b$ the compressed \texttt{MOPED} vector. The first component of the compressed vector is given by \citep{Heavens2017}
\begin{equation}
    b_1 = \frac{C^{-1}\mu_1}{\sqrt{\mu^T_1C^{-1}\mu_1}},
\end{equation}
and the rest of the components
\begin{equation}\label{eq:moped_vec_a}
    b_\alpha = \frac{C^{-1}\mu_\alpha - \sum_\beta^{\alpha-1} (\mu^T_\alpha b_\beta)b_\beta}{\sqrt{\mu^T_\alpha C^{-1}\mu_\alpha - \sum_\beta^{\alpha-1} (\mu^T_\alpha b_\beta)^2}},
\end{equation}
where the inverse covariance $C^{-1}$, which in our case, is estimated from 16000 footprints at the fiducial cosmology and the parameter derivative
\begin{equation}
    \mu_\alpha = \frac{\partial x}{\partial \theta_\alpha},
\end{equation}
estimated from the specific set of \texttt{CosmoGridV1} simulations run for derivative computation around the fiducial cosmology. With \texttt{MOPED}, the dimensionality of the compressed vector $b$ is then the number of parameters varied on the grid, which in our case is 17.

In each case (for 2PCF or the 2PCF + i3PCF data vectors), we train a normalizing flow on the likelihood of these 17 MOPED coefficients, which form our training data $x_{\text{train}}$. However we provide as labels only 9 parameters that are of interest to us in the final posterior inference $\Omega_m, \Omega_b, H_0, n_s, \sigma_8, w_0, A_{\text{IA}}, M_c^0$, and $\nu$. Not specifying the remainder of the 8 systematic parameters (4 multiplicative shear bias and 4 photometric redshift uncertainties) as labels in the NLE amounts to implicitly marginalizing over them (for the final posterior inference) while learning the likelihood function \cite{Alsing2019b}.

For MCMC sampling of the posterior, we use an ensemble sampler implemented in \texttt{emcee} \cite{Foreman2013} with 1,000 parallel workers, each running for 13,000 steps. The first 10,000 steps are discarded as burn-in, so as to have the chains stationary around the true posterior. Further, we only use every third step of the final chain as thinning in order to reduce the autocorrelation length. Overall, we have one million samples in total per posterior. To evaluate how closely a posterior distribution obtained by MCMC sampling with the learned likelihood function (with MAFs) matches the true underlying posterior, one can use \textit{coverage tests}. Such tests verify that the inferred credible intervals contain the true parameter values with the expected frequency (e.g.~a 68\% credible interval should contain the true value in 68\% of repeated inferences). To do this, we use the \textit{Tests of Accuracy with Random Points} (\texttt{TARP}) coverage test. We refer the reader to Ref.~\cite{Lemos2023} for more details on this algorithm.

\section{Results}\label{sec:results}

In this section, we present a series of tests to validate our SBI pipeline and evaluate its performance. We first test whether the likelihood of our data vectors are compatible with a Gaussian distribution in Section~\ref{sec:results_gaussianity}. In Section~\ref{sec:including_systematics} we check at the data level for impacts of two important systematic effects: source galaxy clustering and reduced shear. We then perform coverage tests to ensure that our inferred credible intervals obtained as output from our SBI pipeline are not systematically underestimating or overestimating the true posteriors in Section~\ref{sec:coverage_results}. We then apply the pipeline to mock cosmic shear data vectors to demonstrate its ability to recover fiducial cosmological parameters and present the improvement in parameter constraints from adding the i3PCFs relative to only using 2PCFs in Section~\ref{sec:fiducial_post}.

\subsection{Gaussianity assumption of the likelihood}\label{sec:results_gaussianity}

We have investigated the widely used Gaussianity assumption of the likelihood for the cosmic shear 2PCF and the i3PCFs for our DES Y3-like survey setup. We followed a method similar to Refs.~\cite{Friedrich2021, Lehman2025}, where we calculated the $\chi^2$ distribution of the individual fiducial data vectors with respect to the mean (from the set of 16,000 fiducial mocks). If the true sampling distribution (the likelihood function) is indeed Gaussian, then the computed distribution of the $\chi^2$ from the measured data vectors would indeed also follow a true $\chi^2$ probability density function. Deviations between the two would point to a breakdown in the Gaussian likelihood assumption. We advocate that most higher-order statistics that rely on simulations to compute the covariance matrix should perform this test to easily check whether they are compatible with the Gaussian likelihood assumption. 

To ensure the measurements used for covariance and mean estimation are independent from the measurements used for calculating the $\chi^2$ values, we split the 16,000 fiducial measurements into 200 sets of 80 that are based on the same underlying full-sky simulation. We then estimate the covariance and mean from these 15,920 data vectors and calculate $\chi^2$ for the remaining 80. We do this 200 times for each set to give us 16,000 $\chi^2$ values. For each mock we calculate the $\chi^2$ following
\begin{equation}\label{eq:chi2}
    \chi^2 = (d - \left\langle d \right\rangle)^t \hat{C}^{-1} (d - \left\langle d \right\rangle),
\end{equation}
where the inverse covariance $\hat{C}^{-1}$ is multiplied by the Hartlap factor \cite{Hartlap2007}:
\begin{equation}
    \hat{C}^{-1} = \frac{n - p - 2}{n - 1} \hat{C}^{-1}_* \text{ for } p < n-2.
\end{equation}
Here $\hat{C}_*$ is the sample covariance matrix estimated directly from the data, $n$ is the number of measurements, and $p$ is the number of degrees of freedom (in our case, the number of data points). 

\begin{figure}[htbp!]
    \centering
    \begin{subfigure}{0.5 \textwidth}
        \includegraphics[width=\linewidth]{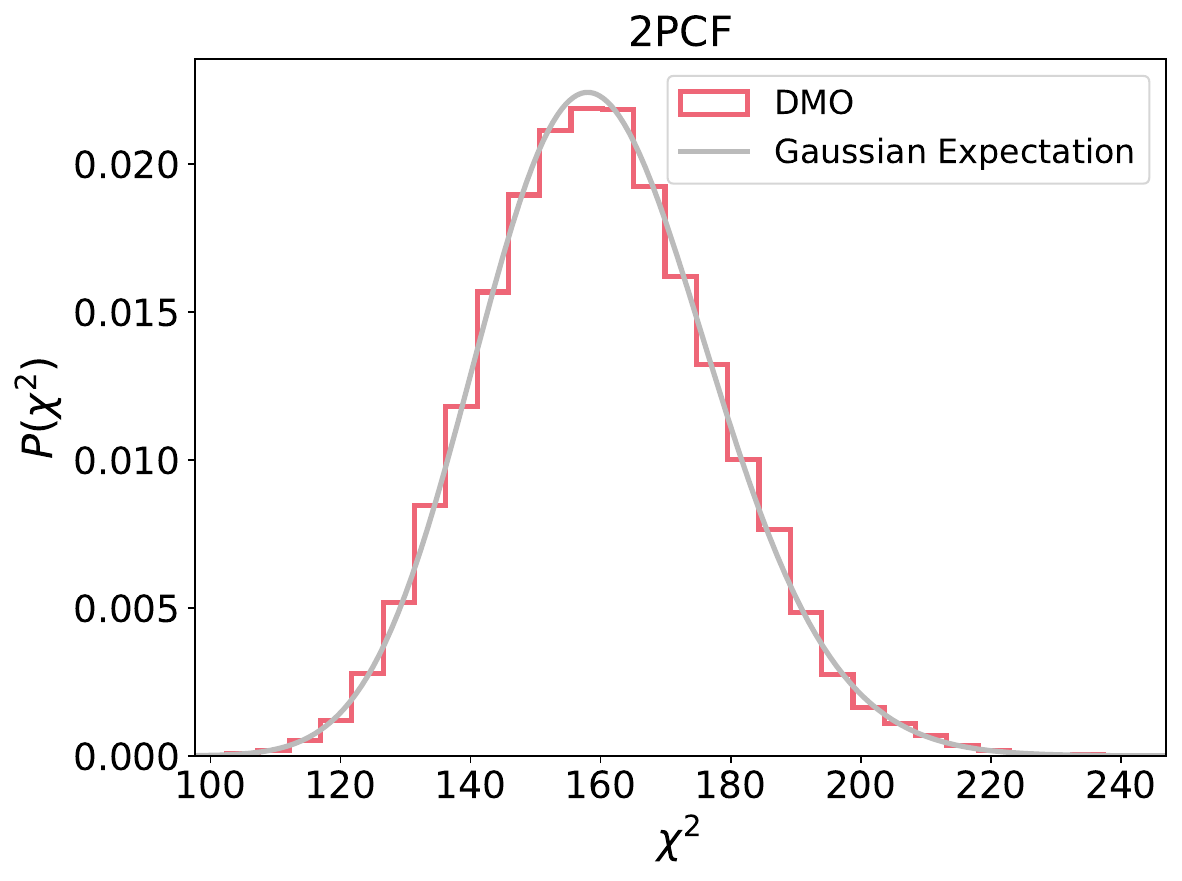}
        \caption{2PCF}
        \label{fig:chi2_2PCF}
    \end{subfigure}%
    \begin{subfigure}{0.5 \textwidth}
        \includegraphics[width=\linewidth]{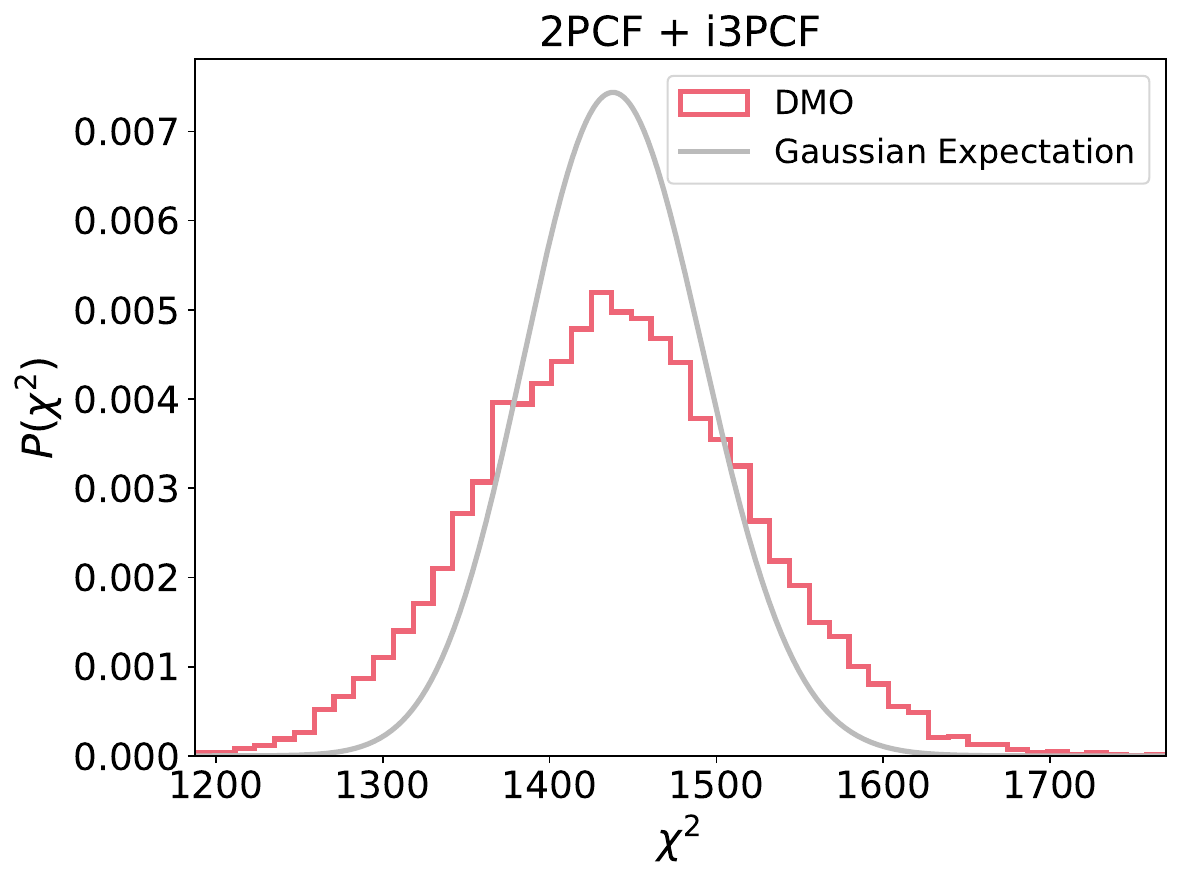}
        \caption{2PCF + i3PCF (all filters)}
        \label{fig:chi2_full}
    \end{subfigure}
    \caption{Histogram of $\chi^2$ values for the 16,000 fiducial measured full data vectors, without \texttt{MOPED} compression, comparing dark matter only simulations (red) with the expected $\chi^2$ distribution for a likelihood assumed to be Gaussian (black).}
    \label{fig:chi2}
\end{figure}

Figure~\ref{fig:chi2} shows histograms of these $\chi^2$ values for measurements evaluated at the fiducial cosmology using dark matter only simulations. To ensure a realistic representation of the sampling distribution, these simulated measurements include standard shape noise as well as the full suite of systematic effects described in Section \ref{sec:forward_model}. Figure~\ref{fig:chi2_2PCF} shows the results for the 2PCF only and Figure~\ref{fig:chi2_full} for the combined data vector of 2PCF with the i3PCF (using all filter sizes [$50', 70', 90', 110', 130'$]). We see that for the 2PCF, the distribution closely follows a $\chi^2$ distribution with the degrees of freedom given by the length of the data vector. This indicates that the likelihood of the 2PCF is compatible with a Gaussian distribution. On the other hand, for the full data vector (2PCF + i3PCF), we find that the computed $\chi^2$ histogram deviates from what one would expect from the Gaussian expectation. Therefore, we conclude that the likelihood of the full 2PCF + i3PCF using all 5 filter sizes cannot be well approximated by a Gaussian function.

Specifically, in Appendix~\ref{sec:gaussianity} Figure~\ref{fig:chi2_i3PCF} we also show the $\chi^2$ for the individual filter sizes of the i3PCF and the combined i3PCF data vector (without the 2PCF). There, we find that the likelihood of the i3PCF data vector becomes increasingly non-Gaussian for larger and larger filter sizes. This can be attributed to the fact that for larger i3PCF filter scales (and for large modes in general), the effective number of quasi-independent patches (modes) within a fixed survey area over which one averages to estimate the i3PCFs, becomes fewer. This results in a departure from the ``Gaussianization" one would expect from the central-limit theorem when averaging over many modes, and therefore, a more skewed sampling distribution of the i3PCF data vector for larger filter scales.

However, when compressing the data vector using \texttt{MOPED}, the likelihood becomes Gaussian again, even using all i3PCF filters (see Figures~\ref{fig:chi2_moped_2PCF} and \ref{fig:chi2_moped_full}). We have also investigated the maximum number of filters that can be added to the 2PCF that is still compatible with the Gaussian assumption and have found that the likelihood of 2PCF + i3PCF for filter sizes $50', 70'$ (without any \texttt{MOPED} compression) to still be approximated as a Gaussian (see Figure~\ref{fig:chi2_gausstest_1}). Looking at the data vector with the i3PCF $90'$ filter size added, one can see the deviation from the $\chi^2$ distribution (see Figure~\ref{fig:chi2_gausstest_2}). 

The key advantage of SBI is its ability to move beyond the Gaussian likelihood assumption and to capture the true sampling distribution directly. For our combined analysis, we therefore adopt the \texttt{MOPED} compressed 2PCF + i3PCF data vector across all filter sizes. While this compressed data vector is consistent with the Gaussian assumption, we choose it as our summary statistic in our SBI framework primarily for computational efficiency in training the NFs for NLE.

\subsection{Impact of reduced shear and source galaxy clustering}\label{sec:including_systematics}

To test for the impact of source galaxy clustering (Section~\ref{sec:systematics_sc}) and reduced shear (Section~\ref{sec:systematics_red}), we have created additional simulation sets that include these effects at the fiducial cosmology. In this section, we only show the tomographic bin combination 1,1,4, since it is most affected by these systematic results (the largest relative difference). For the effect on all correlations see Appendix \ref{sec:systematics_full}.

\begin{figure}[htbp!]
    \centering
    \includegraphics[width=1.\linewidth]{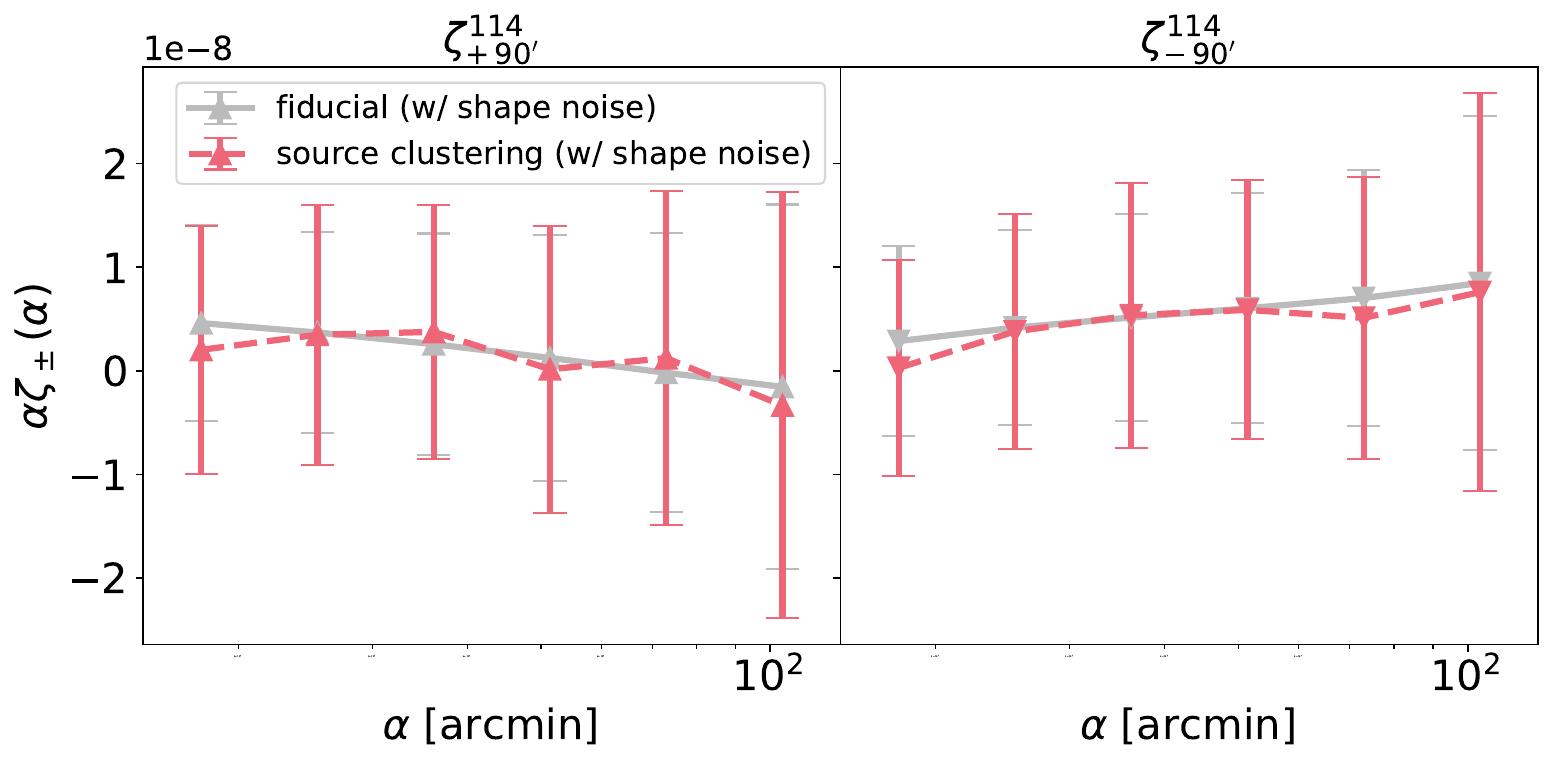}
    \caption{The effect of source clustering on the i3PCFs ($\zeta_+$ (\textcolor{r}{$\triangle$})and $\zeta_-$ (\textcolor{r}{$\triangledown$})) for filter $90'$ compared to the fiducial measurement ($\zeta_+$ (\textcolor{k}{$\triangle$}) and $\zeta_-$ (\textcolor{k}{$\triangledown$})). Shown are the cross-correlations of tomographic bins 1, 1, and 4, as these i3PCFs are most impacted. A linear source galaxy bias $b_g = 1$ was used for this test.}
    \label{fig:i3PCF_source_clustering} 
\end{figure}

\begin{figure}[htbp!]
    \centering
    \includegraphics[width=1.\linewidth]{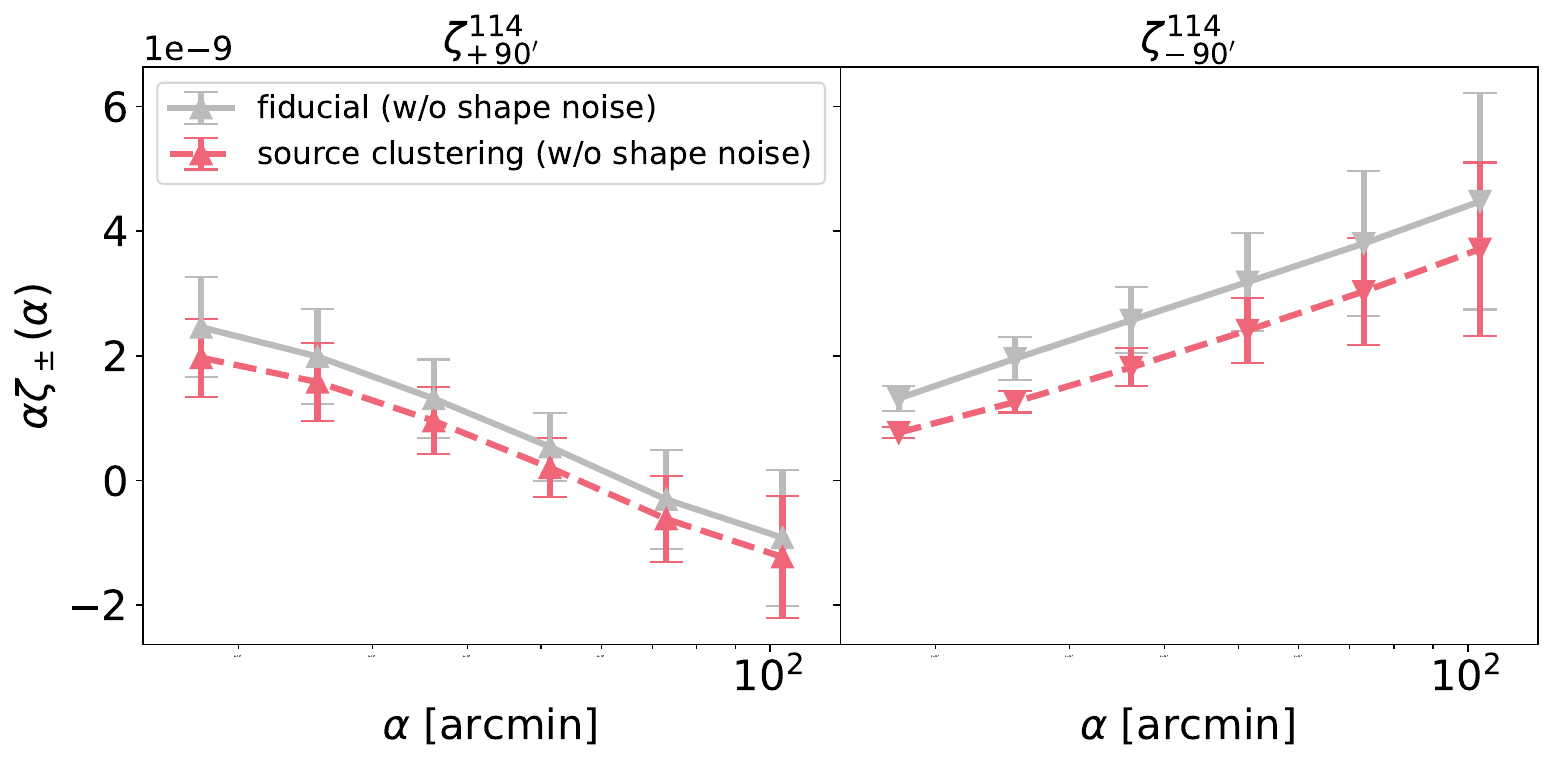}
    \caption{Same as Figure \ref{fig:i3PCF_source_clustering}, but we consider the effect of source clustering without the shape noise term.}
    \label{fig:i3PCF_source_clustering_noiseless}
\end{figure}

The source clustering simulation set is made up of 400 footprints with a linear source galaxy bias $b_g = 1$ \cite{Gatti2024}. These include shape noise and are created following Eq.~\eqref{eq:mock_sc}. In Figure~\ref{fig:i3PCF_source_clustering}, we can see that while this stays far below the variance due to shape noise, source clustering still causes a slight suppression of the i3PCF. To quantify this, we calculate $\chi^2 = 31.8$ for the full 2PCF + i3PCF combined data vector. Dividing by the degrees of freedom yields a $\chi^2_\text{red}=0.022$. We therefore conclude that the impact of SC on the i3PCF can be safely neglected for inference in stage-III surveys.

Investigating the impact of SC further, we also tested a third set of 400 simulation footprints with source clustering, but without shape noise. Looking at Eq.~\eqref{eq:mock_sc}, this means that only the first term is relevant. From Figure~\ref{fig:i3PCF_source_clustering_noiseless} we can see that the SC signal is still slightly suppressed. However, this is still far below cosmic variance, with a $\chi^2_\text{red}=5.6\times10^{-4}$ for the i3PCF only. This slight difference is caused by the overlap in the lensing kernel with the source galaxy distribution (see Figure~\ref{fig:DES_nz}), leading to suppression that is most prominent in the lower tomographic bins.

\begin{figure}[htbp!]
    \centering
    \includegraphics[width=1.\linewidth]{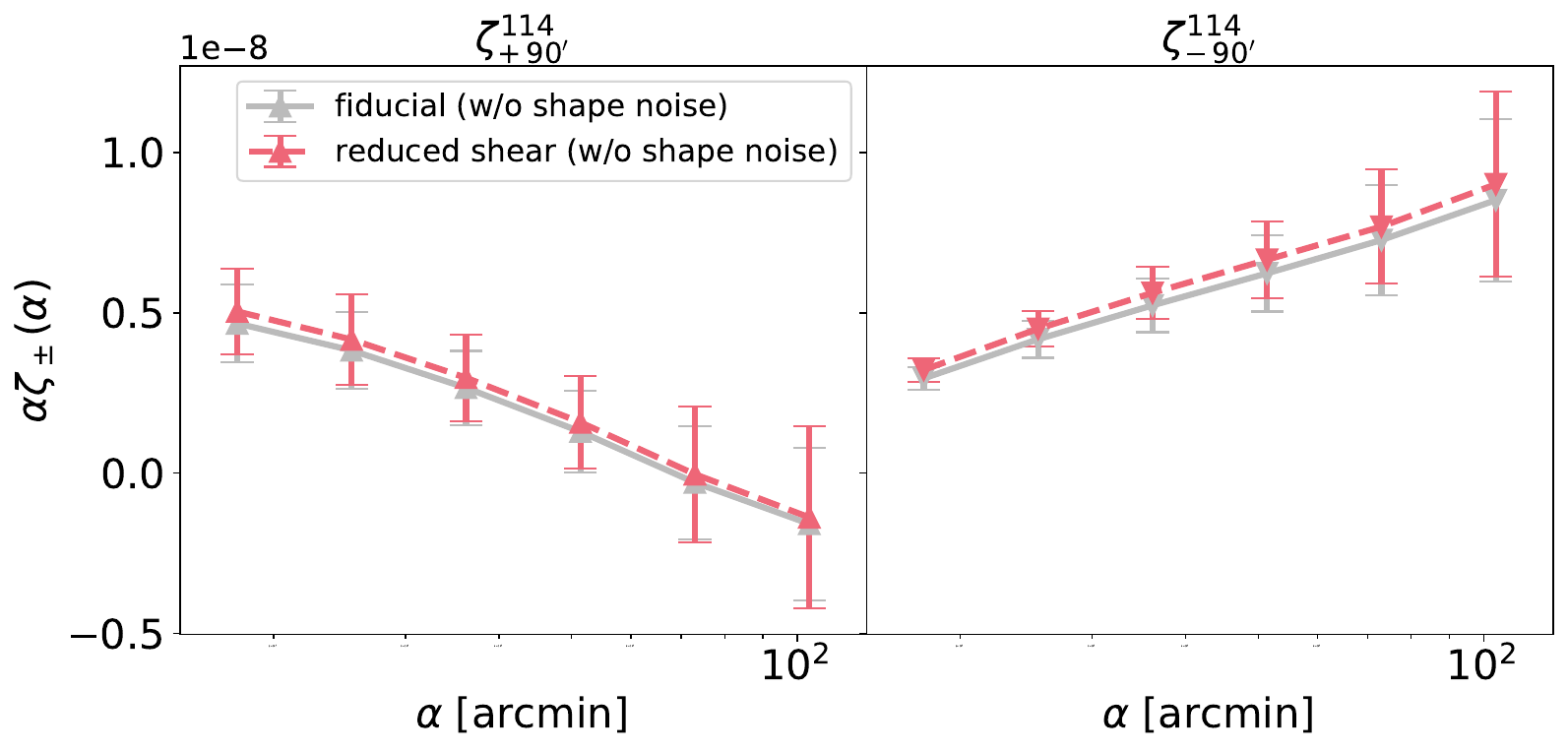}
    \caption{Measurements of the i3PCF at the noiseless fiducial maps using the approximation $\gamma / (1-\kappa) \approx \gamma$ ($\zeta_+$ (\textcolor{k}{$\triangle$}) and $\zeta_-$ (\textcolor{k}{$\triangledown$})), compared to the full calculation using reduced shear ($\zeta_+$ (\textcolor{r}{$\triangle$}) and $\zeta_-$ (\textcolor{r}{$\triangledown$})). We show the cross-correlation of the tomographic bins 1, 1, and 4, as these i3PCFs are most impacted.}
    \label{fig:i3PCF_reduced_shear}
\end{figure}

Figure~\ref{fig:i3PCF_reduced_shear} compares the measured i3PCF with $90'$ filter radius, with and without reduced shear. The reduced shear measurements come from 40 footprints without shape noise. We see that the effect is far below cosmic variance and can thus be safely neglected. The effect of reduced shear leads to a $\chi^2_\text{red} = 7.5\times10^{-4}$ on the i3PCF alone and $\chi^2_\text{red} = 1.74\times10^{-3}$ on the combined 2PCF + i3PCF data vector.

\subsection{Coverage tests}\label{sec:coverage_results}

To assess the accuracy and reliability of the inferred posterior distributions from our SBI pipeline, we performed the \texttt{TARP} \cite{Lemos2023} coverage test. We used $N=500$ measurements from footprints of the wide cosmology grid (see Section~\ref{sec:CosmoGrid}), meaning they are distributed following our prior. These 500 measurements were excluded from the training set, so the NFs did not see these during training. We then ran our full inference pipeline on each of these 500 measurements for both the 2PCF and 2PCF + i3PCF models to obtain the posterior distributions for the cosmological parameters. We calculate the coverage for 100 reference points in each posterior sample to obtain a bootstrap estimate of the variance.

\begin{figure}[htbp!]
    \centering
    \begin{subfigure}{0.5 \textwidth}
        \includegraphics[width=\linewidth]{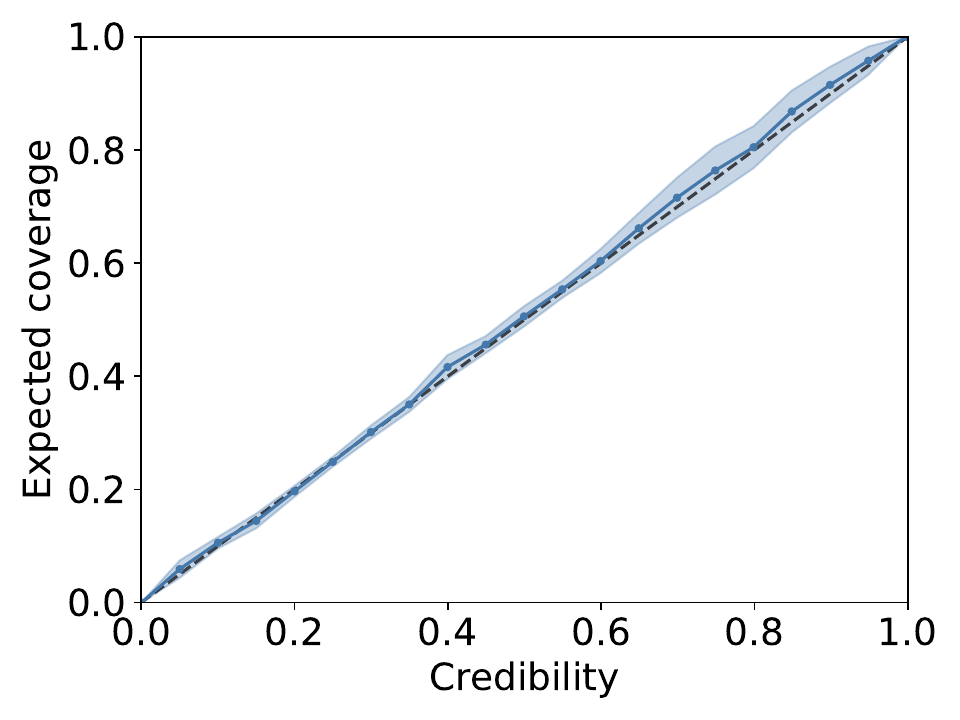}
        \caption{Coverage using 2PCF data vector.}
        \label{fig:tarp_2PCF}
    \end{subfigure}%
    \begin{subfigure}{0.5 \textwidth}
        \includegraphics[width=\linewidth]{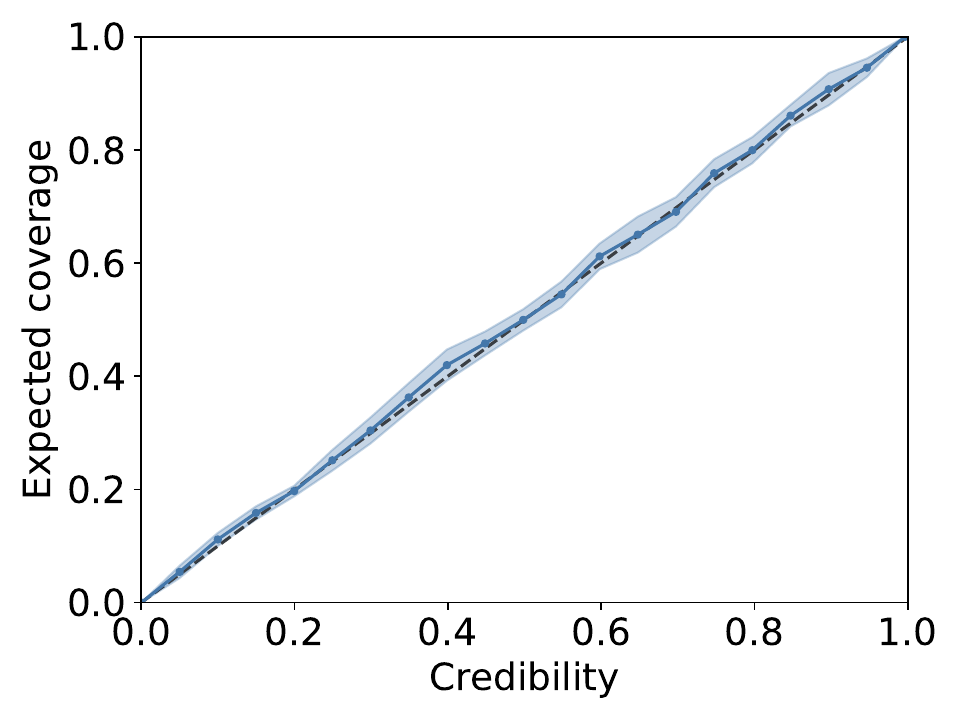}
        \caption{Coverage using 2PCF + i3PCF data vector.}
        \label{fig:tarp_i3PCF}
    \end{subfigure}
    \caption{Probability-probability plot from the \texttt{TARP} coverage test of our SBI pipeline. The $x$-axis shows the credible level $\alpha$, and the $y$-axis shows the empirical coverage probability $P(<\alpha)$, defined as the fraction of the 500 mock inferences for which the true parameter value is enclosed within the $\alpha$-level credible interval. The one-to-one line (black dashed) indicates perfect statistical coverage. The shaded area includes the $1\sigma$ error from the bootstrap estimate.}
    \label{fig:tarp}
\end{figure}

Figure~\ref{fig:tarp} shows the results of our coverage tests for the full 9-dimensional parameter space (with the remaining 8 systematic parameters implicitly marginalized over). The empirical coverage probability of our pipeline closely follows the diagonal line, including the ideal case within 1$\sigma$.\footnote{We further note that the variance using bootstrap estimates is often underestimated \cite{Friedrich2016}.} This demonstrates that the posteriors produced by our method are statistically well-calibrated and that our posteriors are not systematically underestimating or overestimating the true posterior widths.

\begin{figure}[htbp!]
    \centering
    \includegraphics[width=\linewidth]{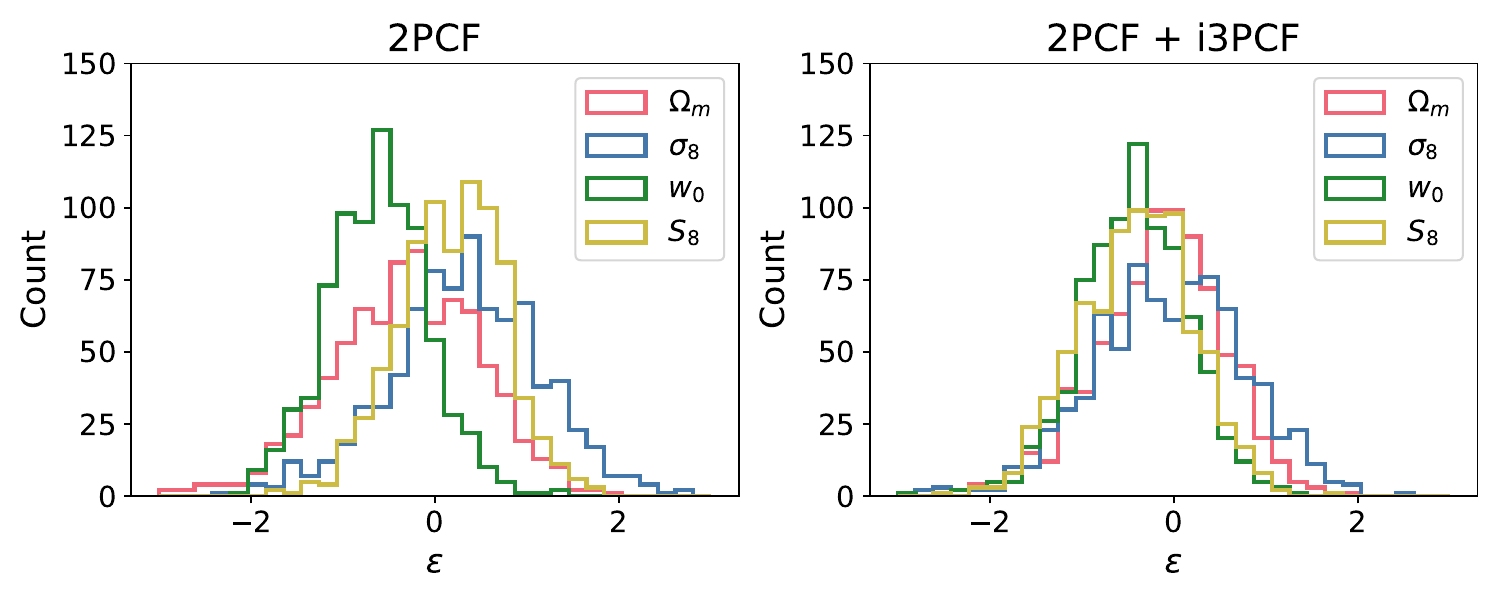}
    \caption{Histogram of the distribution of the difference between the posterior median and the true cosmology divided by half the width of the 68\% credible interval, $\varepsilon$ (see Eq.~\eqref{eq:bias}), for 100 fiducial simulations. We show these for $\Omega_m, \sigma_8, w_0,$ and $S_8$ parameters for 2PCF alone (left panel) and 2PCF + i3PCF (right panel) joint constraints.}
    \label{fig:bias_hist}
\end{figure}

In addition to the \texttt{TARP} test, which tests the coverage globally across the parameter space, we have also tested for the local coverage and performance of the SBI posteriors on noisy mocks at the fiducial cosmology for an intuitive check using
\begin{equation}\label{eq:bias}
    \varepsilon \equiv \frac{\theta_{\text{median}} - \theta_{\text{truth}}}{\frac{1}{2} \left| \theta_{84\%} - \theta_{16\%} \right|},
\end{equation}
where $\theta_{\text{truth}}$ is the known true parameter value from the simulation. This metric, $\varepsilon$, quantifies the bias of the posterior median in units of the statistical uncertainty (the half-width of the 68\% central credible interval). Figure~\ref{fig:bias_hist} shows a histogram of $\varepsilon$ for 800 fiducial noisy simulations. We can see that at the fiducial cosmology, $\Omega_m, \sigma_8$ and $S_8 = \sigma_8 \sqrt{\Omega_m/0.3}$ are recovered well both using the 2PCF alone and jointly with the i3PCF. However, $w_0$ is biased low by $0.62 \epsilon$ with respect to zero when only using the 2PCF. We presume that this may be caused by the irregularly shaped prior for $w_0$ (a resulting feature of the parameter space used for the \texttt{CosmoGridV1} suite) and the 2PCF not able to fully constrain it. Adding the i3PCF shifts the median bias to $0.41 \epsilon$, depicting that adding higher-order information helps to debias the constraints to some extent. We note, though, that this test only the 1D marginal posterior distributions and can suffer from projection effects.

\subsection{Simulated inference and parameter improvement on mock data}\label{sec:fiducial_post}

\begin{figure}[htbp!]
    \centering
    \includegraphics[width=\linewidth]{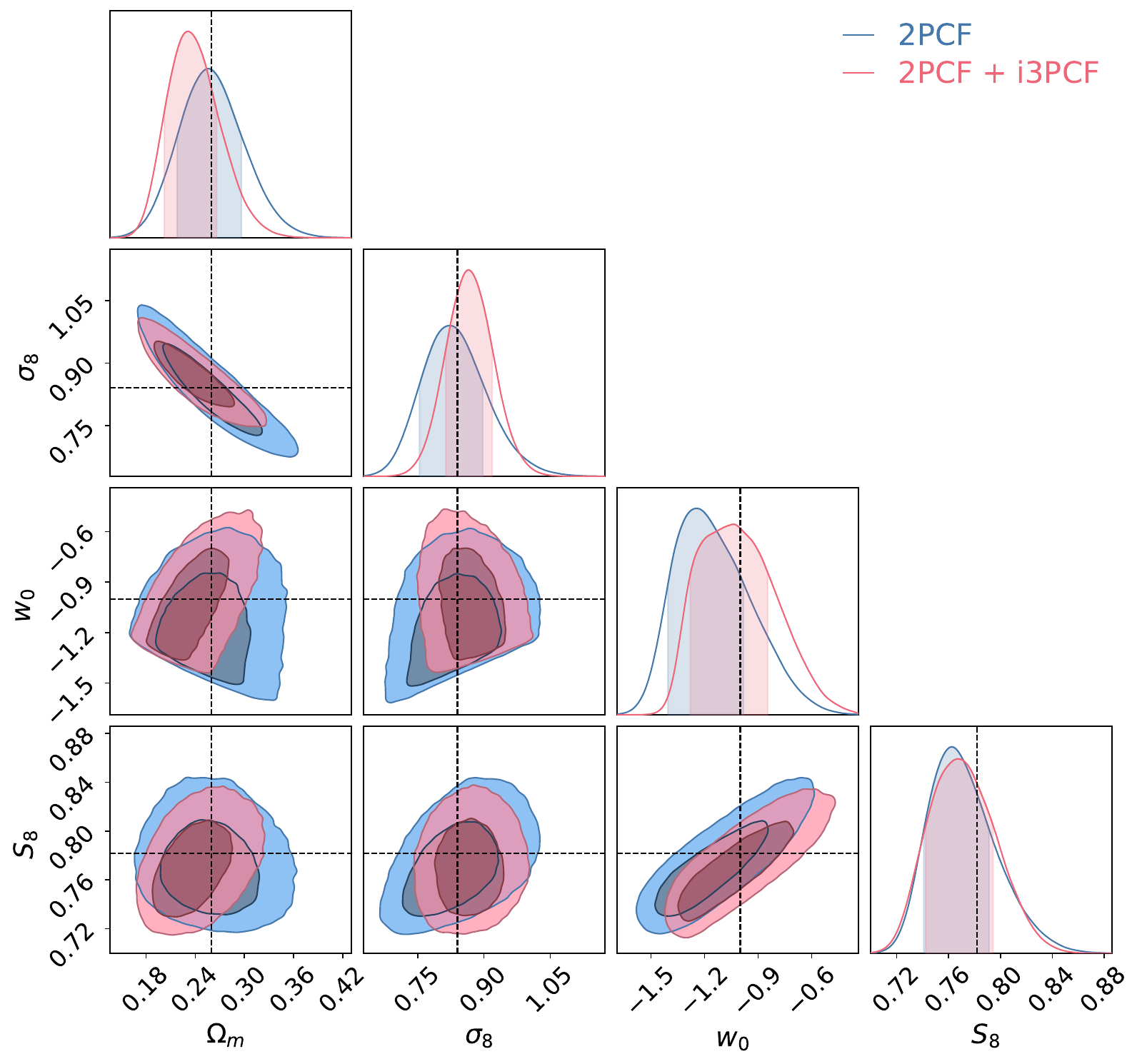}
    \caption{Posterior distributions for the cosmological parameters, inferred from a single mock noisy data vector. The model using 2PCFs is shown in blue, and the model additionally using the i3PCF is shown in red. The 1D marginalized posteriors are shown on the diagonal, and the 2D joint posteriors are shown as 68\% and 95\% credible contours. The true fiducial parameter values are indicated by the dashed black lines.}
    \label{fig:post_fiducial}
\end{figure}

To demonstrate the performance of the pipeline in a realistic scenario, we apply it to a single mock data vector generated at the fiducial cosmology, with a noise realization corresponding to a DES Y3-like survey. Figure~\ref{fig:post_fiducial} presents the posterior distribution for the parameters $\Omega_m$, $\sigma_8$, and $w_0$, inferred by our pipeline. We are not showing the baryonic and intrinsic alignment parameters, as the posterior in these parameters is not tightly constrained and closely resembles the prior. For the full posterior, including a chain sampling the prior probability, see Appendix~\ref{sec:full_post}. We find that the true fiducial cosmology for $\Omega_m$, $\sigma_8$, and $S_8$ is recovered without bias, as can also be seen in Table~\ref{tab:model_params}.

The dark energy equation of state, $w_0$, is only weakly constrained, as is common for weak lensing analyses without complementary probes (such as galaxy clustering). The posterior for $w_0$ is broad and shows some flattening towards the upper bound of our prior, indicating a `projection effect' where the data cannot rule out values beyond the chosen prior range. The added information from the i3PCF does therefore not decrease the posterior width significantly, but reduces the projection effects.

\begin{table}[htbp]
    \centering
    \begin{tabular}{ccccc}
        \hline
		Model & $\Omega_m$ & $\sigma_8$ & $w_0$ & $S_8$ \\ 
		\hline
        Fiducial Cosmology & $0.26$ & $0.84$ & $-1.0$ & $0.78$ \\
		2PCF & $0.259^{+0.041}_{-0.037}$ & $0.831^{+0.076}_{-0.067}$ & $-1.17^{+0.25}_{-0.19}$ & $0.769^{+0.029}_{-0.022}$ \\ 
		2PCF + i3PCF & $0.237^{+0.035}_{-0.029}$ & $0.867^{+0.053}_{-0.051}$ & $-1.02^{+0.24}_{-0.21}$ & $0.77^{+0.027}_{-0.024}$ \\ 
        \hline
    \end{tabular}
    \caption{The results of each inference pipeline for the cosmological parameters $\Omega_m, \sigma_8$, $w_0$ and $S_8$. The `true' fiducial cosmology is shown in the first line.}
    \label{tab:model_params}
\end{table}

To quantify the advantages of adding the i3PCF to the 2PCF, we quantify the relative improvements on the constraints on some parameters $\theta$ obtained from inference on the joint data vector relative to the inference on 2PCF alone. Therefore, we run both (2PCF and joint) inference pipelines on the same 800 independent fiducial noisy footprints. We then calculate the figure of merit (FoM) $= 1/\sqrt{\det\left| \text{Cov} (\theta) \right|}$ for each posterior sample. Using these we evaluate the relative improvement for each footprint using:

\begin{equation}
    \delta_\text{FoM} \equiv \frac{\text{FoM}(\text{2PCF + i3PCF}) - \text{FoM}(\text{2PCF})}{\text{FoM}(\text{2PCF})}.
\end{equation}

\begin{figure}[htbp!]
    \centering
    \includegraphics[width=0.6\linewidth]{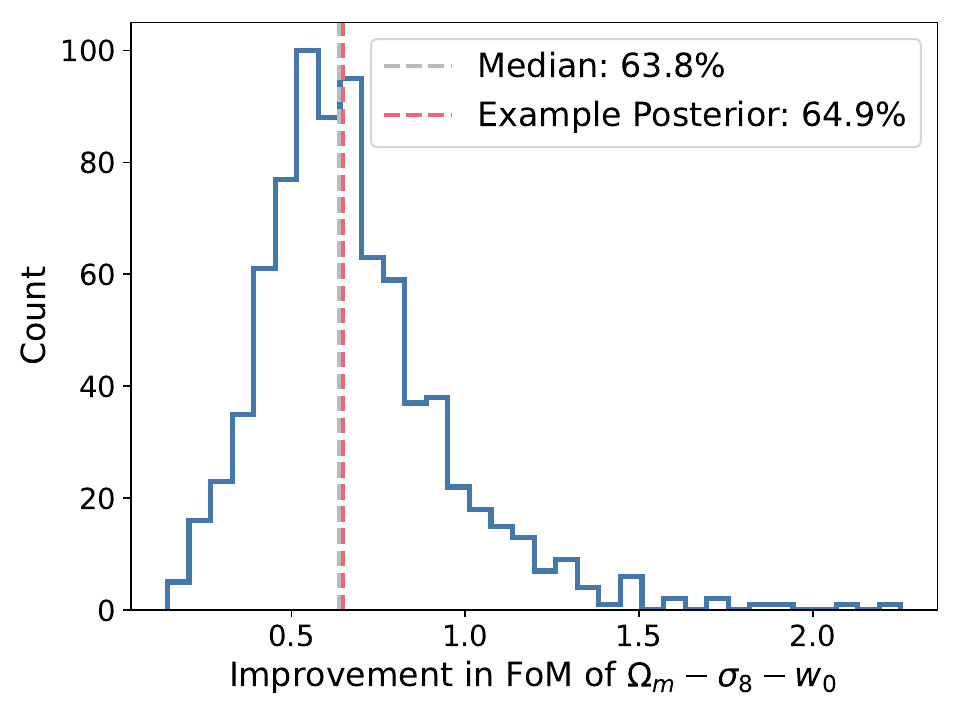}
    \caption{A histogram of the improvement in the FoM for 800 fiducial simulations. We also show the median improvement in black, and the improvement seen in the posterior of the example simulated inference shown in Figure~\ref{fig:post_fiducial} in Section~\ref{sec:fiducial_post} in red.} 
    \label{fig:HOS_improvements}
\end{figure}

\begin{figure}[htbp!]
    \centering
    \includegraphics[width=\linewidth]{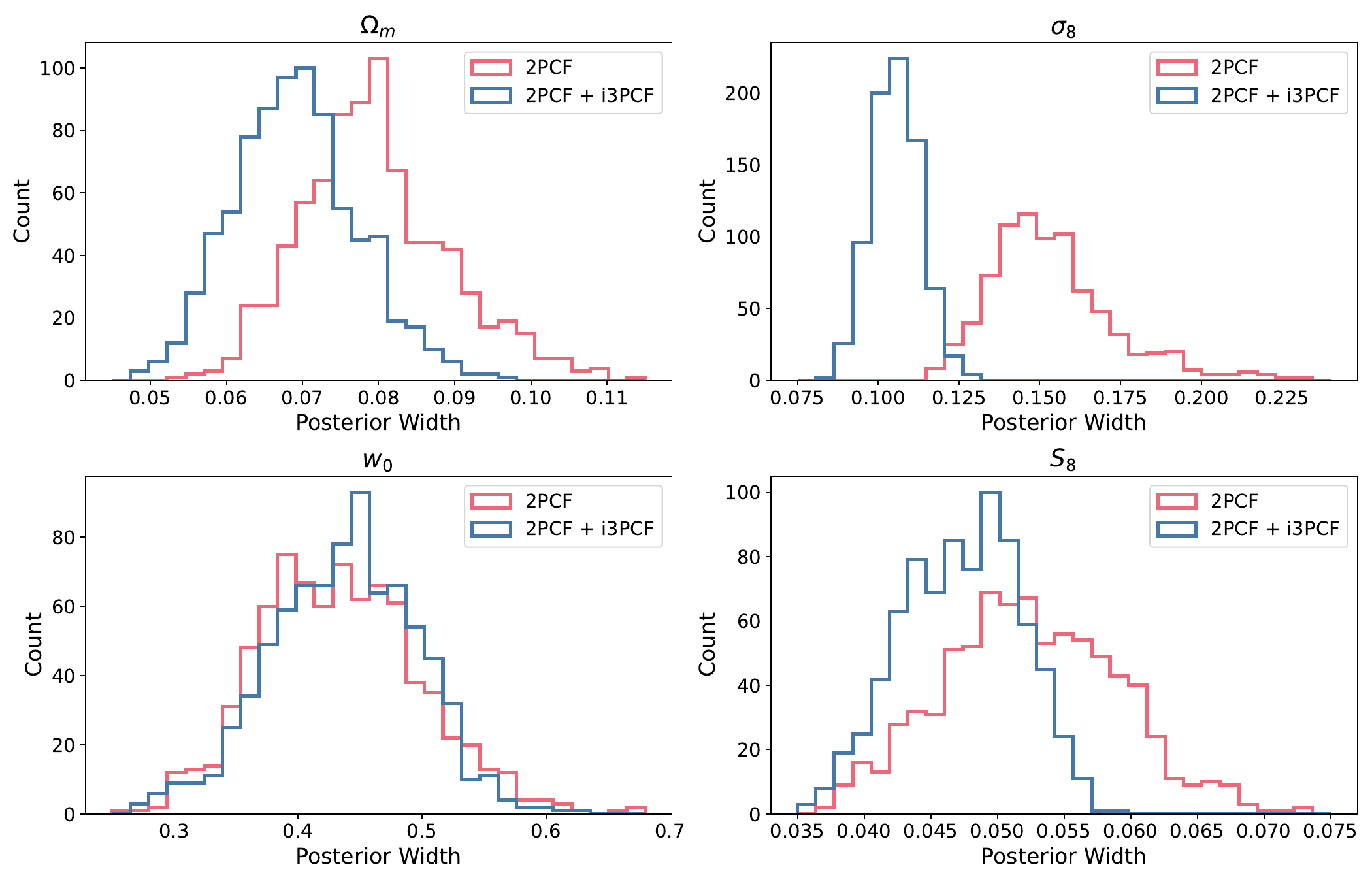}
    \caption{Histograms for the distribution of the width of the 68\% credible interval for $\Omega_m, \sigma_8, w_0,$ and $S_8$ for both 2PCF alone and 2PCF + i3PCF joint analyses.}
    \label{fig:posterior_widths}
\end{figure}

Figure~\ref{fig:HOS_improvements} shows a histogram of the improvements in the FoM from adding the i3PCF. As can be seen the improvement strongly varies for different noise realisations. The improvements for the posterior shown in Figure~\ref{fig:post_fiducial} are close to the median values for the improvements, showing a typical result one would get from adding the i3PCF.
The analysis demonstrates a substantial and consistently positive impact from including the i3PCF, with a median FoM improvement of 63.8\%. The histogram in Figure~\ref{fig:HOS_improvements} shows a wide distribution of these improvements, and crucially, all are positive. The observed variance is in line with expectations from statistical fluctuations. In any given realization, random noise and the influence of priors can sometimes lead to an unusually strong constraint for the 2PCF-only analysis. In these instances, the additional information from the i3PCF provides a smaller, yet still positive, improvement. The key result is the distribution's strong positive skew, which underlines that, on average and in every case, the inclusion of the i3PCF provides a significant improvement in constraining power relative to the 2PCF alone.
To directly visualize the impact of the i3PCF, Figure~\ref{fig:posterior_widths} shows the distribution of the 68\% credible interval widths for $\Omega_m, \sigma_8, w_0,$ and $S_8$ across the 800 fiducial simulations. For all except $w_0$, the distribution for the joint analysis is clearly shifted towards smaller values compared to the 2PCF-only case depicting the tightening of parameter constraints from adding the i3PCF. The improvement is most pronounced for $\sigma_8$, where the median interval width shrinks by approximately 31.0\%.

To investigate the improvement for the individual parameters, we also calculate the relative improvement in the width of the 68\% credible interval $\theta_{68\%}$:

\begin{equation}
    \delta(\theta_{68\%}) \equiv \frac{\theta_{68\%}(\text{2PCF + i3PCF}) - \theta_{68\%}(\text{2PCF})}{\theta_{68\%}(\text{2PCF})}.
\end{equation}

\begin{figure}[htbp!]
    \centering
    \includegraphics[width=\linewidth]{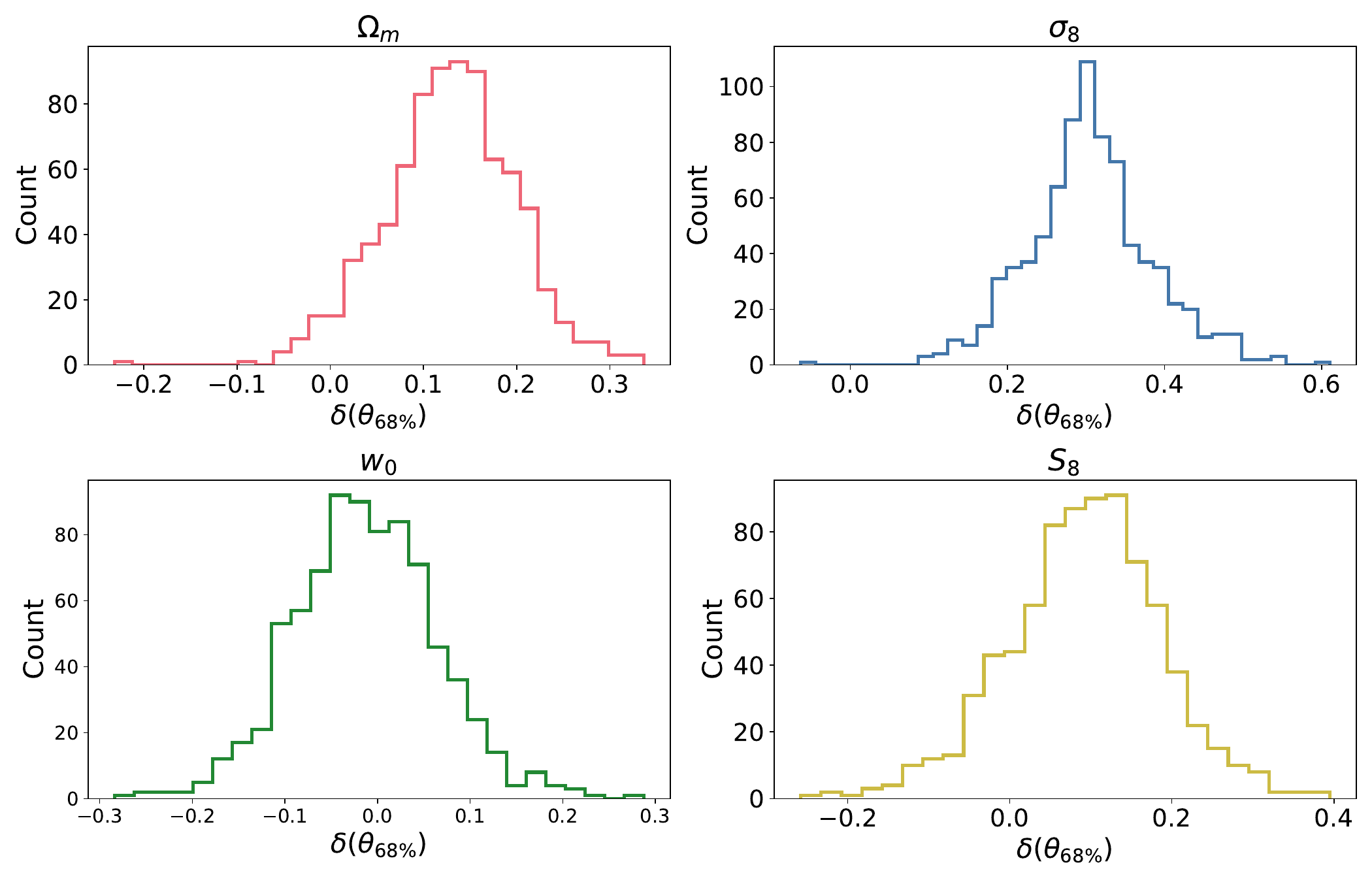}
    \caption{Histograms for the distribution of improvements in the width of the 68\% credible interval for $\Omega_m, \sigma_8, w_0,$ and $S_8$.}
    \label{fig:perc_improvements}
\end{figure}

Figure~\ref{fig:perc_improvements} shows that the median improvement is positive for all parameters, with the most prominent improvement in $\sigma_8$. This is in line with theoretical expectations as the i3PCF at first order on large scales is proportional to $\sigma_8^4$ while the 2PCF is sensitive to $\sigma_8^2$, hence the former having increased sensitivity. Compared to the FoM for the 3-D posterior, the improvement in these 1-D marginals scatters much more between realisations. This is due to random noise along with the impact of priors (e.g.~inference preferring values close to prior boundary), which can lead to the individual 2PCF + i3PCF constraints aligning in a way that happens to degrade the joint constraints. In these marginals, it can also happen that constraints become worse for individual parameters. This might be related to projection effects from the prior, as this most often happens for $w_0$ (see the prior for $w_0$ in Appendix~\ref{sec:full_post}).

\section{Conclusions}\label{sec:conclusions}

Maximizing the scientific return from ongoing Stage-IV cosmic shear surveys like Euclid and Vera Rubin Observatory's LSST is a primary goal for cosmological research over the next decade that will require the extensive analyses of higher order statistics (HOS) alongside 2PCFs. However, key challenges lie in the modelling of systematic effects in HOS. Moreover the likelihood of cosmic shear HOS can deviate significantly from the widely-used Gaussian assumption, implying that standard analyses based on a Gaussian likelihood approximation are suboptimal and risk yielding biased cosmological constraints. 

In this work, we have presented \texttt{SBi3PCF}, a simulation-based inference (SBI) framework for analysing the integrated 3-point shear correlation function (i3PCF), a particular cosmic shear HOS. Using neural likelihood estimation, our pipeline learns the full likelihood of the i3PCF jointly with the 2PCF from forward simulations of the cosmic shear field, including systematic effects (detailed in Section~\ref{sec:forward_model}), thereby avoiding any assumptions about its functional form. We have subjected our pipeline to a series of rigorous tests to demonstrate its validity and performance. Our main findings are as follows:

\begin{enumerate}[label=(\roman*)]
    \item While the likelihood of the 2PCF is well-approximated by a Gaussian distribution, the joint likelihood of the 2PCF + i3PCF deviates significantly from Gaussianity, especially for larger aperture filter sizes ($>90'$). This highlights the need for likelihood-free approaches like SBI for robust HOS analysis with the i3PCF (Section~\ref{sec:results_gaussianity}).
    \item The effect of source galaxy clustering and reduced shear on the i3PCF is negligible for a DES Y3-like survey. The correlation between the signal and the variance of the noise, as well as the effect of the non-isotropic $n(z)$, remains far below the noise level. The total impact of source galaxy clustering only leads to a $\chi^2_\text{red}=0.022$ and the effect of reduced shear to a $\chi^2_\text{red}=0.002$.
    \item Our SBI pipeline, built on a masked autoregressive flow for performing neural likelihood estimation, passes several statistical stress-tests. The \texttt{TARP} coverage test confirms that our posteriors are statistically well-calibrated, ensuring the credible intervals we report are reliable. In addition, local coverage tests conducted at the fiducial cosmology provide further evidence supporting the reliability of our SBI framework  (Section~\ref{sec:coverage_results}).
    \item When applied to a mock DES Y3-like data vector, our pipeline successfully recovers the true fiducial cosmological parameters without any significant bias, demonstrating the accuracy of our entire analysis from simulation to inference. (Section~\ref{sec:fiducial_post}).
    \item The inclusion of the i3PCF to the 2PCF yields a significant gain in constraining power. Across 800 independent noise realizations, we find a median 63.8\% improvement in the figure of merit for the $\Omega_m - \sigma_8 - w_0$ parameter space. The constraints on individual parameters, particularly $\sigma_8$, are tightened substantially (Section~\ref{sec:fiducial_post}).
\end{enumerate}

We show that an SBI pipeline for the i3PCF is feasible, and that it brings improvements when jointly analysed alongside classical cosmic shear 2PCF while including all relevant systematic effects for a Stage-III survey. Looking forward, this framework serves as a pathfinder for several exciting future research directions:

\begin{enumerate}[label=(\roman*)]
    \item The immediate next step is to apply the \texttt{SBi3PCF} pipeline to real observational data. Given that our simulations and systematic effects were tailored to the Dark Energy Survey, an analysis of the DES Y3 data is a natural and compelling goal. This would provide the first cosmological constraints from the i3PCF using a fully simulation-based approach (Gebauer et al.~in preparation) and serve as a test and comparison platform for a standard likelihood-based i3PCF analysis (Halder et al.~in preparation).
    \item Our framework is a crucial stepping stone for next-generation surveys like Euclid and LSST. The unprecedented statistical power of these surveys will make the inclusion of higher-order statistics essential. This will require scaling up our methodology, including the generation of even larger and more complex simulation suites to cover an expanded parameter space of cosmology and systematics.
    \item Our SBI pipeline is highly flexible. It can be extended to incorporate other higher-order statistics, such as the weak lensing convergence PDF. Furthermore, we can also create galaxy count maps from \texttt{CosmoGridV1}, allowing for our pipeline to be adapted for exploring exciting directions for combined, multi-probe analyses to break cosmological parameter degeneracies and obtain even tighter cosmological constraints.
\end{enumerate}
By combining simulation-based inference with analytical methods for validation, our paper contributes to the growing efforts in the weak lensing community towards pushing for robust HOS analyses, moving beyond Gaussian statistics, to more fully exploit the rich cosmological information encoded in the large-scale structure of the Universe.

\acknowledgments
We would like to thank Oliver Friedrich, Marco Gatti, Zhengyangguang Gong, Daniel Gruen, Joachim Harnois-Deraps, Jed Homer, Mike Jarvis, Tomasz Kacprzak, Kai Lehman, Alexander Reeves, Arne Thomsen, Beatriz Tucci, and Cora Uhlemann for fruitful discussions at various stages of this project. We acknowledge support from the Excellence Cluster ORIGINS, which is funded by the Deutsche Forschungsgemeinschaft (DFG, German Research Foundation) under Germany’s Excellence Strategy - EXC-2094-390783311. Some of the numerical calculations have been carried out on the ORIGINS computing facilities of the Computational Center for Particle and Astrophysics (C2PAP). DG was also supported by the European Union (ERC StG, LSS\_BeyondAverage, 101075919). AH has been supported by funding from the European Research Council (ERC) under the European Union’s Horizon 2020 research and innovation programmes (grant agreement no. 101018897 CosmicExplorer).

DG and AH are deeply indebted to Stella Seitz for introducing and inspiring them to pursue research in the field of gravitational lensing, in which she made several foundational contributions. She recognized early on the potential of wide-area weak lensing surveys, advanced statistical tools, and machine learning methods, all of which are central to this work. Stella was deeply enthusiastic and actively involved in this project, but sadly did not live to see its final form. May she rest in peace, knowing that her legacy will live on through this and future works.\\
\\
\newline\textbf{Data availability:} The numerical data underlying the analysis of this paper may be shared upon
request to the authors.

\appendix
\section{CosmoFuse}\label{sec:cosmofuse}

For this work we had to measure the i3PCF in 5 filter sizes on $\mathcal{O}(10^5)$ mock DES Y3 footprints. In order to do these measurements in a feasible amount of time we have developed \href{https://github.com/D-Gebauer/CosmoFuse}{\texttt{CosmoFuse} \faGithub}, a Python package which leverages just-in-time (JIT) compilation and GPU computing using \texttt{CUDA} kernels.
To calculate the shear i3PCF, we have to measure the aperture mass and the local 2PCF within small patches (see Eq.~\eqref{eq:i3PCF}). For both, we need to transform the global shear $\left(\gamma_1,\gamma_2\right)$ to the relative shear $\left( \gamma_t, \gamma_x \right)$ following Eq.~\eqref{eq:relative_shear}. We first find all pairs within an angular separation bin and keep their indices. We then calculate the relative position angle $\vartheta$ for all pairs and compute $\cos\left(2\vartheta\right)$ and $\sin\left(\vartheta\right)$. By saving these together with the \texttt{HEALPIX} pixel indices of the pairs, we can then reduce the computation of the local 2PCFs to a two step process: (i) a simple multiplication of the shear values with the precomputed angles, which can be vectorized and evaluated very efficiently on GPUs, (ii) a sum over all pairs within an angular bin, a fast operation to perform. For a fixed survey geometry and order of thousands of simulations from which to estimate the i3PCF, \texttt{CosmoFuse} can lead to $\mathcal{O}(10^4)$ times quicker computations compared to the rerunning of classical tree codes on CPUs for every simulation. However, it requires significantly more memory, which limits the current maximum resolution that can be used. For a DES Y3 footprint, we needed $\sim 20$GB of VRAM \texttt{NSIDE}$=512$ to load and manipulate all pre-computed pairs.

\begin{figure}[htbp!]
    \centering
    \includegraphics[width=1\linewidth]{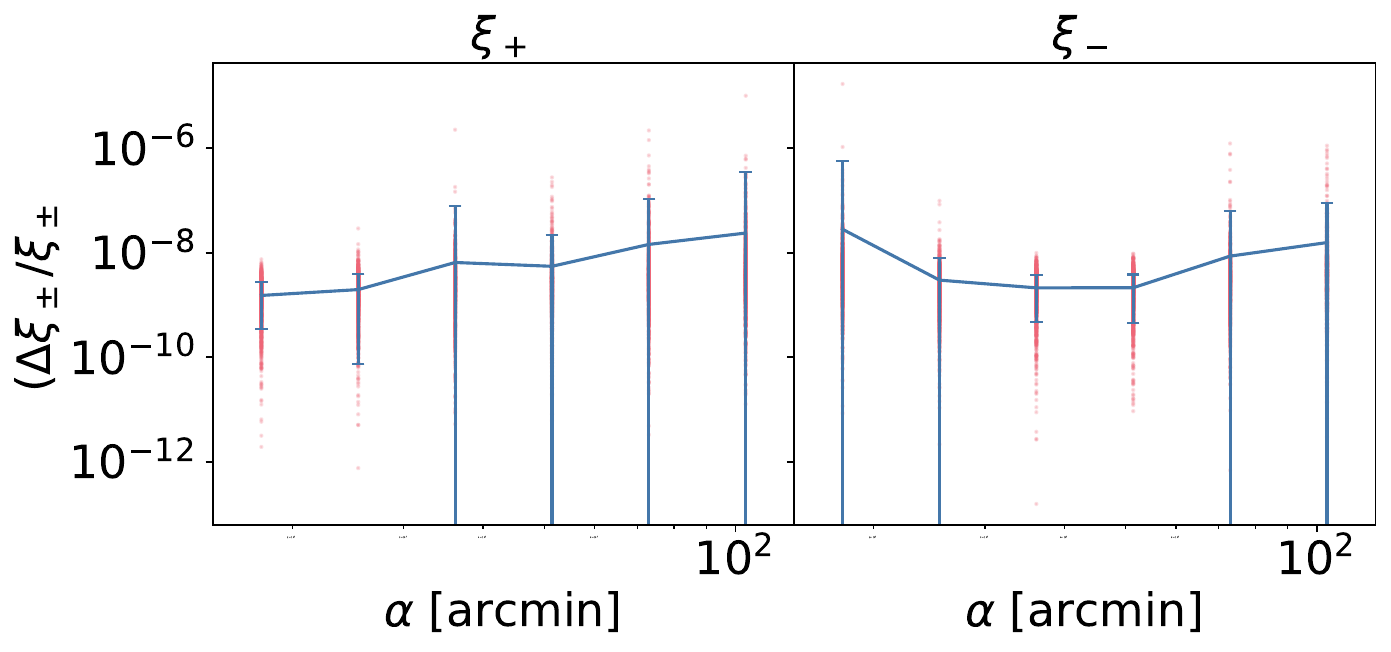}
    \caption{Validation of \texttt{CosmoFuse} against \texttt{TreeCorr}. We show the relative errors for the local 2PCFs $\xi_+$ and $\xi_-$ evaluated with \texttt{CosmoFuse} in 1000 patches with $90'$ radius on a fiducial DES Y3-like shear map. The blue line and shaded region indicate the mean and standard deviation of the relative difference across all patches, while orange points show the individual patch measurements.}
    \label{fig:cosmofuse_val}
\end{figure}

To validate our implementation, we compared \texttt{CosmoFuse} measurements against the widely-used \texttt{TreeCorr} package \cite{Jarvis2019}. Figure~\ref{fig:cosmofuse_val} shows the relative errors for the local 2PCF measurements in 1000 patches with $90'$ radius. It is calculated as

\begin{equation}
    \frac{\Delta \xi_\pm}{\xi_\pm}= \frac{\xi_{\text{CF}} - \xi_{\text{TC}}}{\xi_{\text{TC}}}.
\end{equation}

\noindent The mean relative difference is $\sim 10^{-8}$, with the largest per patch deviations $\lesssim 10^{-5}$. This is well within the acceptable tolerance for our analysis and demonstrates that \texttt{CosmoFuse} produces results consistent with established codes while offering significant computational advantages for the purposes of estimating local 2PCFs in patches. We believe that the code is useful not only for i3PCF estimation but also for multitudes of tests performed at the 2PCF level that require estimating local position-dependent 2PCFs in simulated mocks repeatedly, like testing for spatial variations of survey systematics.

\section{Gaussianity assumption of the individual i3PCFs}\label{sec:gaussianity}

\begin{figure}[htbp!]
\begin{subfigure}{.475\linewidth}
  \includegraphics[width=\linewidth]{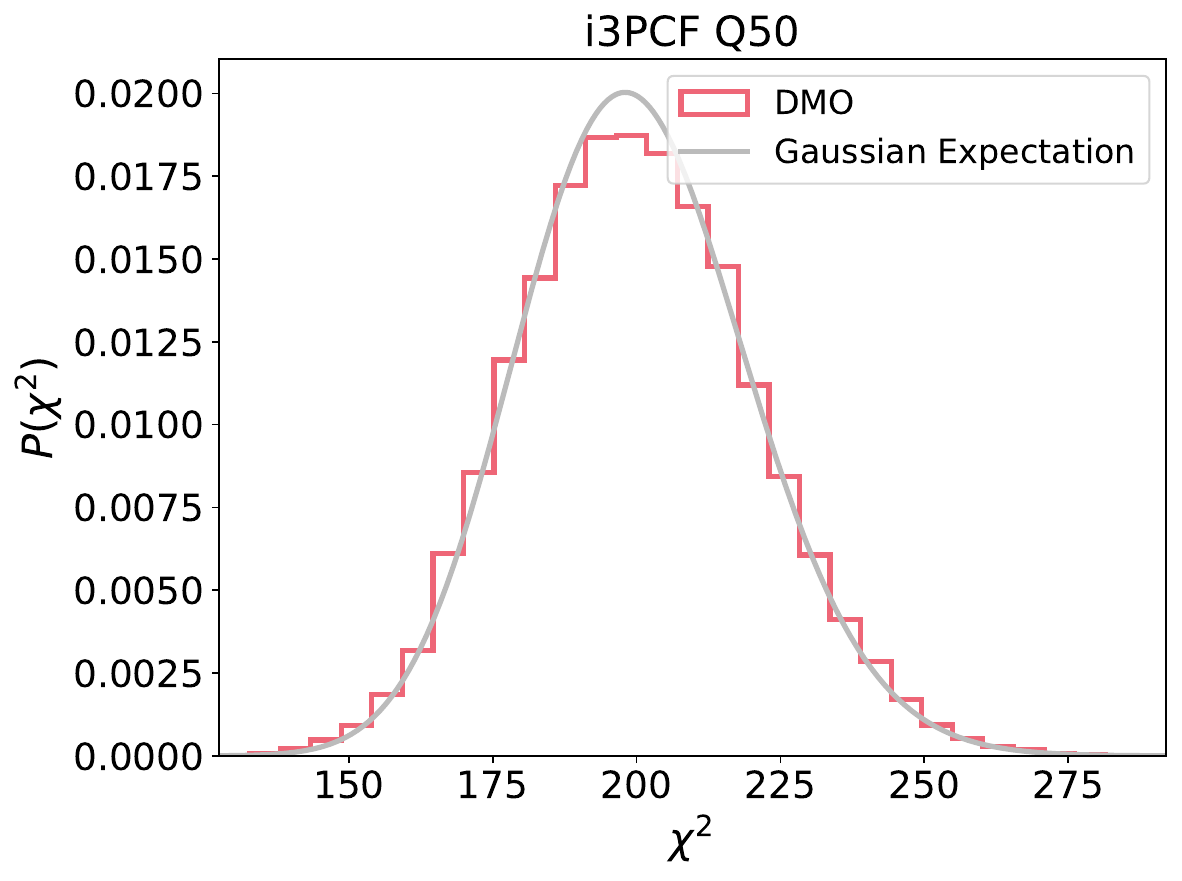}
  \caption{i3PCF with filter radius $\theta = 50'$.}
  \label{fig:chi2_i3PCF_50}
\end{subfigure}\hfill
\begin{subfigure}{.475\linewidth}
  \includegraphics[width=\linewidth]{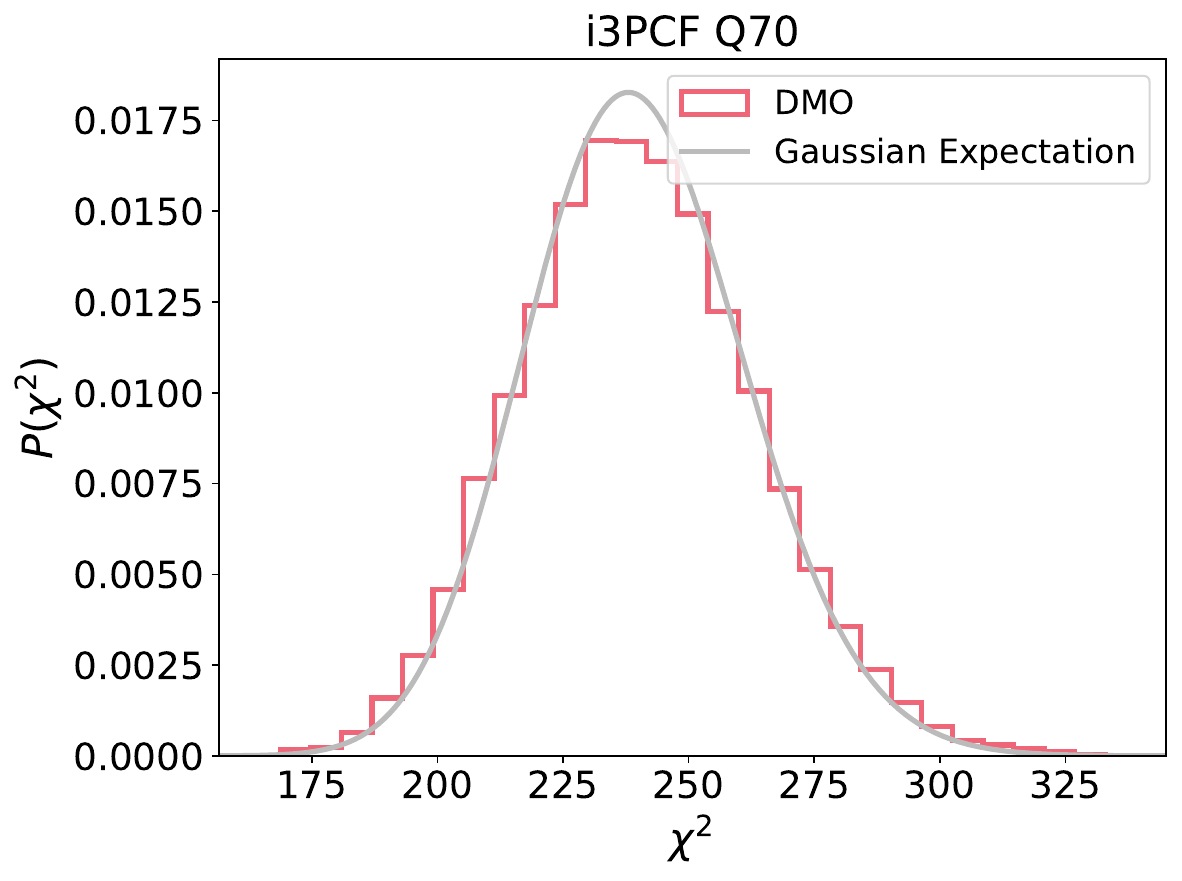}
  \caption{i3PCF with filter radius $\theta = 70'$.}
  \label{fig:chi2_i3PCF_70}
\end{subfigure}

\medskip
\begin{subfigure}{.475\linewidth}
  \includegraphics[width=\linewidth]{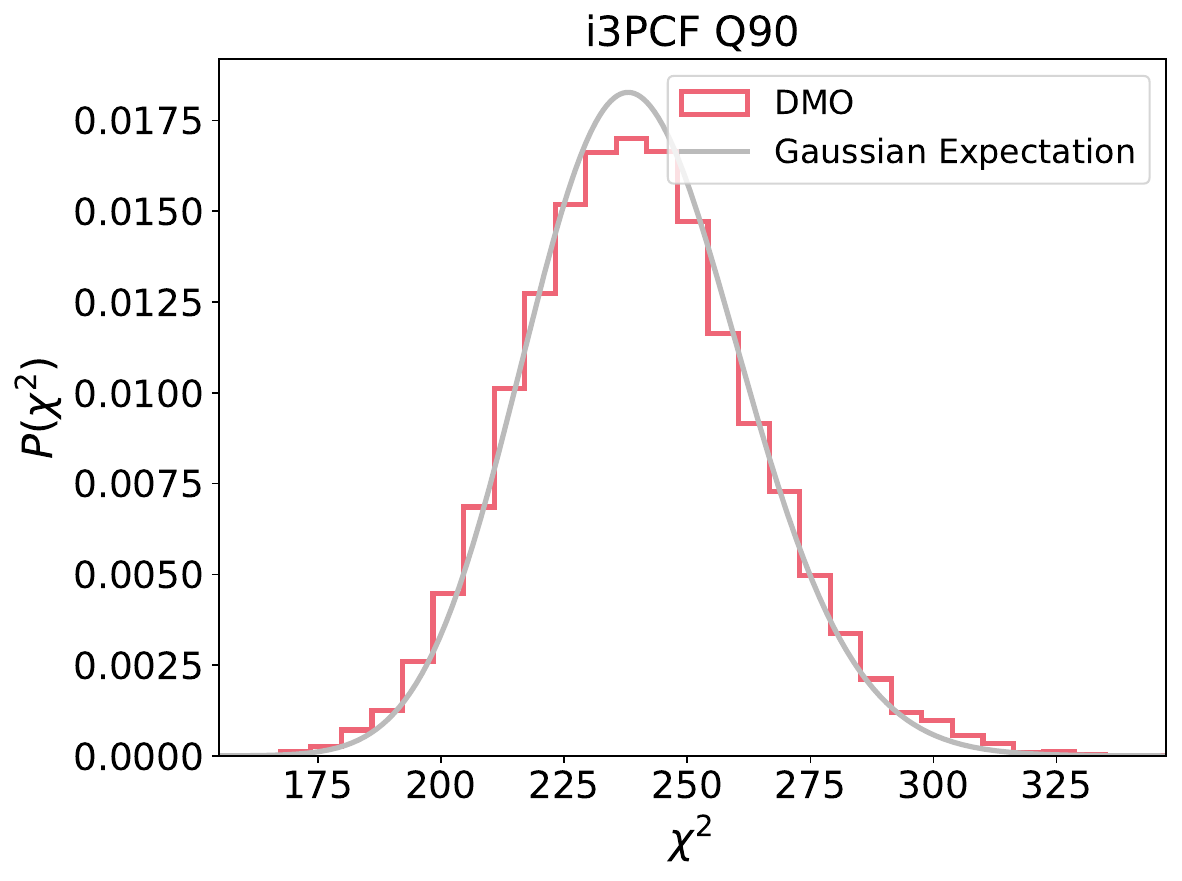}
  \caption{i3PCF with filter radius $\theta = 90'$.}
  \label{fig:chi2_i3PCF_90}
\end{subfigure}\hfill
\begin{subfigure}{.475\linewidth}
  \includegraphics[width=\linewidth]{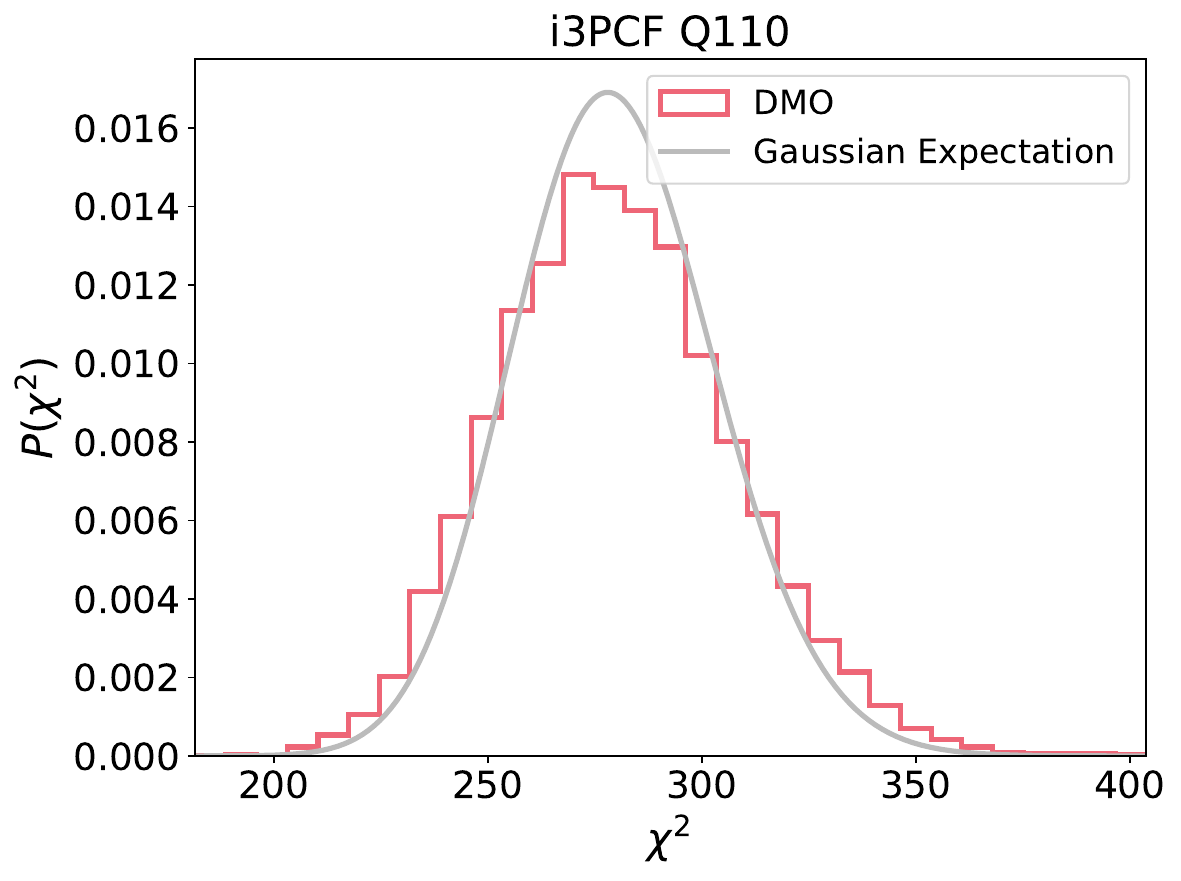}
  \caption{i3PCF with filter radius $\theta = 110'$.}
  \label{fig:chi2_i3PCF_110}
\end{subfigure}

\medskip
\begin{subfigure}{.475\linewidth}
  \includegraphics[width=\linewidth]{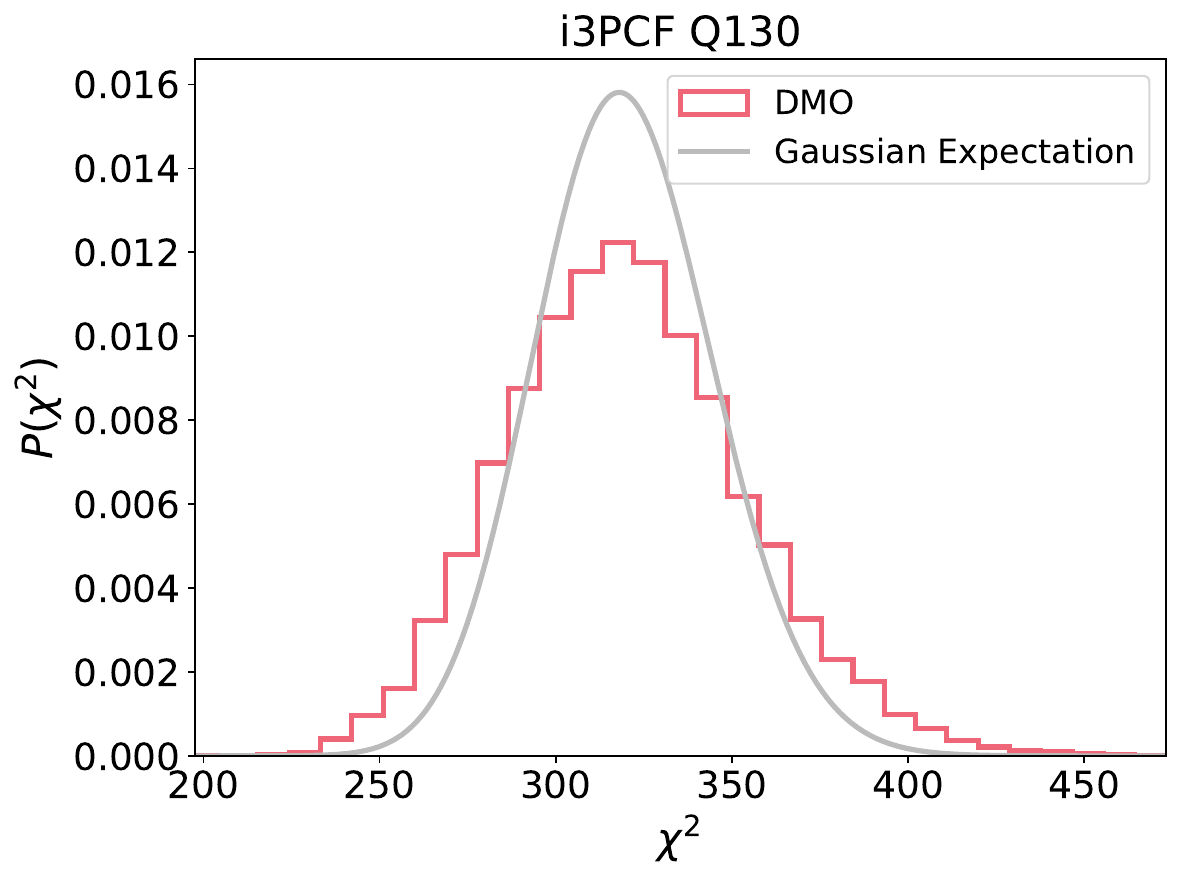}
  \caption{i3PCF with filter radius $\theta = 130'$.}
  \label{fig:chi2_i3PCF_130}
\end{subfigure}\hfill
\begin{subfigure}{.475\linewidth}
  \includegraphics[width=\linewidth]{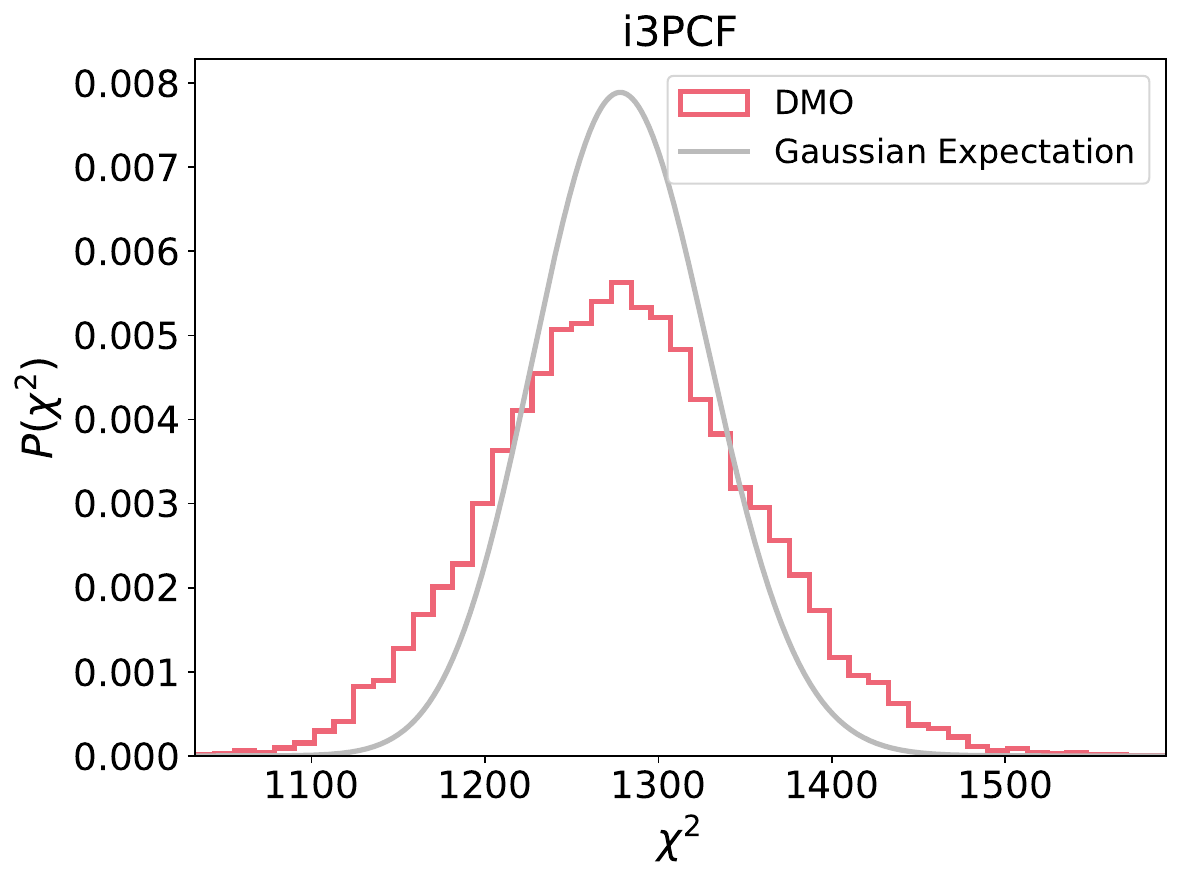}
  \caption{Combined i3PCFs with all 5 filters.}
  \label{fig:chi2_i3PCF_combined}
\end{subfigure}

\caption{Histograms of $\chi^2$ values for the 16,000 fiducial measurements using dark matter only simulations (red). The black line shows the expected $\chi^2$ distribution if the likelihood is Gaussian.}
\label{fig:chi2_i3PCF}
\end{figure}

In Section~\ref{sec:results_gaussianity} we investigated the Gaussianity assumption for the joint data vector of 2PCF + i3PCF measurements. Here, we provide additional details on the Gaussianity of the individual i3PCF measurements at each filter scale. Figure~\ref{fig:chi2_i3PCF} shows histograms of $\chi^2$ values (see Eq.~\eqref{eq:chi2}) computed from 16,000 fiducial measurements using baryonified, as well as dark-matter-only, simulations. If the sampling distribution were perfectly Gaussian, the histogram should follow a $\chi^2$ distribution with degrees of freedom equal to the data vector dimension (shown as black curves). For the smaller filter scales ($\theta = 50', 70'$), the sampling distribution closely follows the expected $\chi^2$ distribution, suggesting that the Gaussianity assumption is reasonable. However, as the filter size increases ($\theta \geq 90'$), we observe systematic deviations from Gaussianity. This effect becomes even more pronounced when combining the filters (Figure~\ref{fig:chi2_i3PCF_combined}).

\begin{figure}[htbp!]
\begin{subfigure}{.475\linewidth}
  \includegraphics[width=\linewidth]{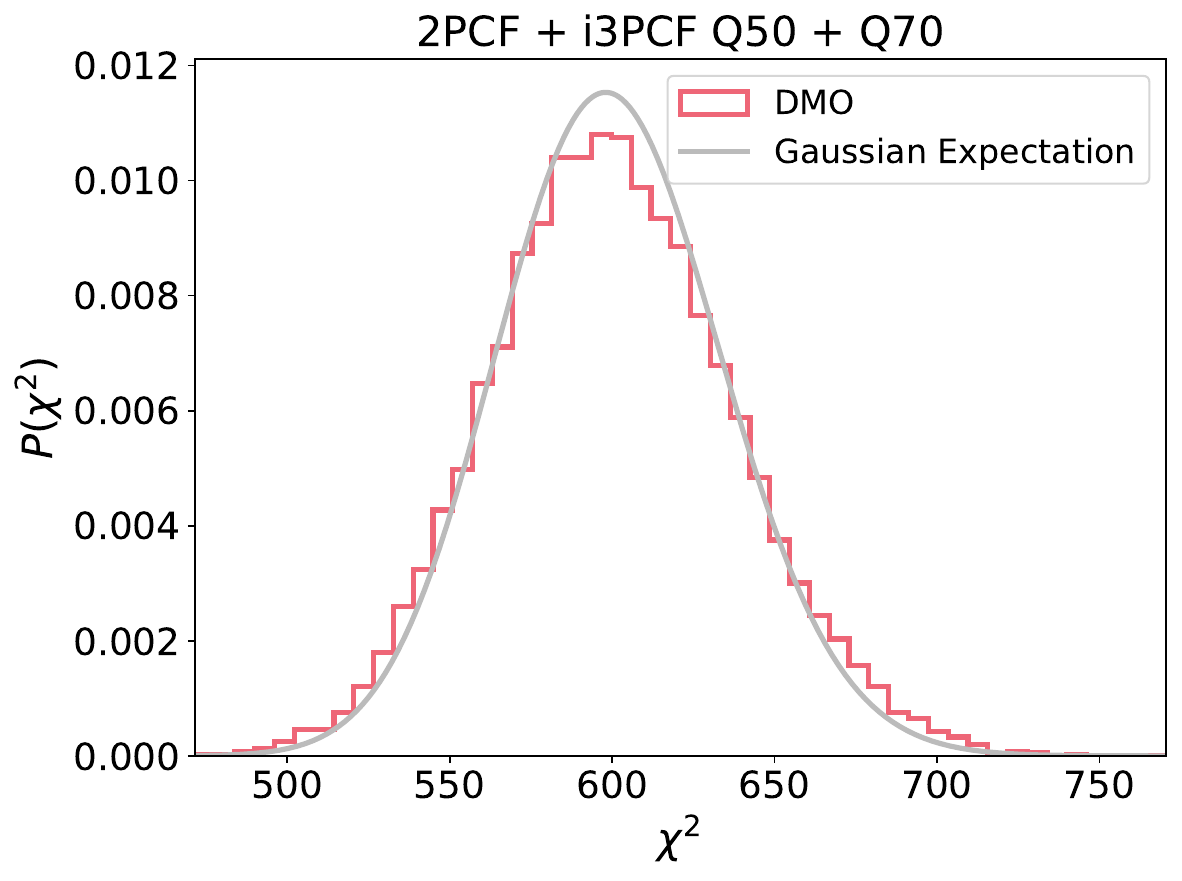}
  \caption{2PCF combined with i3PCF with filter radii $\theta = [50', 70']$.}
  \label{fig:chi2_gausstest_1}
\end{subfigure}\hfill
\begin{subfigure}{.475\linewidth}
  \includegraphics[width=\linewidth]{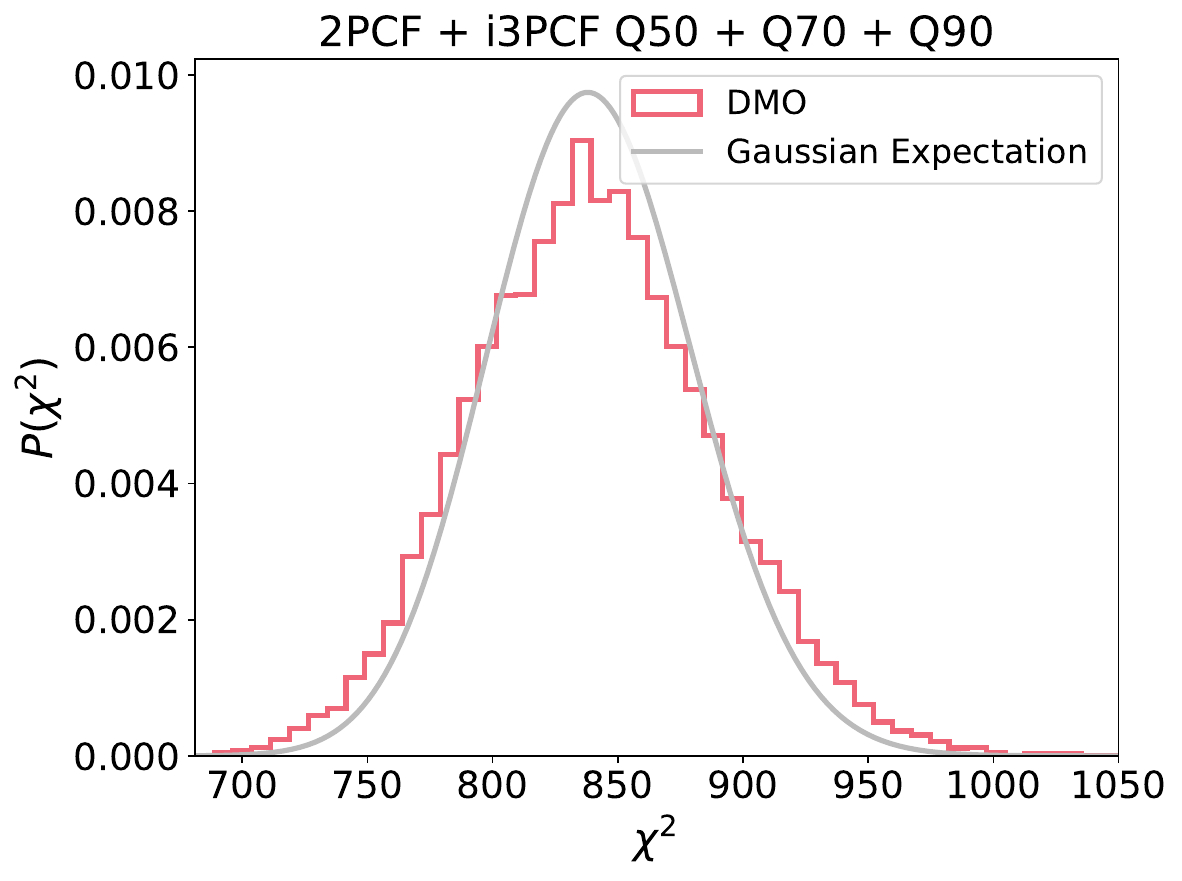}
  \caption{2PCF combined with i3PCF with filter radii $\theta = [50', 70', 90']$.}
  \label{fig:chi2_gausstest_2}
\end{subfigure}

\medskip
\begin{subfigure}{.475\linewidth}
  \includegraphics[width=\linewidth]{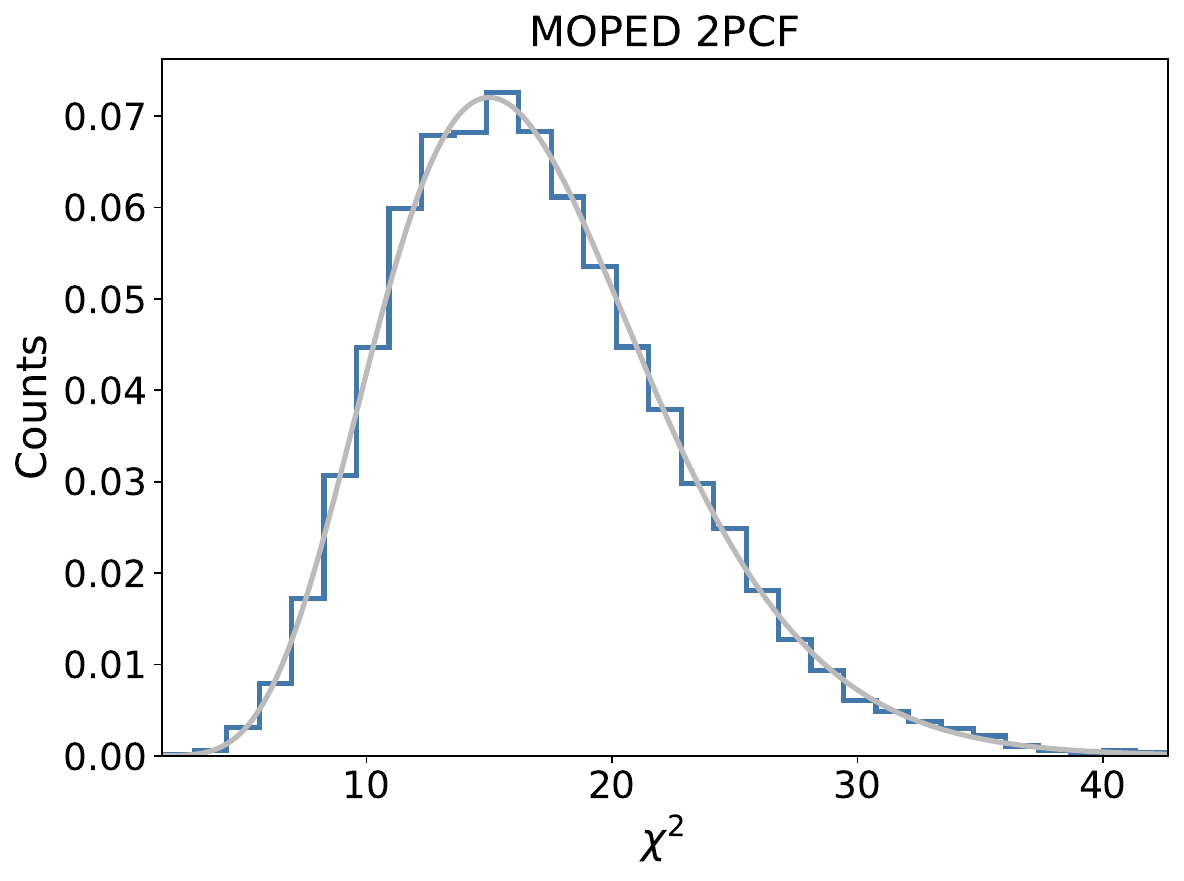}
  \caption{2PCF \texttt{MOPED} coefficients.}
  \label{fig:chi2_moped_2PCF}
\end{subfigure}\hfill
\begin{subfigure}{.475\linewidth}
  \includegraphics[width=\linewidth]{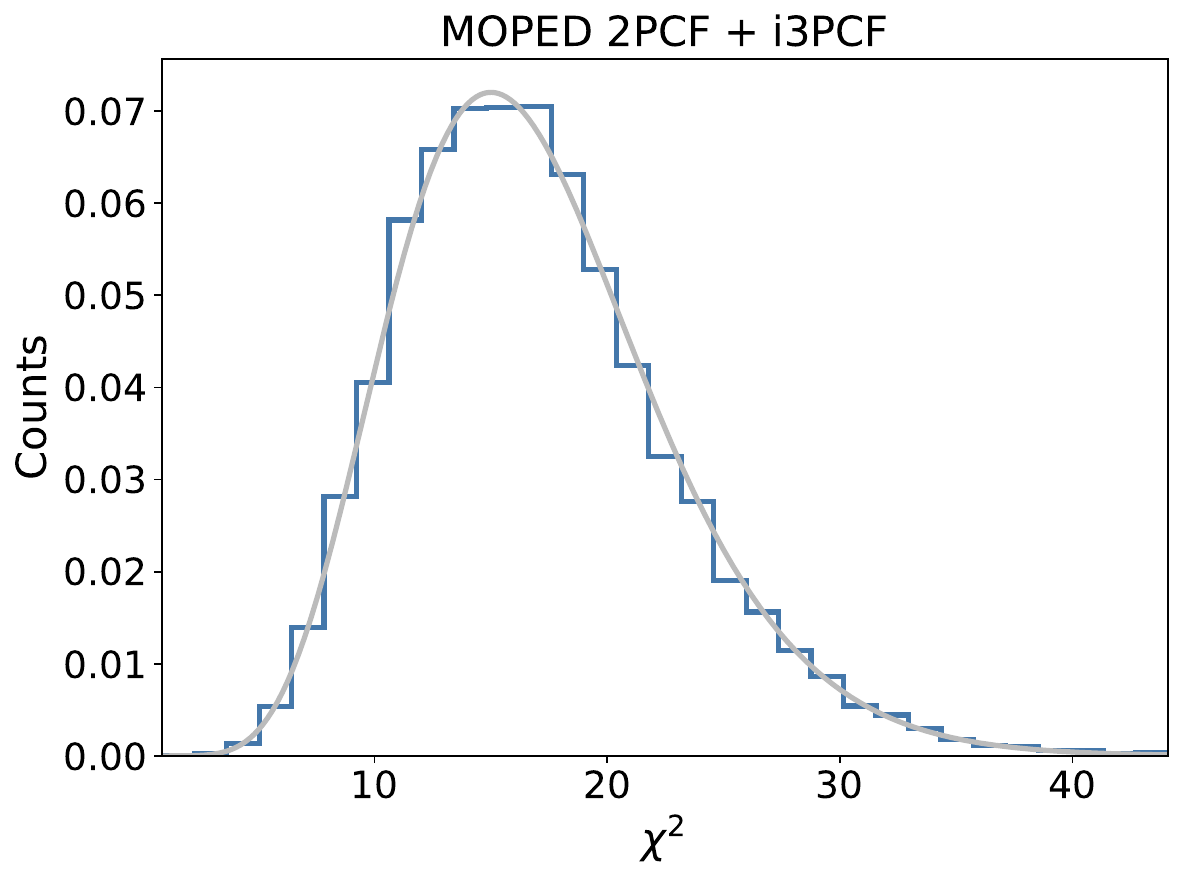}
  \caption{2PCF$ + $i3PCF (all filters) \texttt{MOPED} coefficients.}
  \label{fig:chi2_moped_full}
\end{subfigure}

\caption{Histograms of $\chi^2$ values for the 16,000 fiducial measurements using dark matter only simulations (red). The black line shows the expected $\chi^2$ distribution if the likelihood is Gaussian.}
\label{fig:chi2_extra}
\end{figure}

Figure~\ref{fig:chi2_extra} shows additional diagnostics. Panels \ref{fig:chi2_gausstest_1} and \ref{fig:chi2_gausstest_2} demonstrate that combining 2PCF with progressively larger i3PCF filter scales leads to increasingly non-Gaussian joint distributions. Panels \ref{fig:chi2_moped_2PCF} and \ref{fig:chi2_moped_full} show $\chi^2$ distributions for \texttt{MOPED}-compressed data vectors, where the transformation has removed the non-Gaussian information. This shows that \texttt{MOPED} can be used to ``Gaussianise" the data vector \cite{Gatti2022}.

\section{Full posterior and effects of prior}\label{sec:full_post}

\begin{figure}[!htbp]
    \centering
    \includegraphics[width=\linewidth]{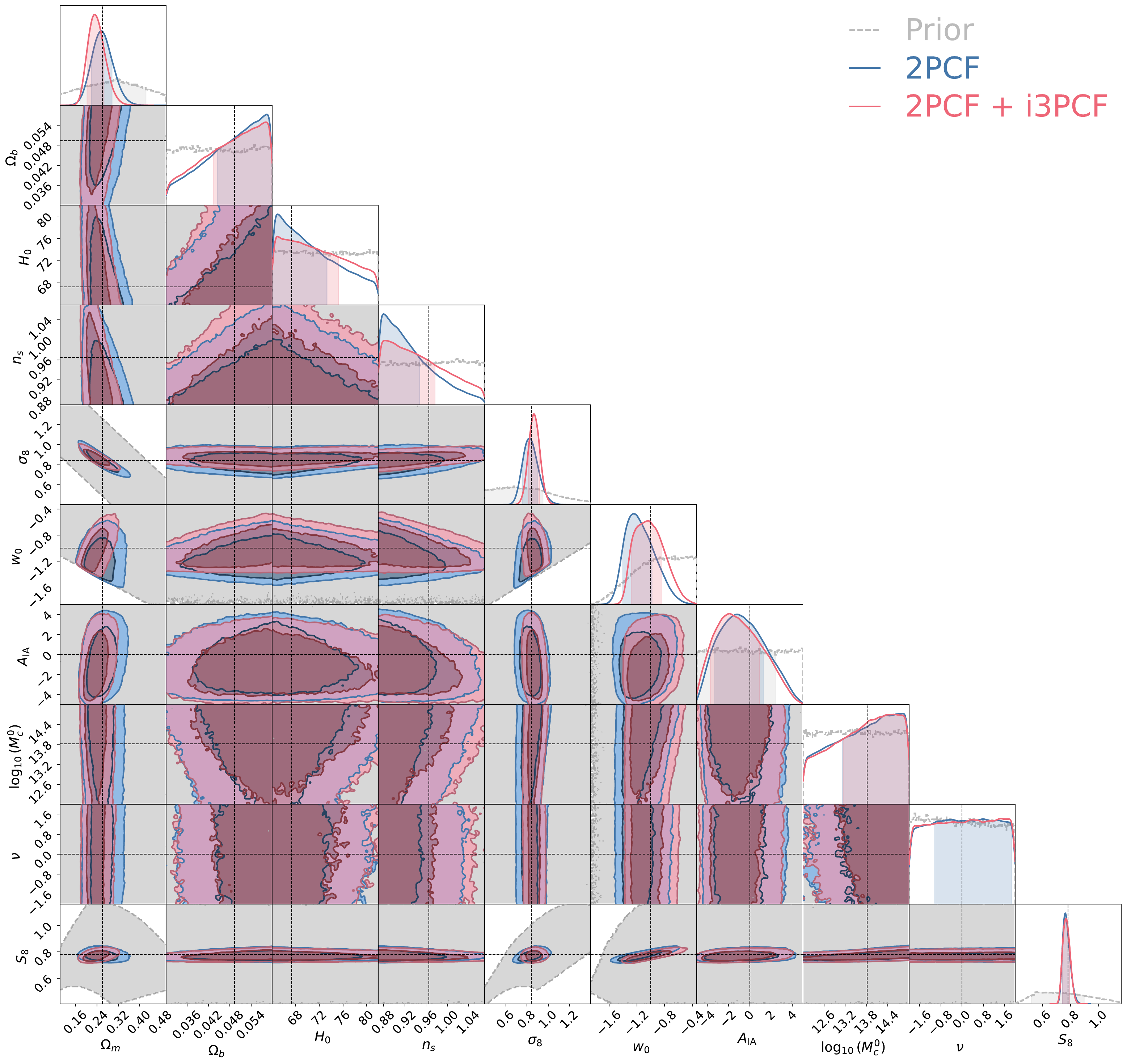}
    \caption{The full posterior of the 9 parameters explicitly learned by the flow and the derived $S_8$. Blue shows the constraints using only 2PCFs and red with added i3PCFs. We also ran a chain to sample the prior probability, shown in grey.}
    \label{fig:full_posterior}
\end{figure}

In this appendix, we present the full posterior distributions for all nine cosmological and systematic parameters that were explicitly learned by our SBI pipeline, in addition to the derived structure growth parameter $S_8$. This complements the results for the main cosmological parameters discussed in Section~\ref{sec:fiducial_post}. Figure~\ref{fig:full_posterior} shows the complete 1D and 2D marginalized posterior distributions. The plot compares the constraints from the 2PCF-only analysis (blue) and the joint 2PCF + i3PCF analysis (red) against the sampled prior distribution (grey). As noted in the main text, the inclusion of the i3PCF significantly tightens the constraints on the matter density parameter, $\Omega_m$, and the amplitude of matter fluctuations, $\sigma_8$. The posterior for the dark energy equation of state, $w_0$, shows a noticeable projection effect, where the distribution flattens against the upper bound of the prior. While the i3PCF provides only a modest improvement in the width of the $w_0$ marginal, it helps to pull the posterior away from the prior boundary, reducing this projection effect. For the systematic parameters, the intrinsic alignment amplitude $A_{\text{IA}}$ and the baryonification parameters $\log M_c^0$ and $\nu$, the posteriors closely follow the prior distributions. This indicates that, as expected, the cosmic shear data used in this analysis does not provide strong constraining power on these parameters, and their posteriors are dominated by the choice of prior.

\section{Systematic effects}\label{sec:systematics_full}

\begin{figure}[!htbp]
    \centering
    \includegraphics[width=\linewidth]{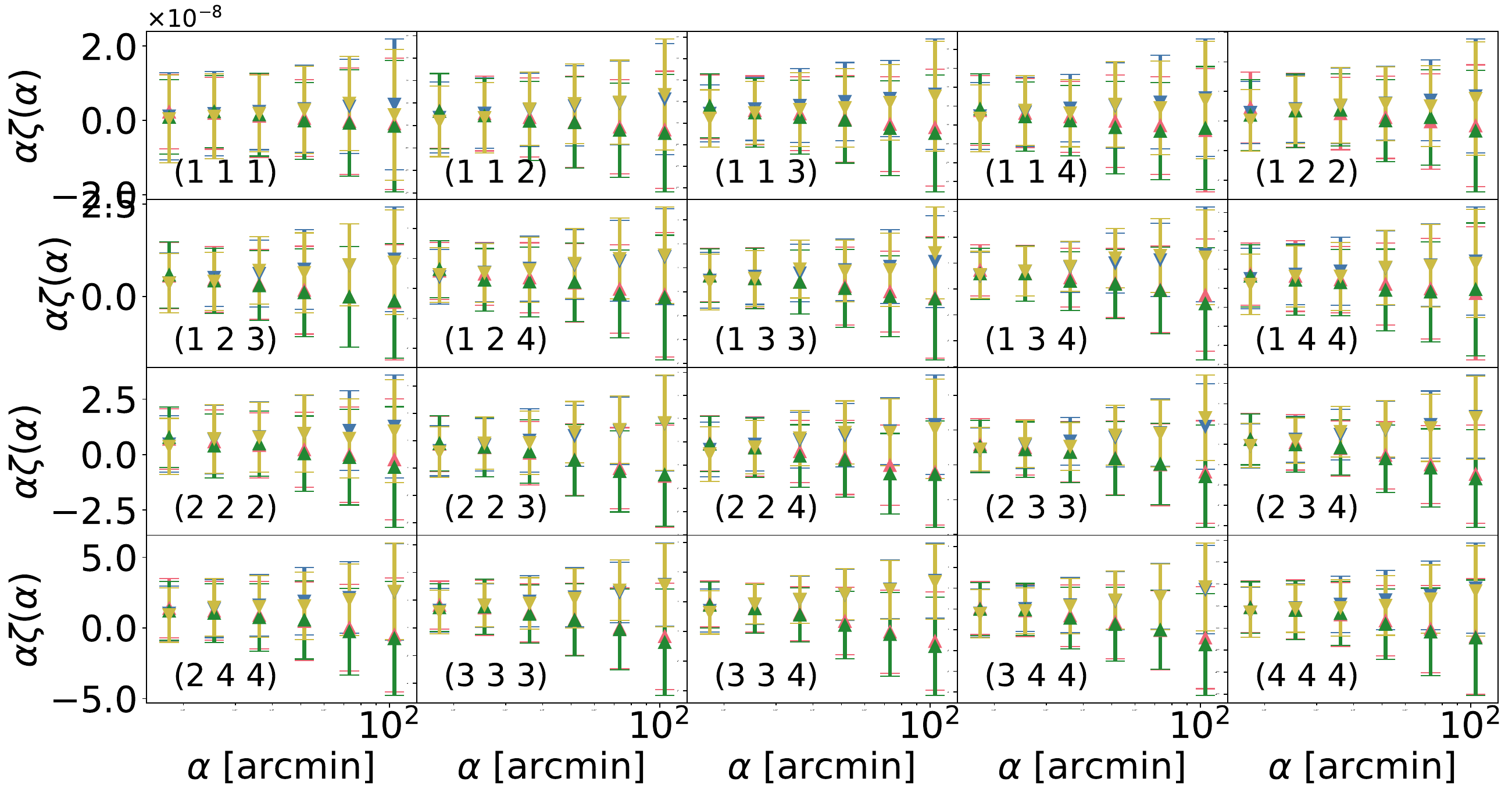}
    \caption{The effect of source clustering on the i3PCFs with a $90'$ filter size across all tomographic bin combinations. The measurements including source clustering ($\zeta_+$ (\textcolor{g}{$\triangle$}) and $\zeta_-$ (\textcolor{y}{$\triangledown$})) are compared to the fiducial measurement without this effect ($\zeta_+$ (\textcolor{r}{$\triangle$}) and $\zeta_-$ (\textcolor{b}{$\triangledown$})). The error bars represent the variance from shape noise.}
    \label{fig:sc_full}
\end{figure}

\begin{figure}[!htbp]
    \centering
    \includegraphics[width=\linewidth]{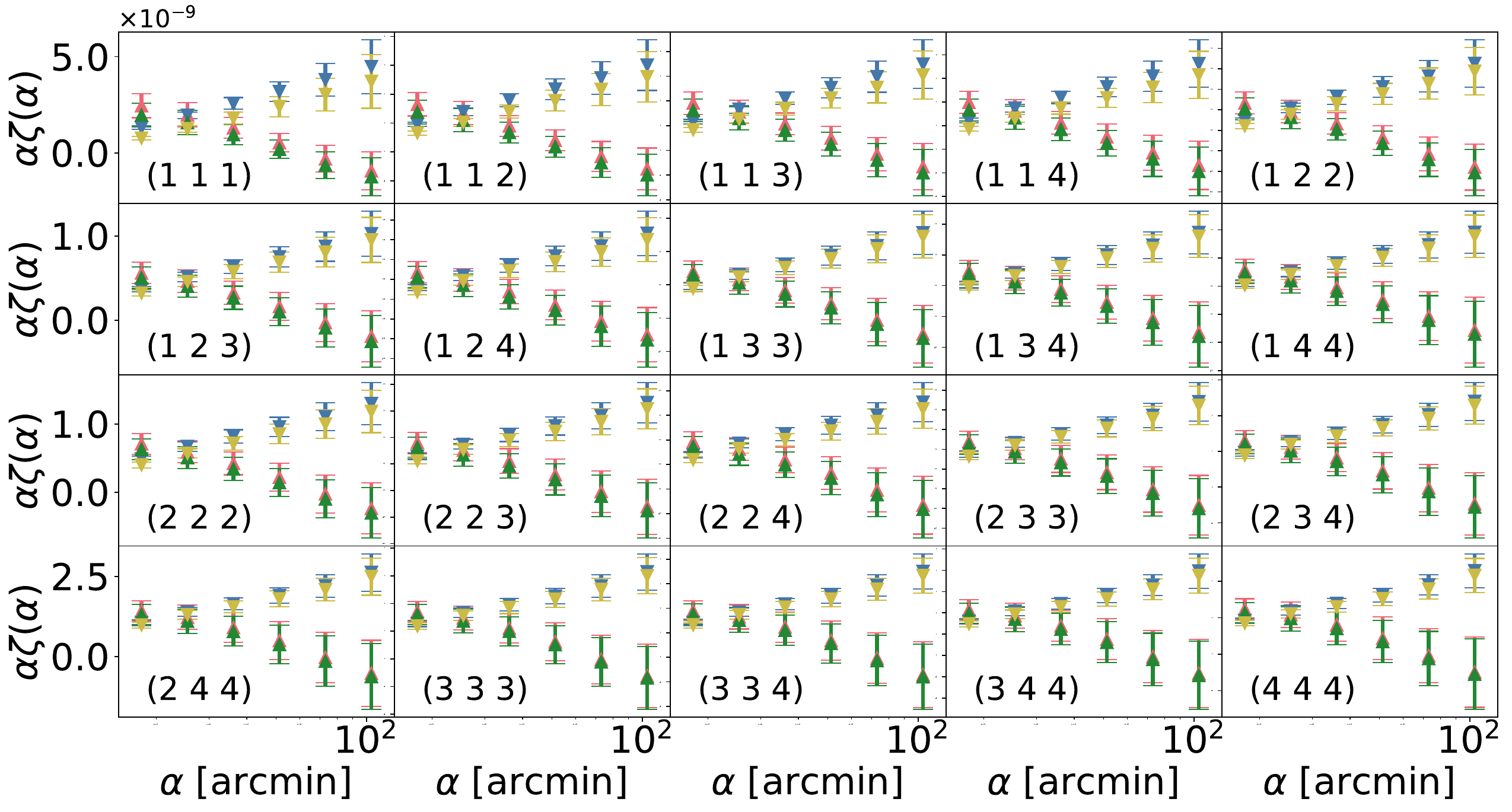}
    \caption{The effect of source clustering on the i3PCFs with a $90'$ filter size across all tomographic bin combinations, shown without the shape noise term. The measurements including source clustering ($\zeta_+$ (\textcolor{g}{$\triangle$}) and $\zeta_-$ (\textcolor{y}{$\triangledown$})) are compared to the noiseless fiducial measurement ($\zeta_+$ (\textcolor{r}{$\triangle$}) and $\zeta_-$ (\textcolor{b}{$\triangledown$})). The error bars represent the cosmic variance for a DES Y3-like survey.}
    \label{fig:sc_noiseless_full}
\end{figure}

Here we present the full tomographic results for the impact of source galaxy clustering and reduced shear on the i3PCF, complementing the results shown in Section~\ref{sec:including_systematics}. Figure~\ref{fig:sc_full} shows the impact of source clustering on the i3PCF across all tomographic bin combinations in the presence of shape noise. The systematic offset introduced by source clustering is orders of magnitude smaller than the statistical uncertainties from shape noise, showing that its effect is negligible in a realistic survey setting. To better isolate the systematic effect, Figure~\ref{fig:sc_noiseless_full} shows the same comparison but for noiseless simulations where the error bars represent only cosmic variance. In this case, a slight but coherent suppression of the i3PCF signal is visible. However, this suppression remains within the $1\sigma$ cosmic variance limits for a DES Y3-like survey, confirming that it does not introduce a significant bias to the measurement in our setup.

\begin{figure}[!htbp]
    \centering
    \includegraphics[width=\linewidth]{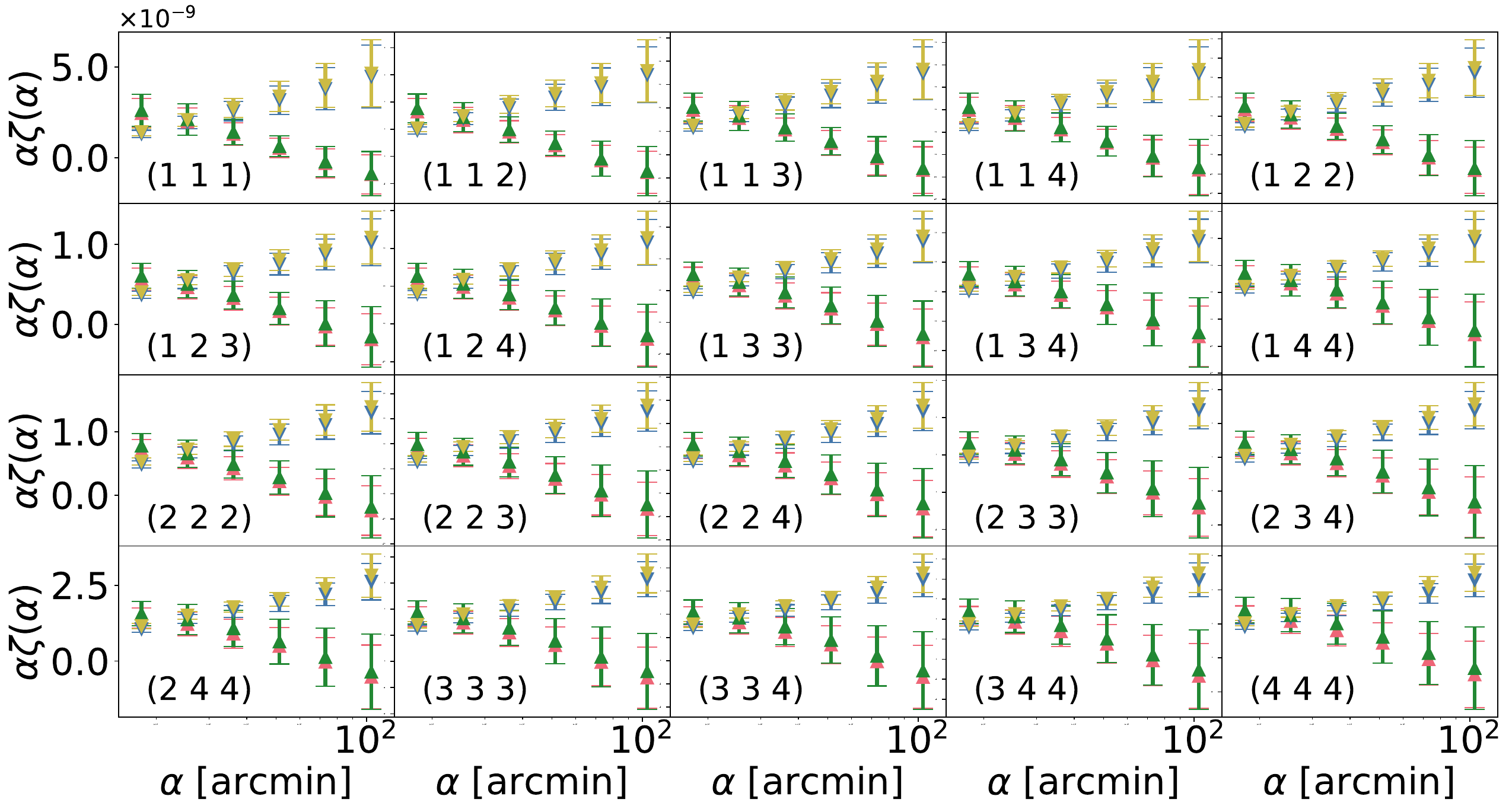}
    \caption{The effect of reduced shear on the i3PCFs with a $90'$ filter size across all tomographic bin combinations. The full calculation using reduced shear ($\zeta_+$ (\textcolor{g}{$\triangle$}) and $\zeta_-$ (\textcolor{y}{$\triangledown$})) is compared to the measurement using the uncorrected shear ($\zeta_+$ (\textcolor{r}{$\triangle$}) and $\zeta_-$ (\textcolor{b}{$\triangledown$})). The error bars represent the cosmic variance for a DES Y3-like survey, as both measurements are noiseless.}
    \label{fig:rs_full}
\end{figure}

Finally, Figure~\ref{fig:rs_full} illustrates the impact of using the full reduced shear formalism instead of the uncorrected shear. The plot shows that the difference between the two is tiny for all tomographic combinations and is completely subdominant to the cosmic variance. This validates the use of the shear, instead of the reduced shear, as a very accurate approximation in our work.

\bibliographystyle{JHEP}
\bibliography{biblio.bib}

@article{Abellan2025,
	title        = {{How to embed any likelihood into SBI: Application to Planck + Stage IV galaxy surveys and Dynamical Dark Energy}},
	author       = {{Abell{\'a}n}, Guillermo Franco and {Anau Montel}, Noemi and {Savchenko}, Oleg and {Weniger}, Christoph},
	year         = {2025},
	month        = jul,
	journal      = {arXiv e-prints},
	pages        = {arXiv:2507.22990},
	doi          = {10.48550/arXiv.2507.22990},
	eid          = {arXiv:2507.22990},
	archiveprefix = {arXiv},
	eprint       = {2507.22990},
	primaryclass = {astro-ph.CO},
	adsurl       = {https://ui.adsabs.harvard.edu/abs/2025arXiv250722990A}
}

@article{Alsing2019,
	title        = {{Fast likelihood-free cosmology with neural density estimators and active learning}},
	author       = {{Alsing}, Justin and {Charnock}, Tom and {Feeney}, Stephen and {Wandelt}, Benjamin},
	year         = {2019},
	month        = sep,
	journal      = {Monthly Notices of the Royal Astronomical Society},
	volume       = {488},
	number       = {3},
	pages        = {4440--4458},
	doi          = {10.1093/mnras/stz1960},
	archiveprefix = {arXiv},
	eprint       = {1903.00007},
	primaryclass = {astro-ph.CO},
	adsurl       = {https://ui.adsabs.harvard.edu/abs/2019MNRAS.488.4440A}
}

@article{Amon2022,
	title        = {{A non-linear solution to the S$_{8}$ tension?}},
	author       = {{Amon}, Alexandra and {Efstathiou}, George},
	year         = {2022},
	month        = nov,
	volume       = {516},
	number       = {4},
	pages        = {5355--5366},
	doi          = {10.1093/mnras/stac2429},
	archiveprefix = {arXiv},
	eprint       = {2206.11794},
	primaryclass = {astro-ph.CO},
	adsurl       = {https://ui.adsabs.harvard.edu/abs/2022MNRAS.516.5355A}
}

@article{Anbajagane2023,
	title        = {{Beyond the 3rd moment: a practical study of using lensing convergence CDFs for cosmology with DES Y3}},
	author       = {{Anbajagane}, D. and {Chang}, C. and {Banerjee}, A. and {Abel}, T. and {Gatti}, M. and {Ajani}, V. and {Alarcon}, A. and {Amon}, A. and {Baxter}, E.~J. and {Bechtol}, K. and {Becker}, M.~R. and {Bernstein}, G.~M. and {Campos}, A. and {Carnero Rosell}, A. and {Carrasco Kind}, M. and {Chen}, R. and {Choi}, A. and {Davis}, C. and {DeRose}, J. and {Diehl}, H.~T. and {Dodelson}, S. and {Doux}, C. and {Drlica-Wagner}, A. and {Eckert}, K. and {Elvin-Poole}, J. and {Everett}, S. and {Fert{\'e}}, A. and {Gruen}, D. and {Gruendl}, R.~A. and {Harrison}, I. and {Hartley}, W.~G. and {Huff}, E.~M. and {Jain}, B. and {Jarvis}, M. and {Jeffrey}, N. and {Kacprzak}, T. and {Kokron}, N. and {Kuropatkin}, N. and {Leget}, P. -F. and {MacCrann}, N. and {McCullough}, J. and {Myles}, J. and {Navarro-Alsina}, A. and {Pandey}, S. and {Prat}, J. and {Raveri}, M. and {Rollins}, R.~P. and {Roodman}, A. and {Rykoff}, E.~S. and {S{\'a}nchez}, C. and {Secco}, L.~F. and {Sevilla-Noarbe}, I. and {Sheldon}, E. and {Shin}, T. and {Troxel}, M.~A. and {Tutusaus}, I. and {Whiteway}, L. and {Yanny}, B. and {Yin}, B. and {Zhang}, Y. and {Abbott}, T.~M.~C. and {Allam}, S. and {Aguena}, M. and {Alves}, O. and {Andrade-Oliveira}, F. and {Annis}, J. and {Bacon}, D. and {Blazek}, J. and {Brooks}, D. and {Cawthon}, R. and {da Costa}, L.~N. and {Pereira}, M.~E.~S. and {Davis}, T.~M. and {Desai}, S. and {Doel}, P. and {Ferrero}, I. and {Frieman}, J. and {Giannini}, G. and {Gutierrez}, G. and {Hinton}, S.~R. and {Hollowood}, D.~L. and {Honscheid}, K. and {James}, D.~J. and {Kuehn}, K. and {Lahav}, O. and {Marshall}, J.~L. and {Mena-Fern{\'a}ndez}, J. and {Menanteau}, F. and {Miquel}, R. and {Palmese}, A. and {Pieres}, A. and {Plazas Malag{\'o}n}, A.~A. and {Reil}, K. and {Sanchez}, E. and {Smith}, M. and {Swanson}, M.~E.~C. and {Tarle}, G. and {Wiseman}, P. and {DES Collaboration}},
	year         = {2023},
	month        = dec,
	volume       = {526},
	number       = {4},
	pages        = {5530--5554},
	doi          = {10.1093/mnras/stad3118},
	archiveprefix = {arXiv},
	eprint       = {2308.03863},
	primaryclass = {astro-ph.CO},
	adsurl       = {https://ui.adsabs.harvard.edu/abs/2023MNRAS.526.5530A}
}

@article{Anbajagane2024,
	title        = {{Map-level baryonification: Efficient modelling of higher-order correlations in the weak lensing and thermal Sunyaev-Zeldovich fields}},
	author       = {{Anbajagane}, Dhayaa and {Pandey}, Shivam and {Chang}, Chihway},
	year         = {2024},
	month        = dec,
	journal      = {The Open Journal of Astrophysics},
	volume       = {7},
	pages        = {108},
	doi          = {10.33232/001c.126788},
	eid          = {108},
	archiveprefix = {arXiv},
	eprint       = {2409.03822},
	primaryclass = {astro-ph.CO},
	adsurl       = {https://ui.adsabs.harvard.edu/abs/2024OJAp....7E.108A}
}

@article{Anbajagane2025,
	title        = {{The DECADE cosmic shear project IV: cosmological constraints from 107 million galaxies across 5,400 deg$^2$ of the sky}},
	author       = {{Anbajagane}, D. and {Chang}, C. and {Drlica-Wagner}, A. and {Tan}, C.~Y. and {Adamow}, M. and {Gruendl}, R.~A. and {Secco}, L.~F. and {Zhang}, Z. and {Becker}, M.~R. and {Ferguson}, P.~S. and {Chicoine}, N. and {Herron}, K. and {Alarcon}, A. and {Teixeira}, R. and {Suson}, D. and {Alsina}, A.~N. and {Amon}, A. and {Andrade-Oliveira}, F. and {Blazek}, J. and {Bom}, C.~R. and {Camacho}, H. and {Carballo-Bello}, J.~A. and {Carnero Rosell}, A. and {Cawthon}, R. and {Cerny}, W. and {Choi}, A. and {Choi}, Y. and {Dodelson}, S. and {Doux}, C. and {Eckert}, K. and {Elvin-Poole}, J. and {Esteves}, J. and {Gatti}, M. and {Giannini}, G. and {Gruen}, D. and {Hartley}, W.~G. and {Herner}, K. and {Huff}, E.~M. and {James}, D.~J. and {Jarvis}, M. and {Krause}, E. and {Kuropatkin}, N. and {Mart{\'\i}nez-V{\'a}zquez}, C.~E. and {Massana}, P. and {Mau}, S. and {McCullough}, J. and {Medina}, G.~E. and {Mutlu-Pakdil}, B. and {Myles}, J. and {Navabi}, M. and {No{\"e}l}, N.~E.~D. and {Pace}, A.~B. and {Porredon}, A. and {Prat}, J. and {Raveri}, M. and {Riley}, A.~H. and {Rykoff}, E.~S. and {Sakowska}, J.~D. and {Samuroff}, S. and {Sanchez-Cid}, D. and {Sand}, D.~J. and {Santana-Silva}, L. and {Sevilla-Noarbe}, I. and {Shin}, T. and {Soares-Santos}, M. and {Stringfellow}, G.~S. and {To}, C. and {Tong}, A. and {Troxel}, M.~A. and {Vivas}, A.~K. and {Yamamoto}, M. and {Yanny}, B. and {Yin}, B. and {Zhang}, Y. and {Zuntz}, J.},
	year         = {2025},
	month        = feb,
	journal      = {arXiv e-prints},
	pages        = {arXiv:2502.17677},
	doi          = {10.48550/arXiv.2502.17677},
	eid          = {arXiv:2502.17677},
	archiveprefix = {arXiv},
	eprint       = {2502.17677},
	primaryclass = {astro-ph.CO},
	adsurl       = {https://ui.adsabs.harvard.edu/abs/2025arXiv250217677A}
}

@article{Anbajagane2025catalog,
	title        = {{The DECADE cosmic shear project I: A new weak lensing shape catalog of 107 million galaxies}},
	author       = {{Anbajagane}, D. and {Chang}, C. and {Zhang}, Z. and {Tan}, C.~Y. and {Adamow}, M. and {Secco}, L.~F. and {Becker}, M.~R. and {Ferguson}, P.~S. and {Drlica-Wagner}, A. and {Gruendl}, R.~A. and {Herron}, K. and {Tong}, A. and {Troxel}, M.~A. and {Sanchez-Cid}, D. and {Sevilla-Noarbe}, I. and {Chicoine}, N. and {Teixeira}, R. and {Alarcon}, A. and {Suson}, D. and {Alsina}, A.~N. and {Amon}, A. and {Bom}, C.~R. and {Carballo-Bello}, J.~A. and {Cerny}, W. and {Choi}, A. and {Choi}, Y. and {Doux}, C. and {Eckert}, K. and {Gatti}, M. and {Gruen}, D. and {James}, D.~J. and {Jarvis}, M. and {Kuropatkin}, N. and {Mart{\'\i}nez-V{\'a}zquez}, C.~E. and {Massana}, P. and {Mau}, S. and {McCullough}, J. and {Medina}, G.~E. and {Mutlu-Pakdil}, B. and {Navabi}, M. and {No{\"e}l}, N.~E.~D. and {Pace}, A.~B. and {Prat}, J. and {Raveri}, M. and {Riley}, A.~H. and {Rykoff}, E.~S. and {Sakowska}, J.~D. and {Sand}, D.~J. and {Santana-Silva}, L. and {Shin}, T. and {Soares-Santos}, M. and {Stringfellow}, G.~S. and {Vivas}, A.~K. and {Yamamoto}, M.},
	year         = {2025},
	month        = feb,
	journal      = {arXiv e-prints},
	pages        = {arXiv:2502.17674},
	doi          = {10.48550/arXiv.2502.17674},
	eid          = {arXiv:2502.17674},
	archiveprefix = {arXiv},
	eprint       = {2502.17674},
	primaryclass = {astro-ph.CO},
	adsurl       = {https://ui.adsabs.harvard.edu/abs/2025arXiv250217674A}
}

@article{Barreira_2017,
	title        = {Responses in large-scale structure},
	author       = {Barreira, Alexandre and Schmidt, Fabian},
	year         = {2017},
	month        = jun,
	journal      = {Journal of Cosmology and Astroparticle Physics},
	publisher    = {IOP Publishing},
	volume       = {2017},
	number       = {06},
	pages        = {053–053},
	doi          = {10.1088/1475-7516/2017/06/053},
	issn         = {1475-7516},
	url          = {http://dx.doi.org/10.1088/1475-7516/2017/06/053}
}

@article{BartelmannSchneider2001,
	title        = {{Weak gravitational lensing}},
	author       = {{Bartelmann}, M. and {Schneider}, P.},
	year         = {2001},
	month        = jan,
	volume       = {340},
	number       = {4-5},
	pages        = {291--472},
	doi          = {10.1016/S0370-1573(00)00082-X},
	archiveprefix = {arXiv},
	eprint       = {astro-ph/9912508},
	primaryclass = {astro-ph},
	adsurl       = {https://ui.adsabs.harvard.edu/abs/2001PhR...340..291B}
}

@inproceedings{bergstra2011,
	title        = {Algorithms for Hyper-parameter Optimization},
	author       = {Bergstra, James and Bardenet, R{\'e}mi and Bengio, Yoshua and K{\'e}gl, Bal{\'a}zs},
	year         = {2011},
	booktitle    = {Advances in Neural Information Processing Systems},
	publisher    = {Curran Associates, Inc.},
	volume       = {24},
	pages        = {2546--2554},
	editor       = {J. Shawe-Taylor and R. S. Zemel and P. L. Bartlett and F. Pereira and K. Q. Weinberger}
}

@article{Bernardeau2002,
	title        = {{Large-scale structure of the Universe and cosmological perturbation theory}},
	author       = {{Bernardeau}, F. and {Colombi}, S. and {Gazta{\~n}aga}, E. and {Scoccimarro}, R.},
	year         = {2002},
	month        = sep,
	journal      = {Physics Reports},
	volume       = {367},
	number       = {1-3},
	pages        = {1--248},
	doi          = {10.1016/S0370-1573(02)00135-7},
	archiveprefix = {arXiv},
	eprint       = {astro-ph/0112551},
	primaryclass = {astro-ph},
	adsurl       = {https://ui.adsabs.harvard.edu/abs/2002PhR...367....1B}
}

@article{Bernstein2002,
	title        = {{Shapes and Shears, Stars and Smears: Optimal Measurements for Weak Lensing}},
	author       = {{Bernstein}, G.~M. and {Jarvis}, M.},
	year         = {2002},
	month        = feb,
	volume       = {123},
	number       = {2},
	pages        = {583--618},
	doi          = {10.1086/338085},
	archiveprefix = {arXiv},
	eprint       = {astro-ph/0107431},
	primaryclass = {astro-ph},
	adsurl       = {https://ui.adsabs.harvard.edu/abs/2002AJ....123..583B}
}

@article{Bigwood2024,
	title        = {{Weak lensing combined with the kinetic Sunyaev-Zel'dovich effect: a study of baryonic feedback}},
	author       = {{Bigwood}, L. and {Amon}, A. and {Schneider}, A. and {Salcido}, J. and {McCarthy}, I.~G. and {Preston}, C. and {Sanchez}, D. and {Sijacki}, D. and {Schaan}, E. and {Ferraro}, S. and {Battaglia}, N. and {Chen}, A. and {Dodelson}, S. and {Roodman}, A. and {Pieres}, A. and {Fert{\'e}}, A. and {Alarcon}, A. and {Drlica-Wagner}, A. and {Choi}, A. and {Navarro-Alsina}, A. and {Campos}, A. and {Ross}, A.~J. and {Carnero Rosell}, A. and {Yin}, B. and {Yanny}, B. and {S{\'a}nchez}, C. and {Chang}, C. and {Davis}, C. and {Doux}, C. and {Gruen}, D. and {Rykoff}, E.~S. and {Huff}, E.~M. and {Sheldon}, E. and {Tarsitano}, F. and {Andrade-Oliveira}, F. and {Bernstein}, G.~M. and {Giannini}, G. and {Diehl}, H.~T. and {Huang}, H. and {Harrison}, I. and {Sevilla-Noarbe}, I. and {Tutusaus}, I. and {Elvin-Poole}, J. and {McCullough}, J. and {Zuntz}, J. and {Blazek}, J. and {DeRose}, J. and {Cordero}, J. and {Prat}, J. and {Myles}, J. and {Eckert}, K. and {Bechtol}, K. and {Herner}, K. and {Secco}, L.~F. and {Gatti}, M. and {Raveri}, M. and {Kind}, M. Carrasco and {Becker}, M.~R. and {Troxel}, M.~A. and {Jarvis}, M. and {MacCrann}, N. and {Friedrich}, O. and {Alves}, O. and {Leget}, P. -F. and {Chen}, R. and {Rollins}, R.~P. and {Wechsler}, R.~H. and {Gruendl}, R.~A. and {Cawthon}, R. and {Allam}, S. and {Bridle}, S.~L. and {Pandey}, S. and {Everett}, S. and {Shin}, T. and {Hartley}, W.~G. and {Fang}, X. and {Zhang}, Y. and {Aguena}, M. and {Annis}, J. and {Bacon}, D. and {Bertin}, E. and {Bocquet}, S. and {Brooks}, D. and {Carretero}, J. and {Castander}, F.~J. and {da Costa}, L.~N. and {Pereira}, M.~E.~S. and {De Vicente}, J. and {Desai}, S. and {Doel}, P. and {Ferrero}, I. and {Flaugher}, B. and {Frieman}, J. and {Garc{\'\i}a-Bellido}, J. and {Gaztanaga}, E. and {Gutierrez}, G. and {Hinton}, S.~R. and {Hollowood}, D.~L. and {Honscheid}, K. and {Huterer}, D. and {James}, D.~J. and {Kuehn}, K. and {Lahav}, O. and {Lee}, S. and {Marshall}, J.~L. and {Mena-Fern{\'a}ndez}, J. and {Miquel}, R. and {Muir}, J. and {Paterno}, M. and {Plazas Malag{\'o}n}, A.~A. and {Porredon}, A. and {Romer}, A.~K. and {Samuroff}, S. and {Sanchez}, E. and {Sanchez Cid}, D. and {Smith}, M. and {Soares-Santos}, M. and {Suchyta}, E. and {Swanson}, M.~E.~C. and {Tarle}, G. and {To}, C. and {Weaverdyck}, N. and {Weller}, J. and {Wiseman}, P. and {Yamamoto}, M.},
	year         = {2024},
	month        = oct,
	volume       = {534},
	number       = {1},
	pages        = {655--682},
	doi          = {10.1093/mnras/stae2100},
	archiveprefix = {arXiv},
	eprint       = {2404.06098},
	primaryclass = {astro-ph.CO},
	adsurl       = {https://ui.adsabs.harvard.edu/abs/2024MNRAS.534..655B}
}

@article{Blas2011,
	title        = {{The Cosmic Linear Anisotropy Solving System (CLASS). Part II: Approximation schemes}},
	author       = {{Blas}, Diego and {Lesgourgues}, Julien and {Tram}, Thomas},
	year         = {2011},
	month        = jul,
	volume       = {2011},
	number       = {7},
	pages        = {034},
	doi          = {10.1088/1475-7516/2011/07/034},
	eid          = {034},
	archiveprefix = {arXiv},
	eprint       = {1104.2933},
	primaryclass = {astro-ph.CO},
	adsurl       = {https://ui.adsabs.harvard.edu/abs/2011JCAP...07..034B}
}

@article{BridleKing2007,
	title        = {Dark energy constraints from cosmic shear power spectra: impact of intrinsic alignments on photometric redshift requirements},
	author       = {Sarah Bridle and Lindsay King},
	year         = {2007},
	month        = {12},
	journal      = {New Journal of Physics},
	volume       = {9},
	number       = {12},
	pages        = {444},
	doi          = {10.1088/1367-2630/9/12/444},
	url          = {https://dx.doi.org/10.1088/1367-2630/9/12/444}
}

@article{Burger2024,
	title        = {{KiDS-1000 cosmology: Combined second- and third-order shear statistics}},
	author       = {{Burger}, Pierre A. and {Porth}, Lucas and {Heydenreich}, Sven and {Linke}, Laila and {Wielders}, Niek and {Schneider}, Peter and {Asgari}, Marika and {Castro}, Tiago and {Dolag}, Klaus and {Harnois-D{\'e}raps}, Joachim and {Hildebrandt}, Hendrik and {Kuijken}, Konrad and {Martinet}, Nicolas},
	year         = {2024},
	month        = mar,
	volume       = {683},
	pages        = {A103},
	doi          = {10.1051/0004-6361/202347986},
	eid          = {A103},
	archiveprefix = {arXiv},
	eprint       = {2309.08602},
	primaryclass = {astro-ph.CO},
	adsurl       = {https://ui.adsabs.harvard.edu/abs/2024A&A...683A.103B}
}

@article{Castiblanco2024,
	title        = {{Unleashing cosmic shear information with the tomographic weak lensing PDF}},
	author       = {{Castiblanco}, Lina and {Uhlemann}, Cora and {Harnois-D{\'e}raps}, Joachim and {Barthelemy}, Alexandre},
	year         = {2024},
	month        = jul,
	journal      = {The Open Journal of Astrophysics},
	volume       = {7},
	pages        = {59},
	doi          = {10.33232/001c.121302},
	eid          = {59},
	archiveprefix = {arXiv},
	eprint       = {2405.09651},
	primaryclass = {astro-ph.CO},
	adsurl       = {https://ui.adsabs.harvard.edu/abs/2024OJAp....7E..59C}
}

@inproceedings{Papamakarios2019,
	title        = {Sequential Neural Likelihood: Fast Likelihood-free Inference with Autoregressive Flows},
	author       = {Papamakarios, George and Sterratt, David and Murray, Iain},
	year         = {2019},
	month        = {Apr},
	booktitle    = {Proceedings of the Twenty-Second International Conference on Artificial Intelligence and Statistics},
	publisher    = {PMLR},
	series       = {Proceedings of Machine Learning Research},
	volume       = {89},
	pages        = {837--848},
	url          = {https://proceedings.mlr.press/v89/papamakarios19a.html},
	editor       = {Chaudhuri, Kamalika and Sugiyama, Masashi}
}

@article{Cheng2020,
	title        = {{A new approach to observational cosmology using the scattering transform}},
	author       = {Cheng, Sihao and Ting, Yuan-Sen and Ménard, Brice and Bruna, Joan},
	year         = {2020},
	month        = {10},
	journal      = {Monthly Notices of the Royal Astronomical Society},
	volume       = {499},
	number       = {4},
	pages        = {5902--5914},
	doi          = {10.1093/mnras/staa3165},
	issn         = {0035-8711},
	url          = {https://doi.org/10.1093/mnras/staa3165},
	eprint       = {https://academic.oup.com/mnras/article-pdf/499/4/5902/34157889/staa3165.pdf}
}

@article{Chiang2014,
	title        = {Position-dependent power spectrum of the large-scale structure: a novel method to measure the squeezed-limit bispectrum},
	author       = {Chi-Ting Chiang and Christian Wagner and Fabian Schmidt and Eiichiro Komatsu},
	year         = {2014},
	month        = {5},
	journal      = {Journal of Cosmology and Astroparticle Physics},
	volume       = {2014},
	number       = {05},
	pages        = {048},
	doi          = {10.1088/1475-7516/2014/05/048},
	url          = {https://dx.doi.org/10.1088/1475-7516/2014/05/048}
}

@article{Chisari2018,
	title        = {{The impact of baryons on the matter power spectrum from the Horizon-AGN cosmological hydrodynamical simulation}},
	author       = {{Chisari}, N.~E. and {Richardson}, M.~L.~A. and {Devriendt}, J. and {Dubois}, Y. and {Schneider}, A. and {Le Brun}, A.~M.~C. and {Beckmann}, R.~S. and {Peirani}, S. and {Slyz}, A. and {Pichon}, C.},
	year         = {2018},
	month        = nov,
	volume       = {480},
	number       = {3},
	pages        = {3962--3977},
	doi          = {10.1093/mnras/sty2093},
	archiveprefix = {arXiv},
	eprint       = {1801.08559},
	primaryclass = {astro-ph.CO},
	adsurl       = {https://ui.adsabs.harvard.edu/abs/2018MNRAS.480.3962C}
}

@article{Deistler2025,
	title        = {{Simulation-Based Inference: A Practical Guide}},
	author       = {{Deistler}, Michael and {Boelts}, Jan and {Steinbach}, Peter and {Moss}, Guy and {Moreau}, Thomas and {Gloeckler}, Manuel and {Rodrigues}, Pedro L. C. and {Linhart}, Julia and {Lappalainen}, Janne K. and {Miller}, Benjamin Kurt and {Gon{\c{c}}alves}, Pedro J. and {Lueckmann}, Jan-Matthis and {Schr{\"o}der}, Cornelius and {Macke}, Jakob H.},
	year         = {2025},
	month        = aug,
	journal      = {arXiv e-prints},
	pages        = {arXiv:2508.12939},
	doi          = {10.48550/arXiv.2508.12939},
	eid          = {arXiv:2508.12939},
	archiveprefix = {arXiv},
	eprint       = {2508.12939},
	primaryclass = {stat.ML},
	adsurl       = {https://ui.adsabs.harvard.edu/abs/2025arXiv250812939D}
}

@article{DES_overview,
	title        = {{The Dark Energy Survey: more than dark energy - an overview}},
	author       = {{Dark Energy Survey Collaboration} and {Abbott}, T. and {Abdalla}, F.~B. and {Aleksi{\'c}}, J. and {Allam}, S. and {Amara}, A. and {Bacon}, D. and {Balbinot}, E. and {Banerji}, M. and {Bechtol}, K. and {Benoit-L{\'e}vy}, A. and {Bernstein}, G.~M. and {Bertin}, E. and {Blazek}, J. and {Bonnett}, C. and {Bridle}, S. and {Brooks}, D. and {Brunner}, R.~J. and {Buckley-Geer}, E. and {Burke}, D.~L. and {Caminha}, G.~B. and {Capozzi}, D. and {Carlsen}, J. and {Carnero-Rosell}, A. and {Carollo}, M. and {Carrasco-Kind}, M. and {Carretero}, J. and {Castander}, F.~J. and {Clerkin}, L. and {Collett}, T. and {Conselice}, C. and {Crocce}, M. and {Cunha}, C.~E. and {D'Andrea}, C.~B. and {da Costa}, L.~N. and {Davis}, T.~M. and {Desai}, S. and {Diehl}, H.~T. and {Dietrich}, J.~P. and {Dodelson}, S. and {Doel}, P. and {Drlica-Wagner}, A. and {Estrada}, J. and {Etherington}, J. and {Evrard}, A.~E. and {Fabbri}, J. and {Finley}, D.~A. and {Flaugher}, B. and {Foley}, R.~J. and {Fosalba}, P. and {Frieman}, J. and {Garc{\'\i}a-Bellido}, J. and {Gaztanaga}, E. and {Gerdes}, D.~W. and {Giannantonio}, T. and {Goldstein}, D.~A. and {Gruen}, D. and {Gruendl}, R.~A. and {Guarnieri}, P. and {Gutierrez}, G. and {Hartley}, W. and {Honscheid}, K. and {Jain}, B. and {James}, D.~J. and {Jeltema}, T. and {Jouvel}, S. and {Kessler}, R. and {King}, A. and {Kirk}, D. and {Kron}, R. and {Kuehn}, K. and {Kuropatkin}, N. and {Lahav}, O. and {Li}, T.~S. and {Lima}, M. and {Lin}, H. and {Maia}, M.~A.~G. and {Makler}, M. and {Manera}, M. and {Maraston}, C. and {Marshall}, J.~L. and {Martini}, P. and {McMahon}, R.~G. and {Melchior}, P. and {Merson}, A. and {Miller}, C.~J. and {Miquel}, R. and {Mohr}, J.~J. and {Morice-Atkinson}, X. and {Naidoo}, K. and {Neilsen}, E. and {Nichol}, R.~C. and {Nord}, B. and {Ogando}, R. and {Ostrovski}, F. and {Palmese}, A. and {Papadopoulos}, A. and {Peiris}, H.~V. and {Peoples}, J. and {Percival}, W.~J. and {Plazas}, A.~A. and {Reed}, S.~L. and {Refregier}, A. and {Romer}, A.~K. and {Roodman}, A. and {Ross}, A. and {Rozo}, E. and {Rykoff}, E.~S. and {Sadeh}, I. and {Sako}, M. and {S{\'a}nchez}, C. and {Sanchez}, E. and {Santiago}, B. and {Scarpine}, V. and {Schubnell}, M. and {Sevilla-Noarbe}, I. and {Sheldon}, E. and {Smith}, M. and {Smith}, R.~C. and {Soares-Santos}, M. and {Sobreira}, F. and {Soumagnac}, M. and {Suchyta}, E. and {Sullivan}, M. and {Swanson}, M. and {Tarle}, G. and {Thaler}, J. and {Thomas}, D. and {Thomas}, R.~C. and {Tucker}, D. and {Vieira}, J.~D. and {Vikram}, V. and {Walker}, A.~R. and {Wechsler}, R.~H. and {Weller}, J. and {Wester}, W. and {Whiteway}, L. and {Wilcox}, H. and {Yanny}, B. and {Zhang}, Y. and {Zuntz}, J.},
	year         = {2016},
	month        = aug,
	volume       = {460},
	number       = {2},
	pages        = {1270--1299},
	doi          = {10.1093/mnras/stw641},
	archiveprefix = {arXiv},
	eprint       = {1601.00329},
	primaryclass = {astro-ph.CO},
	adsurl       = {https://ui.adsabs.harvard.edu/abs/2016MNRAS.460.1270D}
}

@article{DES_Y3_data,
	title        = {{Dark Energy Survey Year 3 Results: Photometric Data Set for Cosmology}},
	author       = {{Sevilla-Noarbe}, I. and {Bechtol}, K. and {Carrasco Kind}, M. and {Carnero Rosell}, A. and {Becker}, M.~R. and {Drlica-Wagner}, A. and {Gruendl}, R.~A. and {Rykoff}, E.~S. and {Sheldon}, E. and {Yanny}, B. and {Alarcon}, A. and {Allam}, S. and {Amon}, A. and {Benoit-L{\'e}vy}, A. and {Bernstein}, G.~M. and {Bertin}, E. and {Burke}, D.~L. and {Carretero}, J. and {Choi}, A. and {Diehl}, H.~T. and {Everett}, S. and {Flaugher}, B. and {Gaztanaga}, E. and {Gschwend}, J. and {Harrison}, I. and {Hartley}, W.~G. and {Hoyle}, B. and {Jarvis}, M. and {Johnson}, M.~D. and {Kessler}, R. and {Kron}, R. and {Kuropatkin}, N. and {Leistedt}, B. and {Li}, T.~S. and {Menanteau}, F. and {Morganson}, E. and {Ogando}, R.~L.~C. and {Palmese}, A. and {Paz-Chinch{\'o}n}, F. and {Pieres}, A. and {Pond}, C. and {Rodriguez-Monroy}, M. and {Smith}, J. Allyn and {Stringer}, K.~M. and {Troxel}, M.~A. and {Tucker}, D.~L. and {de Vicente}, J. and {Wester}, W. and {Zhang}, Y. and {Abbott}, T.~M.~C. and {Aguena}, M. and {Annis}, J. and {Avila}, S. and {Bhargava}, S. and {Bridle}, S.~L. and {Brooks}, D. and {Brout}, D. and {Castander}, F.~J. and {Cawthon}, R. and {Chang}, C. and {Conselice}, C. and {Costanzi}, M. and {Crocce}, M. and {da Costa}, L.~N. and {Pereira}, M.~E.~S. and {Davis}, T.~M. and {Desai}, S. and {Dietrich}, J.~P. and {Doel}, P. and {Eckert}, K. and {Evrard}, A.~E. and {Ferrero}, I. and {Fosalba}, P. and {Garc{\'\i}a-Bellido}, J. and {Gerdes}, D.~W. and {Giannantonio}, T. and {Gruen}, D. and {Gutierrez}, G. and {Hinton}, S.~R. and {Hollowood}, D.~L. and {Honscheid}, K. and {Huff}, E.~M. and {Huterer}, D. and {James}, D.~J. and {Jeltema}, T. and {Kuehn}, K. and {Lahav}, O. and {Lidman}, C. and {Lima}, M. and {Lin}, H. and {Maia}, M.~A.~G. and {Marshall}, J.~L. and {Martini}, P. and {Melchior}, P. and {Miquel}, R. and {Mohr}, J.~J. and {Morgan}, R. and {Neilsen}, E. and {Plazas}, A.~A. and {Romer}, A.~K. and {Roodman}, A. and {Sanchez}, E. and {Scarpine}, V. and {Schubnell}, M. and {Serrano}, S. and {Smith}, M. and {Suchyta}, E. and {Tarle}, G. and {Thomas}, D. and {To}, C. and {Varga}, T.~N. and {Wechsler}, R.~H. and {Weller}, J. and {Wilkinson}, R.~D. and {DES Collaboration}},
	year         = {2021},
	month        = jun,
	volume       = {254},
	number       = {2},
	pages        = {24},
	doi          = {10.3847/1538-4365/abeb66},
	eid          = {24},
	archiveprefix = {arXiv},
	eprint       = {2011.03407},
	primaryclass = {astro-ph.CO},
	adsurl       = {https://ui.adsabs.harvard.edu/abs/2021ApJS..254...24S}
}

@article{Deshpande2024,
	title        = {{Euclid preparation. XXXVI. Modelling the weak lensing angular power spectrum}},
	author       = {{Euclid Collaboration} and {Deshpande}, A.~C. and {Kitching}, T. and {Hall}, A. and {Brown}, M.~L. and {Aghanim}, N. and {Amendola}, L. and {Andreon}, S. and {Auricchio}, N. and {Baldi}, M. and {Bardelli}, S. and {Bender}, R. and {Bonino}, D. and {Branchini}, E. and {Brescia}, M. and {Brinchmann}, J. and {Camera}, S. and {Candini}, G.~P. and {Capobianco}, V. and {Carbone}, C. and {Cardone}, V.~F. and {Carretero}, J. and {Casas}, S. and {Castander}, F.~J. and {Castellano}, M. and {Cavuoti}, S. and {Cimatti}, A. and {Cledassou}, R. and {Congedo}, G. and {Conselice}, C.~J. and {Conversi}, L. and {Corcione}, L. and {Courbin}, F. and {Courtois}, H.~M. and {Cropper}, M. and {Da Silva}, A. and {Degaudenzi}, H. and {Douspis}, M. and {Dubath}, F. and {Duncan}, C.~A.~J. and {Dupac}, X. and {Farina}, M. and {Farrens}, S. and {Ferriol}, S. and {Fosalba}, P. and {Frailis}, M. and {Franceschi}, E. and {Fumana}, M. and {Galeotta}, S. and {Garilli}, B. and {Gillis}, B. and {Giocoli}, C. and {Grazian}, A. and {Grupp}, F. and {Haugan}, S.~V.~H. and {Hoekstra}, H. and {Holmes}, W. and {Hornstrup}, A. and {Hudelot}, P. and {Jahnke}, K. and {Keih{\"a}nen}, E. and {Kermiche}, S. and {Kilbinger}, M. and {Kunz}, M. and {Kurki-Suonio}, H. and {Ligori}, S. and {Lilje}, P.~B. and {Lindholm}, V. and {Lloro}, I. and {Maiorano}, E. and {Mansutti}, O. and {Marggraf}, O. and {Markovic}, K. and {Martinet}, N. and {Marulli}, F. and {Massey}, R. and {Mei}, S. and {Mellier}, Y. and {Meneghetti}, M. and {Meylan}, G. and {Moscardini}, L. and {Niemi}, S. -M. and {Nightingale}, J.~W. and {Nutma}, T. and {Padilla}, C. and {Paltani}, S. and {Pasian}, F. and {Pedersen}, K. and {Pettorino}, V. and {Pires}, S. and {Polenta}, G. and {Pollack}, J. and {Poncet}, M. and {Popa}, L.~A. and {Raison}, F. and {Renzi}, A. and {Rhodes}, J. and {Riccio}, G. and {Romelli}, E. and {Roncarelli}, M. and {Rossetti}, E. and {Saglia}, R. and {Sapone}, D. and {Sartoris}, B. and {Schneider}, P. and {Schrabback}, T. and {Secroun}, A. and {Seidel}, G. and {Serrano}, S. and {Sirignano}, C. and {Sirri}, G. and {Stanco}, L. and {Tallada-Cresp{\'\i}}, P. and {Taylor}, A.~N. and {Tereno}, I. and {Toledo-Moreo}, R. and {Torradeflot}, F. and {Tutusaus}, I. and {Valentijn}, E.~A. and {Valenziano}, L. and {Vassallo}, T. and {Wang}, Y. and {Weller}, J. and {Zacchei}, A. and {Zamorani}, G. and {Zoubian}, J. and {Zucca}, E. and {Boucaud}, A. and {Bozzo}, E. and {Colodro-Conde}, C. and {Di Ferdinando}, D. and {Fabbian}, G. and {Graci{\'a}-Carpio}, J. and {Mauri}, N. and {Scottez}, V. and {Tenti}, M. and {Akrami}, Y. and {Baccigalupi}, C. and {Balaguera-Antol{\'\i}nez}, A. and {Ballardini}, M. and {Bernardeau}, F. and {Biviano}, A. and {Blanchard}, A. and {Borlaff}, A.~S. and {Burigana}, C. and {Cabanac}, R. and {Cappi}, A. and {Carvalho}, C.~S. and {Castignani}, G. and {Castro}, T. and {Chambers}, K.~C. and {Cooray}, A.~R. and {Coupon}, J. and {Davini}, S. and {de la Torre}, S. and {De Lucia}, G. and {Desprez}, G. and {Dole}, H. and {Escartin}, J.~A. and {Escoffier}, S. and {Ferrero}, I. and {Finelli}, F. and {Garcia-Bellido}, J. and {George}, K. and {Giacomini}, F. and {Gozaliasl}, G. and {Hildebrandt}, H. and {Kajava}, J.~J.~E. and {Kansal}, V. and {Kirkpatrick}, C.~C. and {Legrand}, L. and {Loureiro}, A. and {Macias-Perez}, J. and {Magliocchetti}, M. and {Mainetti}, G. and {Maoli}, R. and {Martinelli}, M. and {Martins}, C.~J.~A.~P. and {Matthew}, S. and {Maurin}, L. and {Metcalf}, R.~B. and {Monaco}, P. and {Morgante}, G. and {Nadathur}, S. and {Nucita}, A.~A. and {Patrizii}, L. and {Peel}, A. and {P{\"o}ntinen}, M. and {Popa}, V. and {Porciani}, C. and {Potter}, D. and {Pourtsidou}, A. and {Reimberg}, P. and {Sakr}, Z. and {S{\'a}nchez}, A.~G. and {Schneider}, A. and {Sefusatti}, E. and {Sereno}, M. and {Shulevski}, A. and {Spurio Mancini}, A. and {Steinwagner}, J. and {Teyssier}, R. and {Viel}, M. and {Zinchenko}, I.~A. and {Fleury}, P.},
	year         = {2024},
	month        = apr,
	journal      = {Astronomy \& Astrophysics},
	volume       = {684},
	pages        = {A138},
	doi          = {10.1051/0004-6361/202346110},
	eid          = {A138},
	archiveprefix = {arXiv},
	eprint       = {2302.04507},
	primaryclass = {astro-ph.CO},
	adsurl       = {https://ui.adsabs.harvard.edu/abs/2024A&A...684A.138E}
}

@article{DESY3_results,
	title        = {{Dark Energy Survey Year 3 results: Cosmological constraints from galaxy clustering and weak lensing}},
	author       = {{Abbott}, T.~M.~C. and {Aguena}, M. and {Alarcon}, A. and {Allam}, S. and {Alves}, O. and {Amon}, A. and {Andrade-Oliveira}, F. and {Annis}, J. and {Avila}, S. and {Bacon}, D. and {Baxter}, E. and {Bechtol}, K. and {Becker}, M.~R. and {Bernstein}, G.~M. and {Bhargava}, S. and {Birrer}, S. and {Blazek}, J. and {Brandao-Souza}, A. and {Bridle}, S.~L. and {Brooks}, D. and {Buckley-Geer}, E. and {Burke}, D.~L. and {Camacho}, H. and {Campos}, A. and {Carnero Rosell}, A. and {Carrasco Kind}, M. and {Carretero}, J. and {Castander}, F.~J. and {Cawthon}, R. and {Chang}, C. and {Chen}, A. and {Chen}, R. and {Choi}, A. and {Conselice}, C. and {Cordero}, J. and {Costanzi}, M. and {Crocce}, M. and {da Costa}, L.~N. and {da Silva Pereira}, M.~E. and {Davis}, C. and {Davis}, T.~M. and {De Vicente}, J. and {DeRose}, J. and {Desai}, S. and {Di Valentino}, E. and {Diehl}, H.~T. and {Dietrich}, J.~P. and {Dodelson}, S. and {Doel}, P. and {Doux}, C. and {Drlica-Wagner}, A. and {Eckert}, K. and {Eifler}, T.~F. and {Elsner}, F. and {Elvin-Poole}, J. and {Everett}, S. and {Evrard}, A.~E. and {Fang}, X. and {Farahi}, A. and {Fernandez}, E. and {Ferrero}, I. and {Fert{\'e}}, A. and {Fosalba}, P. and {Friedrich}, O. and {Frieman}, J. and {Garc{\'\i}a-Bellido}, J. and {Gatti}, M. and {Gaztanaga}, E. and {Gerdes}, D.~W. and {Giannantonio}, T. and {Giannini}, G. and {Gruen}, D. and {Gruendl}, R.~A. and {Gschwend}, J. and {Gutierrez}, G. and {Harrison}, I. and {Hartley}, W.~G. and {Herner}, K. and {Hinton}, S.~R. and {Hollowood}, D.~L. and {Honscheid}, K. and {Hoyle}, B. and {Huff}, E.~M. and {Huterer}, D. and {Jain}, B. and {James}, D.~J. and {Jarvis}, M. and {Jeffrey}, N. and {Jeltema}, T. and {Kovacs}, A. and {Krause}, E. and {Kron}, R. and {Kuehn}, K. and {Kuropatkin}, N. and {Lahav}, O. and {Leget}, P. -F. and {Lemos}, P. and {Liddle}, A.~R. and {Lidman}, C. and {Lima}, M. and {Lin}, H. and {MacCrann}, N. and {Maia}, M.~A.~G. and {Marshall}, J.~L. and {Martini}, P. and {McCullough}, J. and {Melchior}, P. and {Mena-Fern{\'a}ndez}, J. and {Menanteau}, F. and {Miquel}, R. and {Mohr}, J.~J. and {Morgan}, R. and {Muir}, J. and {Myles}, J. and {Nadathur}, S. and {Navarro-Alsina}, A. and {Nichol}, R.~C. and {Ogando}, R.~L.~C. and {Omori}, Y. and {Palmese}, A. and {Pandey}, S. and {Park}, Y. and {Paz-Chinch{\'o}n}, F. and {Petravick}, D. and {Pieres}, A. and {Plazas Malag{\'o}n}, A.~A. and {Porredon}, A. and {Prat}, J. and {Raveri}, M. and {Rodriguez-Monroy}, M. and {Rollins}, R.~P. and {Romer}, A.~K. and {Roodman}, A. and {Rosenfeld}, R. and {Ross}, A.~J. and {Rykoff}, E.~S. and {Samuroff}, S. and {S{\'a}nchez}, C. and {Sanchez}, E. and {Sanchez}, J. and {Sanchez Cid}, D. and {Scarpine}, V. and {Schubnell}, M. and {Scolnic}, D. and {Secco}, L.~F. and {Serrano}, S. and {Sevilla-Noarbe}, I. and {Sheldon}, E. and {Shin}, T. and {Smith}, M. and {Soares-Santos}, M. and {Suchyta}, E. and {Swanson}, M.~E.~C. and {Tabbutt}, M. and {Tarle}, G. and {Thomas}, D. and {To}, C. and {Troja}, A. and {Troxel}, M.~A. and {Tucker}, D.~L. and {Tutusaus}, I. and {Varga}, T.~N. and {Walker}, A.~R. and {Weaverdyck}, N. and {Wechsler}, R. and {Weller}, J. and {Yanny}, B. and {Yin}, B. and {Zhang}, Y. and {Zuntz}, J. and {DES Collaboration}},
	year         = {2022},
	month        = jan,
	volume       = {105},
	number       = {2},
	pages        = {023520},
	doi          = {10.1103/PhysRevD.105.023520},
	eid          = {023520},
	archiveprefix = {arXiv},
	eprint       = {2105.13549},
	primaryclass = {astro-ph.CO},
	adsurl       = {https://ui.adsabs.harvard.edu/abs/2022PhRvD.105b3520A}
}

@article{Euclid_HOWLS,
	title        = {{Euclid preparation. XXVIII. Forecasts for ten different higher-order weak lensing statistics}},
	author       = {{Euclid Collaboration} and {Ajani}, V. and {Baldi}, M. and {Barthelemy}, A. and {Boyle}, A. and {Burger}, P. and {Cardone}, V.~F. and {Cheng}, S. and {Codis}, S. and {Giocoli}, C. and {Harnois-D{\'e}raps}, J. and {Heydenreich}, S. and {Kansal}, V. and {Kilbinger}, M. and {Linke}, L. and {Llinares}, C. and {Martinet}, N. and {Parroni}, C. and {Peel}, A. and {Pires}, S. and {Porth}, L. and {Tereno}, I. and {Uhlemann}, C. and {Vicinanza}, M. and {Vinciguerra}, S. and {Aghanim}, N. and {Auricchio}, N. and {Bonino}, D. and {Branchini}, E. and {Brescia}, M. and {Brinchmann}, J. and {Camera}, S. and {Capobianco}, V. and {Carbone}, C. and {Carretero}, J. and {Castander}, F.~J. and {Castellano}, M. and {Cavuoti}, S. and {Cimatti}, A. and {Cledassou}, R. and {Congedo}, G. and {Conselice}, C.~J. and {Conversi}, L. and {Corcione}, L. and {Courbin}, F. and {Cropper}, M. and {Da Silva}, A. and {Degaudenzi}, H. and {Di Giorgio}, A.~M. and {Dinis}, J. and {Douspis}, M. and {Dubath}, F. and {Dupac}, X. and {Farrens}, S. and {Ferriol}, S. and {Fosalba}, P. and {Frailis}, M. and {Franceschi}, E. and {Galeotta}, S. and {Garilli}, B. and {Gillis}, B. and {Grazian}, A. and {Grupp}, F. and {Hoekstra}, H. and {Holmes}, W. and {Hornstrup}, A. and {Hudelot}, P. and {Jahnke}, K. and {Jhabvala}, M. and {K{\"u}mmel}, M. and {Kitching}, T. and {Kunz}, M. and {Kurki-Suonio}, H. and {Lilje}, P.~B. and {Lloro}, I. and {Maiorano}, E. and {Mansutti}, O. and {Marggraf}, O. and {Markovic}, K. and {Marulli}, F. and {Massey}, R. and {Mei}, S. and {Mellier}, Y. and {Meneghetti}, M. and {Moresco}, M. and {Moscardini}, L. and {Niemi}, S. -M. and {Nightingale}, J. and {Nutma}, T. and {Padilla}, C. and {Paltani}, S. and {Pedersen}, K. and {Pettorino}, V. and {Polenta}, G. and {Poncet}, M. and {Popa}, L.~A. and {Raison}, F. and {Renzi}, A. and {Rhodes}, J. and {Riccio}, G. and {Romelli}, E. and {Roncarelli}, M. and {Rossetti}, E. and {Saglia}, R. and {Sapone}, D. and {Sartoris}, B. and {Schneider}, P. and {Schrabback}, T. and {Secroun}, A. and {Seidel}, G. and {Serrano}, S. and {Sirignano}, C. and {Stanco}, L. and {Starck}, J. -L. and {Tallada-Cresp{\'\i}}, P. and {Taylor}, A.~N. and {Toledo-Moreo}, R. and {Torradeflot}, F. and {Tutusaus}, I. and {Valentijn}, E.~A. and {Valenziano}, L. and {Vassallo}, T. and {Wang}, Y. and {Weller}, J. and {Zamorani}, G. and {Zoubian}, J. and {Andreon}, S. and {Bardelli}, S. and {Boucaud}, A. and {Bozzo}, E. and {Colodro-Conde}, C. and {Di Ferdinando}, D. and {Fabbian}, G. and {Farina}, M. and {Graci{\'a}-Carpio}, J. and {Keih{\"a}nen}, E. and {Lindholm}, V. and {Maino}, D. and {Mauri}, N. and {Neissner}, C. and {Schirmer}, M. and {Scottez}, V. and {Zucca}, E. and {Akrami}, Y. and {Baccigalupi}, C. and {Balaguera-Antol{\'\i}nez}, A. and {Ballardini}, M. and {Bernardeau}, F. and {Biviano}, A. and {Blanchard}, A. and {Borgani}, S. and {Borlaff}, A.~S. and {Burigana}, C. and {Cabanac}, R. and {Cappi}, A. and {Carvalho}, C.~S. and {Casas}, S. and {Castignani}, G. and {Castro}, T. and {Chambers}, K.~C. and {Cooray}, A.~R. and {Coupon}, J. and {Courtois}, H.~M. and {Davini}, S. and {de la Torre}, S. and {De Lucia}, G. and {Desprez}, G. and {Dole}, H. and {Escartin}, J.~A. and {Escoffier}, S. and {Ferrero}, I. and {Finelli}, F. and {Ganga}, K. and {Garcia-Bellido}, J. and {George}, K. and {Giacomini}, F. and {Gozaliasl}, G. and {Hildebrandt}, H. and {Jimenez Mu{\~n}oz}, A. and {Joachimi}, B. and {Kajava}, J.~J.~E. and {Kirkpatrick}, C.~C. and {Legrand}, L. and {Loureiro}, A. and {Magliocchetti}, M. and {Maoli}, R. and {Marcin}, S. and {Martinelli}, M. and {Martins}, C.~J.~A.~P. and {Matthew}, S. and {Maurin}, L. and {Metcalf}, R.~B. and {Monaco}, P. and {Morgante}, G. and {Nadathur}, S. and {Nucita}, A.~A. and {Popa}, V. and {Potter}, D. and {Pourtsidou}, A. and {P{\"o}ntinen}, M.},
	year         = {2023},
	month        = jul,
	volume       = {675},
	pages        = {A120},
	doi          = {10.1051/0004-6361/202346017},
	eid          = {A120},
	archiveprefix = {arXiv},
	eprint       = {2301.12890},
	primaryclass = {astro-ph.CO},
	adsurl       = {https://ui.adsabs.harvard.edu/abs/2023A&A...675A.120E}
}

@article{Euclid_overview,
	title        = {{Euclid preparation. I. The Euclid Wide Survey}},
	author       = {{Euclid Collaboration} and {Scaramella}, R. and {Amiaux}, J. and {Mellier}, Y. and {Burigana}, C. and {Carvalho}, C.~S. and {Cuillandre}, J. -C. and {Da Silva}, A. and {Derosa}, A. and {Dinis}, J. and {Maiorano}, E. and {Maris}, M. and {Tereno}, I. and {Laureijs}, R. and {Boenke}, T. and {Buenadicha}, G. and {Dupac}, X. and {Gaspar Venancio}, L.~M. and {G{\'o}mez-{\'A}lvarez}, P. and {Hoar}, J. and {Lorenzo Alvarez}, J. and {Racca}, G.~D. and {Saavedra-Criado}, G. and {Schwartz}, J. and {Vavrek}, R. and {Schirmer}, M. and {Aussel}, H. and {Azzollini}, R. and {Cardone}, V.~F. and {Cropper}, M. and {Ealet}, A. and {Garilli}, B. and {Gillard}, W. and {Granett}, B.~R. and {Guzzo}, L. and {Hoekstra}, H. and {Jahnke}, K. and {Kitching}, T. and {Maciaszek}, T. and {Meneghetti}, M. and {Miller}, L. and {Nakajima}, R. and {Niemi}, S.~M. and {Pasian}, F. and {Percival}, W.~J. and {Pottinger}, S. and {Sauvage}, M. and {Scodeggio}, M. and {Wachter}, S. and {Zacchei}, A. and {Aghanim}, N. and {Amara}, A. and {Auphan}, T. and {Auricchio}, N. and {Awan}, S. and {Balestra}, A. and {Bender}, R. and {Bodendorf}, C. and {Bonino}, D. and {Branchini}, E. and {Brau-Nogue}, S. and {Brescia}, M. and {Candini}, G.~P. and {Capobianco}, V. and {Carbone}, C. and {Carlberg}, R.~G. and {Carretero}, J. and {Casas}, R. and {Castander}, F.~J. and {Castellano}, M. and {Cavuoti}, S. and {Cimatti}, A. and {Cledassou}, R. and {Congedo}, G. and {Conselice}, C.~J. and {Conversi}, L. and {Copin}, Y. and {Corcione}, L. and {Costille}, A. and {Courbin}, F. and {Degaudenzi}, H. and {Douspis}, M. and {Dubath}, F. and {Duncan}, C.~A.~J. and {Dusini}, S. and {Farrens}, S. and {Ferriol}, S. and {Fosalba}, P. and {Fourmanoit}, N. and {Frailis}, M. and {Franceschi}, E. and {Franzetti}, P. and {Fumana}, M. and {Gillis}, B. and {Giocoli}, C. and {Grazian}, A. and {Grupp}, F. and {Haugan}, S.~V.~H. and {Holmes}, W. and {Hormuth}, F. and {Hudelot}, P. and {Kermiche}, S. and {Kiessling}, A. and {Kilbinger}, M. and {Kohley}, R. and {Kubik}, B. and {K{\"u}mmel}, M. and {Kunz}, M. and {Kurki-Suonio}, H. and {Lahav}, O. and {Ligori}, S. and {Lilje}, P.~B. and {Lloro}, I. and {Mansutti}, O. and {Marggraf}, O. and {Markovic}, K. and {Marulli}, F. and {Massey}, R. and {Maurogordato}, S. and {Melchior}, M. and {Merlin}, E. and {Meylan}, G. and {Mohr}, J.~J. and {Moresco}, M. and {Morin}, B. and {Moscardini}, L. and {Munari}, E. and {Nichol}, R.~C. and {Padilla}, C. and {Paltani}, S. and {Peacock}, J. and {Pedersen}, K. and {Pettorino}, V. and {Pires}, S. and {Poncet}, M. and {Popa}, L. and {Pozzetti}, L. and {Raison}, F. and {Rebolo}, R. and {Rhodes}, J. and {Rix}, H. -W. and {Roncarelli}, M. and {Rossetti}, E. and {Saglia}, R. and {Schneider}, P. and {Schrabback}, T. and {Secroun}, A. and {Seidel}, G. and {Serrano}, S. and {Sirignano}, C. and {Sirri}, G. and {Skottfelt}, J. and {Stanco}, L. and {Starck}, J.~L. and {Tallada-Cresp{\'\i}}, P. and {Tavagnacco}, D. and {Taylor}, A.~N. and {Teplitz}, H.~I. and {Toledo-Moreo}, R. and {Torradeflot}, F. and {Trifoglio}, M. and {Valentijn}, E.~A. and {Valenziano}, L. and {Verdoes Kleijn}, G.~A. and {Wang}, Y. and {Welikala}, N. and {Weller}, J. and {Wetzstein}, M. and {Zamorani}, G. and {Zoubian}, J. and {Andreon}, S. and {Baldi}, M. and {Bardelli}, S. and {Boucaud}, A. and {Camera}, S. and {Di Ferdinando}, D. and {Fabbian}, G. and {Farinelli}, R. and {Galeotta}, S. and {Graci{\'a}-Carpio}, J. and {Maino}, D. and {Medinaceli}, E. and {Mei}, S. and {Neissner}, C. and {Polenta}, G. and {Renzi}, A. and {Romelli}, E. and {Rosset}, C. and {Sureau}, F. and {Tenti}, M. and {Vassallo}, T. and {Zucca}, E. and {Baccigalupi}, C. and {Balaguera-Antol{\'\i}nez}, A. and {Battaglia}, P. and {Biviano}, A. and {Borgani}, S. and {Bozzo}, E. and {Cabanac}, R. and {Cappi}, A.},
	year         = {2022},
	month        = jun,
	volume       = {662},
	pages        = {A112},
	doi          = {10.1051/0004-6361/202141938},
	eid          = {A112},
	archiveprefix = {arXiv},
	eprint       = {2108.01201},
	primaryclass = {astro-ph.CO},
	adsurl       = {https://ui.adsabs.harvard.edu/abs/2022A&A...662A.112E}
}

@article{Euclid_overview2,
	title        = {{Euclid -- The Dark Universe detective}},
	author       = {{Linke}, L.},
	year         = {2024},
	month        = may,
	journal      = {arXiv e-prints},
	pages        = {arXiv:2405.01037},
	doi          = {10.48550/arXiv.2405.01037},
	eid          = {arXiv:2405.01037},
	archiveprefix = {arXiv},
	eprint       = {2405.01037},
	primaryclass = {astro-ph.CO},
	adsurl       = {https://ui.adsabs.harvard.edu/abs/2024arXiv240501037L}
}

@article{Fluri2022,
	title        = {{Full w CDM analysis of KiDS-1000 weak lensing maps using deep learning}},
	author       = {{Fluri}, Janis and {Kacprzak}, Tomasz and {Lucchi}, Aurelien and {Schneider}, Aurel and {Refregier}, Alexandre and {Hofmann}, Thomas},
	year         = {2022},
	month        = apr,
	journal      = {Phys. Rev. D},
	volume       = {105},
	number       = {8},
	pages        = {083518},
	doi          = {10.1103/PhysRevD.105.083518},
	eid          = {083518},
	archiveprefix = {arXiv},
	eprint       = {2201.07771},
	primaryclass = {astro-ph.CO},
	adsurl       = {https://ui.adsabs.harvard.edu/abs/2022PhRvD.105h3518F}
}

@article{Foreman2013,
	title        = {{emcee: The MCMC Hammer}},
	author       = {{Foreman-Mackey}, Daniel and {Hogg}, David W. and {Lang}, Dustin and {Goodman}, Jonathan},
	year         = {2013},
	month        = mar,
	volume       = {125},
	number       = {925},
	pages        = {306},
	doi          = {10.1086/670067},
	archiveprefix = {arXiv},
	eprint       = {1202.3665},
	primaryclass = {astro-ph.IM},
	adsurl       = {https://ui.adsabs.harvard.edu/abs/2013PASP..125..306F}
}

@article{Friedrich2016,
	title        = {{Performance of internal covariance estimators for cosmic shear correlation functions}},
	author       = {{Friedrich}, O. and {Seitz}, S. and {Eifler}, T.~F. and {Gruen}, D.},
	year         = {2016},
	month        = mar,
	volume       = {456},
	number       = {3},
	pages        = {2662--2680},
	doi          = {10.1093/mnras/stv2833},
	archiveprefix = {arXiv},
	eprint       = {1508.00895},
	primaryclass = {astro-ph.CO},
	adsurl       = {https://ui.adsabs.harvard.edu/abs/2016MNRAS.456.2662F}
}

@article{Friedrich2021,
	title        = {{Dark Energy Survey year 3 results: covariance modelling and its impact on parameter estimation and quality of fit}},
	author       = {{Friedrich}, O. and {Andrade-Oliveira}, F. and {Camacho}, H. and {Alves}, O. and {Rosenfeld}, R. and {Sanchez}, J. and {Fang}, X. and {Eifler}, T.~F. and {Krause}, E. and {Chang}, C. and {Omori}, Y. and {Amon}, A. and {Baxter}, E. and {Elvin-Poole}, J. and {Huterer}, D. and {Porredon}, A. and {Prat}, J. and {Terra}, V. and {Troja}, A. and {Alarcon}, A. and {Bechtol}, K. and {Bernstein}, G.~M. and {Buchs}, R. and {Campos}, A. and {Carnero Rosell}, A. and {Carrasco Kind}, M. and {Cawthon}, R. and {Choi}, A. and {Cordero}, J. and {Crocce}, M. and {Davis}, C. and {DeRose}, J. and {Diehl}, H.~T. and {Dodelson}, S. and {Doux}, C. and {Drlica-Wagner}, A. and {Elsner}, F. and {Everett}, S. and {Fosalba}, P. and {Gatti}, M. and {Giannini}, G. and {Gruen}, D. and {Gruendl}, R.~A. and {Harrison}, I. and {Hartley}, W.~G. and {Jain}, B. and {Jarvis}, M. and {MacCrann}, N. and {McCullough}, J. and {Muir}, J. and {Myles}, J. and {Pandey}, S. and {Raveri}, M. and {Roodman}, A. and {Rodriguez-Monroy}, M. and {Rykoff}, E.~S. and {Samuroff}, S. and {S{\'a}nchez}, C. and {Secco}, L.~F. and {Sevilla-Noarbe}, I. and {Sheldon}, E. and {Troxel}, M.~A. and {Weaverdyck}, N. and {Yanny}, B. and {Aguena}, M. and {Avila}, S. and {Bacon}, D. and {Bertin}, E. and {Bhargava}, S. and {Brooks}, D. and {Burke}, D.~L. and {Carretero}, J. and {Costanzi}, M. and {da Costa}, L.~N. and {Pereira}, M.~E.~S. and {De Vicente}, J. and {Desai}, S. and {Evrard}, A.~E. and {Ferrero}, I. and {Frieman}, J. and {Garc{\'\i}a-Bellido}, J. and {Gaztanaga}, E. and {Gerdes}, D.~W. and {Giannantonio}, T. and {Gschwend}, J. and {Gutierrez}, G. and {Hinton}, S.~R. and {Hollowood}, D.~L. and {Honscheid}, K. and {James}, D.~J. and {Kuehn}, K. and {Lahav}, O. and {Lima}, M. and {Maia}, M.~A.~G. and {Menanteau}, F. and {Miquel}, R. and {Morgan}, R. and {Palmese}, A. and {Paz-Chinch{\'o}n}, F. and {Plazas}, A.~A. and {Sanchez}, E. and {Scarpine}, V. and {Serrano}, S. and {Soares-Santos}, M. and {Smith}, M. and {Suchyta}, E. and {Tarle}, G. and {Thomas}, D. and {To}, C. and {Varga}, T.~N. and {Weller}, J. and {Wilkinson}, R.~D. and {Wilkinson}, R.~D. and {DES Collaboration}},
	year         = {2021},
	month        = dec,
	volume       = {508},
	number       = {3},
	pages        = {3125--3165},
	doi          = {10.1093/mnras/stab2384},
	archiveprefix = {arXiv},
	eprint       = {2012.08568},
	primaryclass = {astro-ph.CO},
	adsurl       = {https://ui.adsabs.harvard.edu/abs/2021MNRAS.508.3125F}
}

@article{Friedrich2025,
	title        = {{Bye binormal: analysing the joint PDF of galaxy density and weak lensing convergence}},
	author       = {Friedrich, Oliver and Castiblanco, Lina and Halder, Anik and Uhlemann, Cora},
	year         = {2025},
	month        = {7},
	eprint       = {2507.16957},
	archiveprefix = {arXiv},
	primaryclass = {astro-ph.CO}
}

@article{Gatti2021,
	title        = {{Dark energy survey year 3 results: weak lensing shape catalogue}},
	author       = {{Gatti}, M. and {Sheldon}, E. and {Amon}, A. and {Becker}, M. and {Troxel}, M. and {Choi}, A. and {Doux}, C. and {MacCrann}, N. and {Navarro-Alsina}, A. and {Harrison}, I. and {Gruen}, D. and {Bernstein}, G. and {Jarvis}, M. and {Secco}, L.~F. and {Fert{\'e}}, A. and {Shin}, T. and {McCullough}, J. and {Rollins}, R.~P. and {Chen}, R. and {Chang}, C. and {Pandey}, S. and {Tutusaus}, I. and {Prat}, J. and {Elvin-Poole}, J. and {Sanchez}, C. and {Plazas}, A.~A. and {Roodman}, A. and {Zuntz}, J. and {Abbott}, T.~M.~C. and {Aguena}, M. and {Allam}, S. and {Annis}, J. and {Avila}, S. and {Bacon}, D. and {Bertin}, E. and {Bhargava}, S. and {Brooks}, D. and {Burke}, D.~L. and {Carnero Rosell}, A. and {Carrasco Kind}, M. and {Carretero}, J. and {Castander}, F.~J. and {Conselice}, C. and {Costanzi}, M. and {Crocce}, M. and {da Costa}, L.~N. and {Davis}, T.~M. and {De Vicente}, J. and {Desai}, S. and {Diehl}, H.~T. and {Dietrich}, J.~P. and {Doel}, P. and {Drlica-Wagner}, A. and {Eckert}, K. and {Everett}, S. and {Ferrero}, I. and {Frieman}, J. and {Garc{\'\i}a-Bellido}, J. and {Gerdes}, D.~W. and {Giannantonio}, T. and {Gruendl}, R.~A. and {Gschwend}, J. and {Gutierrez}, G. and {Hartley}, W.~G. and {Hinton}, S.~R. and {Hollowood}, D.~L. and {Honscheid}, K. and {Hoyle}, B. and {Huff}, E.~M. and {Huterer}, D. and {Jain}, B. and {James}, D.~J. and {Jeltema}, T. and {Krause}, E. and {Kron}, R. and {Kuropatkin}, N. and {Lima}, M. and {Maia}, M.~A.~G. and {Marshall}, J.~L. and {Miquel}, R. and {Morgan}, R. and {Myles}, J. and {Palmese}, A. and {Paz-Chinch{\'o}n}, F. and {Rykoff}, E.~S. and {Samuroff}, S. and {Sanchez}, E. and {Scarpine}, V. and {Schubnell}, M. and {Serrano}, S. and {Sevilla-Noarbe}, I. and {Smith}, M. and {Soares-Santos}, M. and {Suchyta}, E. and {Swanson}, M.~E.~C. and {Tarle}, G. and {Thomas}, D. and {To}, C. and {Tucker}, D.~L. and {Varga}, T.~N. and {Wechsler}, R.~H. and {Weller}, J. and {Wester}, W. and {Wilkinson}, R.~D.},
	year         = {2021},
	month        = jul,
	volume       = {504},
	number       = {3},
	pages        = {4312--4336},
	doi          = {10.1093/mnras/stab918},
	archiveprefix = {arXiv},
	eprint       = {2011.03408},
	primaryclass = {astro-ph.CO},
	adsurl       = {https://ui.adsabs.harvard.edu/abs/2021MNRAS.504.4312G}
}

@article{Gatti2022,
	title        = {{Dark Energy Survey Year 3 results: Cosmology with moments of weak lensing mass maps}},
	author       = {{Gatti}, M. and {Jain}, B. and {Chang}, C. and {Raveri}, M. and {Z{\"u}rcher}, D. and {Secco}, L. and {Whiteway}, L. and {Jeffrey}, N. and {Doux}, C. and {Kacprzak}, T. and {Bacon}, D. and {Fosalba}, P. and {Alarcon}, A. and {Amon}, A. and {Bechtol}, K. and {Becker}, M. and {Bernstein}, G. and {Blazek}, J. and {Campos}, A. and {Choi}, A. and {Davis}, C. and {Derose}, J. and {Dodelson}, S. and {Elsner}, F. and {Elvin-Poole}, J. and {Everett}, S. and {Ferte}, A. and {Gruen}, D. and {Harrison}, I. and {Huterer}, D. and {Jarvis}, M. and {Krause}, E. and {Leget}, P.~F. and {Lemos}, P. and {Maccrann}, N. and {Mccullough}, J. and {Muir}, J. and {Myles}, J. and {Navarro}, A. and {Pandey}, S. and {Prat}, J. and {Rollins}, R.~P. and {Roodman}, A. and {Sanchez}, C. and {Sheldon}, E. and {Shin}, T. and {Troxel}, M. and {Tutusaus}, I. and {Yin}, B. and {Aguena}, M. and {Allam}, S. and {Andrade-Oliveira}, F. and {Annis}, J. and {Bertin}, E. and {Brooks}, D. and {Burke}, D.~L. and {Carnero Rosell}, A. and {Carrasco Kind}, M. and {Carretero}, J. and {Cawthon}, R. and {Costanzi}, M. and {da Costa}, L.~N. and {Pereira}, M.~E.~S. and {De Vicente}, J. and {Desai}, S. and {Diehl}, H.~T. and {Dietrich}, J.~P. and {Doel}, P. and {Drlica-Wagner}, A. and {Eckert}, K. and {Evrard}, A.~E. and {Ferrero}, I. and {Garc{\'\i}a-Bellido}, J. and {Gaztanaga}, E. and {Giannantonio}, T. and {Gruendl}, R.~A. and {Gschwend}, J. and {Gutierrez}, G. and {Hinton}, S.~R. and {Hollowood}, D.~L. and {Honscheid}, K. and {James}, D.~J. and {Kuehn}, K. and {Kuropatkin}, N. and {Lahav}, O. and {Lidman}, C. and {Maia}, M.~A.~G. and {Marshall}, J.~L. and {Melchior}, P. and {Menanteau}, F. and {Miquel}, R. and {Morgan}, R. and {Palmese}, A. and {Paz-Chinch{\'o}n}, F. and {Pieres}, A. and {Plazas Malag{\'o}n}, A.~A. and {Reil}, K. and {Rodriguez-Monroyv}, M. and {Romer}, A.~K. and {Sanchez}, E. and {Schubnell}, M. and {Serrano}, S. and {Sevilla-Noarbe}, I. and {Smith}, M. and {Soares-Santos}, M. and {Suchyta}, E. and {Tarle}, G. and {Thomas}, D. and {To}, C. and {Varga}, T.~N. and {DES Collaboration}},
	year         = {2022},
	month        = oct,
	volume       = {106},
	number       = {8},
	pages        = {083509},
	doi          = {10.1103/PhysRevD.106.083509},
	eid          = {083509},
	archiveprefix = {arXiv},
	eprint       = {2110.10141},
	primaryclass = {astro-ph.CO},
	adsurl       = {https://ui.adsabs.harvard.edu/abs/2022PhRvD.106h3509G}
}

@article{Gatti2024,
	title        = {{Detection of the significant impact of source clustering on higher order statistics with DES Year 3 weak gravitational lensing data}},
	author       = {{Gatti}, M. and {Jeffrey}, N. and {Whiteway}, L. and {Ajani}, V. and {Kacprzak}, T. and {Z{\"u}rcher}, D. and {Chang}, C. and {Jain}, B. and {Blazek}, J. and {Krause}, E. and {Alarcon}, A. and {Amon}, A. and {Bechtol}, K. and {Becker}, M. and {Bernstein}, G. and {Campos}, A. and {Chen}, R. and {Choi}, A. and {Davis}, C. and {Derose}, J. and {Diehl}, H.~T. and {Dodelson}, S. and {Doux}, C. and {Eckert}, K. and {Elvin-Poole}, J. and {Everett}, S. and {Ferte}, A. and {Gruen}, D. and {Gruendl}, R. and {Harrison}, I. and {Hartley}, W.~G. and {Herner}, K. and {Huff}, E.~M. and {Jarvis}, M. and {Kuropatkin}, N. and {Leget}, P.~F. and {MacCrann}, N. and {McCullough}, J. and {Myles}, J. and {Navarro-Alsina}, A. and {Pandey}, S. and {Prat}, J. and {Raveri}, M. and {Rollins}, R.~P. and {Roodman}, A. and {Sanchez}, C. and {Secco}, L.~F. and {Sevilla-Noarbe}, I. and {Sheldon}, E. and {Shin}, T. and {Troxel}, M. and {Tutusaus}, I. and {Varga}, T.~N. and {Yanny}, B. and {Yin}, B. and {Zhang}, Y. and {Zuntz}, J. and {Allam}, S.~S. and {Alves}, O. and {Aguena}, M. and {Bacon}, D. and {Bertin}, E. and {Brooks}, D. and {Burke}, D.~L. and {Rosell}, A. Carnero and {Carretero}, J. and {Cawthon}, R. and {da Costa}, L.~N. and {Davis}, T.~M. and {De Vicente}, J. and {Desai}, S. and {Doel}, P. and {Garc{\'\i}a-Bellido}, J. and {Giannini}, G. and {Gutierrez}, G. and {Ferrero}, I. and {Frieman}, J. and {Hinton}, S.~R. and {Hollowood}, D.~L. and {Honscheid}, K. and {James}, D.~J. and {Kuehn}, K. and {Lahav}, O. and {Marshall}, J.~L. and {Mena-Fern{\'a}ndez}, J. and {Miquel}, R. and {Ogando}, R.~L.~C. and {Palmese}, A. and {Pereira}, M.~E.~S. and {Malag{\'o}n}, A.~A. Plazas and {Rodriguez-Monroy}, M. and {Samuroff}, S. and {Sanchez}, E. and {Schubnell}, M. and {Smith}, M. and {Sobreira}, F. and {Suchyta}, E. and {Swanson}, M.~E.~C. and {Tarle}, G. and {Weaverdyck}, N. and {Wiseman}, P. and {DES Collaboration}},
	year         = {2024},
	month        = jan,
	volume       = {527},
	number       = {1},
	pages        = {L115-L121},
	doi          = {10.1093/mnrasl/slad143},
	archiveprefix = {arXiv},
	eprint       = {2307.13860},
	primaryclass = {astro-ph.CO},
	adsurl       = {https://ui.adsabs.harvard.edu/abs/2024MNRAS.527L.115G}
}

@article{Gatti2024b,
	title        = {{Dark Energy Survey Year 3 results: Simulation-based cosmological inference with wavelet harmonics, scattering transforms, and moments of weak lensing mass maps. Validation on simulations}},
	author       = {{Gatti}, M. and {Jeffrey}, N. and {Whiteway}, L. and {Williamson}, J. and {Jain}, B. and {Ajani}, V. and {Anbajagane}, D. and {Giannini}, G. and {Zhou}, C. and {Porredon}, A. and {Prat}, J. and {Yamamoto}, M. and {Blazek}, J. and {Kacprzak}, T. and {Samuroff}, S. and {Alarcon}, A. and {Amon}, A. and {Bechtol}, K. and {Becker}, M. and {Bernstein}, G. and {Campos}, A. and {Chang}, C. and {Chen}, R. and {Choi}, A. and {Davis}, C. and {Derose}, J. and {Diehl}, H.~T. and {Dodelson}, S. and {Doux}, C. and {Eckert}, K. and {Elvin-Poole}, J. and {Everett}, S. and {Ferte}, A. and {Gruen}, D. and {Gruendl}, R. and {Harrison}, I. and {Hartley}, W.~G. and {Herner}, K. and {Huff}, E.~M. and {Jarvis}, M. and {Kuropatkin}, N. and {Leget}, P.~F. and {MacCrann}, N. and {McCullough}, J. and {Myles}, J. and {Navarro-Alsina}, A. and {Pandey}, S. and {Raveri}, M. and {Rollins}, R.~P. and {Roodman}, A. and {Sanchez}, C. and {Secco}, L.~F. and {Sevilla-Noarbe}, I. and {Sheldon}, E. and {Shin}, T. and {Troxel}, M. and {Tutusaus}, I. and {Varga}, T.~N. and {Yanny}, B. and {Yin}, B. and {Zhang}, Y. and {Zuntz}, J. and {Aguena}, M. and {Alves}, O. and {Annis}, J. and {Brooks}, D. and {Carretero}, J. and {Castander}, F.~J. and {Cawthon}, R. and {Costanzi}, M. and {da Costa}, L.~N. and {Pereira}, M.~E.~S. and {Evrard}, A.~E. and {Flaugher}, B. and {Fosalba}, P. and {Frieman}, J. and {Garc{\'\i}a-Bellido}, J. and {Gerdes}, D.~W. and {Gruen}, D. and {Gruendl}, R.~A. and {Gschwend}, J. and {Gutierrez}, G. and {Hollowood}, D.~L. and {Honscheid}, K. and {James}, D.~J. and {Kuehn}, K. and {Lahav}, O. and {Lee}, S. and {Marshall}, J.~L. and {Mena-Fern{\'a}ndez}, J. and {Menanteau}, F. and {Miquel}, R. and {Ogando}, R.~L.~C. and {Pereira}, M.~E.~S. and {Pieres}, A. and {Plazas Malag{\'o}n}, A.~A. and {Sanchez}, E. and {Smith}, M. and {Suchyta}, E. and {Swanson}, M.~E.~C. and {Tarle}, G. and {Weaverdyck}, N. and {Weller}, J. and {Wiseman}, P. and {DES Collaboration}},
	year         = {2024},
	month        = mar,
	volume       = {109},
	number       = {6},
	pages        = {063534},
	doi          = {10.1103/PhysRevD.109.063534},
	eid          = {063534},
	archiveprefix = {arXiv},
	eprint       = {2310.17557},
	primaryclass = {astro-ph.CO},
	adsurl       = {https://ui.adsabs.harvard.edu/abs/2024PhRvD.109f3534G}
}

@article{Gatti2025,
	title        = {{Dark Energy Survey Year 3 results: Simulation-based cosmological inference with wavelet harmonics, scattering transforms, and moments of weak lensing mass maps. II. cosmological results}},
	author       = {{Gatti}, M. and {Campailla}, G. and {Jeffrey}, N. and {Whiteway}, L. and {Porredon}, A. and {Prat}, J. and {Williamson}, J. and {Raveri}, M. and {Jain}, B. and {Ajani}, V. and {Giannini}, G. and {Yamamoto}, M. and {Zhou}, C. and {Blazek}, J. and {Anbajagane}, D. and {Samuroff}, S. and {Kacprzak}, T. and {Alarcon}, A. and {Amon}, A. and {Bechtol}, K. and {Becker}, M. and {Bernstein}, G. and {Campos}, A. and {Chang}, C. and {Chen}, R. and {Choi}, A. and {Davis}, C. and {Derose}, J. and {Diehl}, H.~T. and {Dodelson}, S. and {Doux}, C. and {Eckert}, K. and {Elvin-Poole}, J. and {Everett}, S. and {Ferte}, A. and {Gruen}, D. and {Gruendl}, R. and {Harrison}, I. and {Hartley}, W.~G. and {Herner}, K. and {Huff}, E.~M. and {Jarvis}, M. and {Kuropatkin}, N. and {Leget}, P.~F. and {MacCrann}, N. and {McCullough}, J. and {Myles}, J. and {Navarro-Alsina}, A. and {Pandey}, S. and {Rollins}, R.~P. and {Roodman}, A. and {Sanchez}, C. and {Secco}, L.~F. and {Sevilla-Noarbe}, I. and {Sheldon}, E. and {Shin}, T. and {Troxel}, M. and {Tutusaus}, I. and {Varga}, T.~N. and {Yanny}, B. and {Yin}, B. and {Zhang}, Y. and {Zuntz}, J. and {Abbott}, T.~M.~C. and {Aguena}, M. and {Allam}, S.~S. and {Alves}, O. and {Andrade-Oliveira}, F. and {Bacon}, D. and {Bocquet}, S. and {Brooks}, D. and {Carnero Rosell}, A. and {Carretero}, J. and {da Costa}, L.~N. and {Pereira}, M.~E.~S. and {De Vicente}, J. and {Ferrero}, I. and {Frieman}, J. and {Garc{\'\i}a-Bellido}, J. and {Gaztanaga}, E. and {Gutierrez}, G. and {Hinton}, S.~R. and {Hollowood}, D.~L. and {Honscheid}, K. and {James}, D.~J. and {Kuehn}, K. and {Lahav}, O. and {Lee}, S. and {Marshall}, J.~L. and {Mena-Fern{\'a}ndez}, J. and {Miquel}, R. and {Pieres}, A. and {Plazas Malag{\'o}n}, A.~A. and {Sanchez}, E. and {Sanchez Cid}, D. and {Schubnell}, M. and {Smith}, M. and {Suchyta}, E. and {Tarle}, G. and {Weaverdyck}, N. and {Weller}, J. and {Wiseman}, P. and {(Dark Energy Survey)}},
	year         = {2025},
	month        = mar,
	volume       = {111},
	number       = {6},
	pages        = {063504},
	doi          = {10.1103/PhysRevD.111.063504},
	eid          = {063504},
	archiveprefix = {arXiv},
	eprint       = {2405.10881},
	primaryclass = {astro-ph.CO},
	adsurl       = {https://ui.adsabs.harvard.edu/abs/2025PhRvD.111f3504G}
}

@inproceedings{Germain2015,
	title        = {{MADE}: Masked Autoencoder for Distribution Estimation},
	author       = {Germain, Mathieu and Gregor, Karol and Murray, Iain and Larochelle, Hugo},
	year         = {2015},
	booktitle    = {Proceedings of the 32nd International Conference on Machine Learning},
	publisher    = {PMLR},
	volume       = {37},
	pages        = {1418--1427},
	editor       = {Bach, Francis and Blei, David}
}

@article{Giri2021,
	title        = {Emulation of baryonic effects on the matter power spectrum and constraints from galaxy cluster data},
	author       = {Sambit K. Giri and Aurel Schneider},
	year         = {2021},
	month        = {12},
	journal      = {Journal of Cosmology and Astroparticle Physics},
	publisher    = {IOP Publishing},
	volume       = {2021},
	number       = {12},
	pages        = {046},
	doi          = {10.1088/1475-7516/2021/12/046},
	url          = {https://dx.doi.org/10.1088/1475-7516/2021/12/046}
}

@article{Gomes2025,
	title        = {{Cosmology with second and third-order shear statistics for the Dark Energy Survey: Methods and simulated analysis}},
	author       = {{Gomes}, R.~C.~H. and {Sugiyama}, S. and {Jain}, B. and {Jarvis}, M. and {Anbajagane}, D. and {Gatti}, M. and {Gebauer}, D. and {Gong}, Z. and {Halder}, A. and {Marques}, G.~A. and {Pandey}, S. and {Marshall}, J.~L. and {Allam}, S. and {Alves}, O. and {Andrade-Oliveira}, F. and {Bacon}, D. and {Blazek}, J. and {Bocquet}, S. and {Brooks}, D. and {Carnero Rosell}, A. and {Carretero}, J. and {da Costa}, L.~N. and {Doel}, P. and {Doux}, C. and {Everett}, S. and {Flaugher}, B. and {Frieman}, J. and {Garc{\'\i}a-Bellido}, J. and {Gaztanaga}, E. and {Gruen}, D. and {Gruendl}, R.~A. and {Gutierrez}, G. and {Herner}, K. and {Hinton}, S.~R. and {Hollowood}, D.~L. and {Honscheid}, K. and {Huterer}, D. and {James}, D.~J. and {Jeffrey}, N. and {Mena-Fern{\'a}ndez}, J. and {Miquel}, R. and {Muir}, J. and {Ogando}, R.~L.~C. and {Pereira}, M.~E.~S. and {Pieres}, A. and {Plazas Malag{\'o}n}, A.~A. and {Samuroff}, S. and {Sanchez}, E. and {Sanchez Cid}, D. and {Santiago}, B. and {Sevilla-Noarbe}, I. and {Smith}, M. and {Suchyta}, E. and {Swanson}, M.~E.~C. and {Tarle}, G. and {To}, C. and {Vikram}, V. and {Weaverdyck}, N. and {Weller}, J.},
	year         = {2025},
	month        = mar,
	journal      = {arXiv e-prints},
	pages        = {arXiv:2503.03964},
	doi          = {10.48550/arXiv.2503.03964},
	eid          = {arXiv:2503.03964},
	archiveprefix = {arXiv},
	eprint       = {2503.03964},
	primaryclass = {astro-ph.CO},
	adsurl       = {https://ui.adsabs.harvard.edu/abs/2025arXiv250303964G}
}

@article{Gong2023,
	title        = {Cosmology from the integrated shear 3-point correlation function:  simulated likelihood analyses with machine-learning emulators},
	author       = {Zhengyangguang Gong and Anik Halder and Alexandre Barreira and Stella Seitz and Oliver Friedrich},
	year         = {2023},
	month        = {07},
	journal      = {Journal of Cosmology and Astroparticle Physics},
	publisher    = {IOP Publishing},
	volume       = {2023},
	number       = {07},
	pages        = {040},
	doi          = {10.1088/1475-7516/2023/07/040},
	url          = {https://dx.doi.org/10.1088/1475-7516/2023/07/040}
}

@article{Gong2024,
	title        = {C3NN: Cosmological Correlator Convolutional Neural Network an Interpretable Machine-learning Framework for Cosmological Analyses},
	author       = {Gong, Zhengyangguang and Halder, Anik and Bohrdt, Annabelle and Seitz, Stella and Gebauer, David},
	year         = {2024},
	month        = {08},
	journal      = {The Astrophysical Journal},
	volume       = {971},
	pages        = {156},
	doi          = {10.3847/1538-4357/ad582e}
}

@article{Gorski2005,
	title        = {{HEALPix: A Framework for High-Resolution Discretization and Fast Analysis of Data Distributed on the Sphere}},
	author       = {{G{\'o}rski}, K.~M. and {Hivon}, E. and {Banday}, A.~J. and {Wandelt}, B.~D. and {Hansen}, F.~K. and {Reinecke}, M. and {Bartelmann}, M.},
	year         = {2005},
	month        = apr,
	journal      = {Astrophysical Journal},
	volume       = {622},
	number       = {2},
	pages        = {759--771},
	doi          = {10.1086/427976},
	archiveprefix = {arXiv},
	eprint       = {astro-ph/0409513},
	primaryclass = {astro-ph},
	adsurl       = {https://ui.adsabs.harvard.edu/abs/2005ApJ...622..759G}
}

@article{Halder2021,
	title        = {The integrated three-point correlation function of cosmic shear},
	author       = {Anik Halder and Oliver Friedrich and Stella Seitz and Tamas N. Varga},
	year         = {2021},
	journal      = {Monthly Notices of the Royal Astronomical Society},
	publisher    = {Oxford University Press},
	number       = {2},
	pages        = {2780--2803}
}

@article{Halder2022,
	title        = {Response approach to the integrated shear 3-point correlation function: the impact of baryonic effects on small scales},
	author       = {Anik Halder and Alexandre Barreira},
	year         = {2022},
	journal      = {Monthly Notices of the Royal Astronomical Society},
	publisher    = {Oxford University Press (OUP)},
	number       = {3},
	pages        = {4639--4654}
}

@article{Halder2023,
	title        = {Beyond 3×2-point cosmology: the integrated shear and galaxy 3-point correlation functions},
	author       = {Anik Halder and Zhengyangguang Gong and Alexandre Barreira and Oliver Friedrich and Stella Seitz and Daniel Gruen},
	year         = {2023},
	month        = {10},
	journal      = {Journal of Cosmology and Astroparticle Physics},
	publisher    = {IOP Publishing},
	volume       = {2023},
	number       = {10},
	pages        = {028},
	doi          = {10.1088/1475-7516/2023/10/028},
	url          = {https://dx.doi.org/10.1088/1475-7516/2023/10/028}
}

@article{Hartlap2007,
	title        = {{Why your model parameter confidences might be too optimistic. Unbiased estimation of the inverse covariance matrix}},
	author       = {{Hartlap}, J. and {Simon}, P. and {Schneider}, P.},
	year         = {2007},
	month        = mar,
	volume       = {464},
	number       = {1},
	pages        = {399--404},
	doi          = {10.1051/0004-6361:20066170},
	archiveprefix = {arXiv},
	eprint       = {astro-ph/0608064},
	primaryclass = {astro-ph},
	adsurl       = {https://ui.adsabs.harvard.edu/abs/2007A&A...464..399H}
}

@article{Heavens2000,
	title        = {{Massive lossless data compression and multiple parameter estimation from galaxy spectra}},
	author       = {{Heavens}, Alan F. and {Jimenez}, Raul and {Lahav}, Ofer},
	year         = {2000},
	month        = oct,
	volume       = {317},
	number       = {4},
	pages        = {965--972},
	doi          = {10.1046/j.1365-8711.2000.03692.x},
	archiveprefix = {arXiv},
	eprint       = {astro-ph/9911102},
	primaryclass = {astro-ph},
	adsurl       = {https://ui.adsabs.harvard.edu/abs/2000MNRAS.317..965H}
}

@article{Heavens2017,
	title        = {{Massive data compression for parameter-dependent covariance matrices}},
	author       = {{Heavens}, Alan F. and {Sellentin}, Elena and {de Mijolla}, Damien and {Vianello}, Alvise},
	year         = {2017},
	month        = dec,
	volume       = {472},
	number       = {4},
	pages        = {4244--4250},
	doi          = {10.1093/mnras/stx2326},
	archiveprefix = {arXiv},
	eprint       = {1707.06529},
	primaryclass = {astro-ph.CO},
	adsurl       = {https://ui.adsabs.harvard.edu/abs/2017MNRAS.472.4244H}
}

@article{Heymans2006,
	title        = {{The Shear Testing Programme - I. Weak lensing analysis of simulated ground-based observations}},
	author       = {{Heymans}, Catherine and {Van Waerbeke}, Ludovic and {Bacon}, David and {Berge}, Joel and {Bernstein}, Gary and {Bertin}, Emmanuel and {Bridle}, Sarah and {Brown}, Michael L. and {Clowe}, Douglas and {Dahle}, H{\r{a}}kon and {Erben}, Thomas and {Gray}, Meghan and {Hetterscheidt}, Marco and {Hoekstra}, Henk and {Hudelot}, Patrick and {Jarvis}, Mike and {Kuijken}, Konrad and {Margoniner}, Vera and {Massey}, Richard and {Mellier}, Yannick and {Nakajima}, Reiko and {Refregier}, Alexandre and {Rhodes}, Jason and {Schrabback}, Tim and {Wittman}, David},
	year         = {2006},
	month        = may,
	volume       = {368},
	number       = {3},
	pages        = {1323--1339},
	doi          = {10.1111/j.1365-2966.2006.10198.x},
	archiveprefix = {arXiv},
	eprint       = {astro-ph/0506112},
	primaryclass = {astro-ph},
	adsurl       = {https://ui.adsabs.harvard.edu/abs/2006MNRAS.368.1323H}
}

@article{HirataSeljak2004,
	title        = {Intrinsic alignment-lensing interference as a contaminant of cosmic shear},
	author       = {Hirata, Christopher M. and Seljak, Uro\ifmmode \check{s}\else \v{s}\fi{}},
	year         = {2004},
	month        = {09},
	journal      = {Phys. Rev. D},
	publisher    = {American Physical Society},
	volume       = {70},
	pages        = {063526},
	doi          = {10.1103/PhysRevD.70.063526},
	url          = {https://link.aps.org/doi/10.1103/PhysRevD.70.063526},
	issue        = {6},
	numpages     = {11}
}

@article{Homer2025,
	title        = {{Simulation-based inference has its own Dodelson{\textendash}Schneider effect (but it knows that it does)}},
	author       = {{Homer}, J. and {Friedrich}, O. and {Gruen}, D.},
	year         = {2025},
	month        = jul,
	volume       = {699},
	pages        = {A213},
	doi          = {10.1051/0004-6361/202453339},
	eid          = {A213},
	archiveprefix = {arXiv},
	eprint       = {2412.02311},
	primaryclass = {astro-ph.CO},
	adsurl       = {https://ui.adsabs.harvard.edu/abs/2025A&A...699A.213H}
}

@article{HSC_overview,
	title        = {{The Hyper Suprime-Cam SSP Survey: Overview and survey design}},
	author       = {{Aihara}, Hiroaki and {Arimoto}, Nobuo and {Armstrong}, Robert and {Arnouts}, St{\'e}phane and {Bahcall}, Neta A. and {Bickerton}, Steven and {Bosch}, James and {Bundy}, Kevin and {Capak}, Peter L. and {Chan}, James H.~H. and {Chiba}, Masashi and {Coupon}, Jean and {Egami}, Eiichi and {Enoki}, Motohiro and {Finet}, Francois and {Fujimori}, Hiroki and {Fujimoto}, Seiji and {Furusawa}, Hisanori and {Furusawa}, Junko and {Goto}, Tomotsugu and {Goulding}, Andy and {Greco}, Johnny P. and {Greene}, Jenny E. and {Gunn}, James E. and {Hamana}, Takashi and {Harikane}, Yuichi and {Hashimoto}, Yasuhiro and {Hattori}, Takashi and {Hayashi}, Masao and {Hayashi}, Yusuke and {He{\l}miniak}, Krzysztof G. and {Higuchi}, Ryo and {Hikage}, Chiaki and {Ho}, Paul T.~P. and {Hsieh}, Bau-Ching and {Huang}, Kuiyun and {Huang}, Song and {Ikeda}, Hiroyuki and {Imanishi}, Masatoshi and {Inoue}, Akio K. and {Iwasawa}, Kazushi and {Iwata}, Ikuru and {Jaelani}, Anton T. and {Jian}, Hung-Yu and {Kamata}, Yukiko and {Karoji}, Hiroshi and {Kashikawa}, Nobunari and {Katayama}, Nobuhiko and {Kawanomoto}, Satoshi and {Kayo}, Issha and {Koda}, Jin and {Koike}, Michitaro and {Kojima}, Takashi and {Komiyama}, Yutaka and {Konno}, Akira and {Koshida}, Shintaro and {Koyama}, Yusei and {Kusakabe}, Haruka and {Leauthaud}, Alexie and {Lee}, Chien-Hsiu and {Lin}, Lihwai and {Lin}, Yen-Ting and {Lupton}, Robert H. and {Mandelbaum}, Rachel and {Matsuoka}, Yoshiki and {Medezinski}, Elinor and {Mineo}, Sogo and {Miyama}, Shoken and {Miyatake}, Hironao and {Miyazaki}, Satoshi and {Momose}, Rieko and {More}, Anupreeta and {More}, Surhud and {Moritani}, Yuki and {Moriya}, Takashi J. and {Morokuma}, Tomoki and {Mukae}, Shiro and {Murata}, Ryoma and {Murayama}, Hitoshi and {Nagao}, Tohru and {Nakata}, Fumiaki and {Niida}, Mana and {Niikura}, Hiroko and {Nishizawa}, Atsushi J. and {Obuchi}, Yoshiyuki and {Oguri}, Masamune and {Oishi}, Yukie and {Okabe}, Nobuhiro and {Okamoto}, Sakurako and {Okura}, Yuki and {Ono}, Yoshiaki and {Onodera}, Masato and {Onoue}, Masafusa and {Osato}, Ken and {Ouchi}, Masami and {Price}, Paul A. and {Pyo}, Tae-Soo and {Sako}, Masao and {Sawicki}, Marcin and {Shibuya}, Takatoshi and {Shimasaku}, Kazuhiro and {Shimono}, Atsushi and {Shirasaki}, Masato and {Silverman}, John D. and {Simet}, Melanie and {Speagle}, Joshua and {Spergel}, David N. and {Strauss}, Michael A. and {Sugahara}, Yuma and {Sugiyama}, Naoshi and {Suto}, Yasushi and {Suyu}, Sherry H. and {Suzuki}, Nao and {Tait}, Philip J. and {Takada}, Masahiro and {Takata}, Tadafumi and {Tamura}, Naoyuki and {Tanaka}, Manobu M. and {Tanaka}, Masaomi and {Tanaka}, Masayuki and {Tanaka}, Yoko and {Terai}, Tsuyoshi and {Terashima}, Yuichi and {Toba}, Yoshiki and {Tominaga}, Nozomu and {Toshikawa}, Jun and {Turner}, Edwin L. and {Uchida}, Tomohisa and {Uchiyama}, Hisakazu and {Umetsu}, Keiichi and {Uraguchi}, Fumihiro and {Urata}, Yuji and {Usuda}, Tomonori and {Utsumi}, Yousuke and {Wang}, Shiang-Yu and {Wang}, Wei-Hao and {Wong}, Kenneth C. and {Yabe}, Kiyoto and {Yamada}, Yoshihiko and {Yamanoi}, Hitomi and {Yasuda}, Naoki and {Yeh}, Sherry and {Yonehara}, Atsunori and {Yuma}, Suraphong},
	year         = {2018},
	month        = jan,
	volume       = {70},
	pages        = {S4},
	doi          = {10.1093/pasj/psx066},
	eid          = {S4},
	archiveprefix = {arXiv},
	eprint       = {1704.05858},
	primaryclass = {astro-ph.IM},
	adsurl       = {https://ui.adsabs.harvard.edu/abs/2018PASJ...70S...4A}
}

@article{HSC_results,
	title        = {{Hyper Suprime-Cam Year 3 results: Cosmology from cosmic shear power spectra}},
	author       = {{Dalal}, Roohi and {Li}, Xiangchong and {Nicola}, Andrina and {Zuntz}, Joe and {Strauss}, Michael A. and {Sugiyama}, Sunao and {Zhang}, Tianqing and {Rau}, Markus M. and {Mandelbaum}, Rachel and {Takada}, Masahiro and {More}, Surhud and {Miyatake}, Hironao and {Kannawadi}, Arun and {Shirasaki}, Masato and {Taniguchi}, Takanori and {Takahashi}, Ryuichi and {Osato}, Ken and {Hamana}, Takashi and {Oguri}, Masamune and {Nishizawa}, Atsushi J. and {Malag{\'o}n}, Andr{\'e}s A. Plazas and {Sunayama}, Tomomi and {Alonso}, David and {Slosar}, An{\v{z}}e and {Luo}, Wentao and {Armstrong}, Robert and {Bosch}, James and {Hsieh}, Bau-Ching and {Komiyama}, Yutaka and {Lupton}, Robert H. and {Lust}, Nate B. and {MacArthur}, Lauren A. and {Miyazaki}, Satoshi and {Murayama}, Hitoshi and {Nishimichi}, Takahiro and {Okura}, Yuki and {Price}, Paul A. and {Tait}, Philip J. and {Tanaka}, Masayuki and {Wang}, Shiang-Yu},
	year         = {2023},
	month        = dec,
	volume       = {108},
	number       = {12},
	pages        = {123519},
	doi          = {10.1103/PhysRevD.108.123519},
	eid          = {123519},
	archiveprefix = {arXiv},
	eprint       = {2304.00701},
	primaryclass = {astro-ph.CO},
	adsurl       = {https://ui.adsabs.harvard.edu/abs/2023PhRvD.108l3519D}
}

@article{Jarvis2004,
	title        = {{The skewness of the aperture mass statistic}},
	author       = {Jarvis, M. and Bernstein, G. and Jain, B.},
	year         = {2004},
	month        = {07},
	journal      = {Monthly Notices of the Royal Astronomical Society},
	volume       = {352},
	number       = {1},
	pages        = {338--352},
	doi          = {10.1111/j.1365-2966.2004.07926.x},
	issn         = {0035-8711},
	url          = {https://doi.org/10.1111/j.1365-2966.2004.07926.x},
	eprint       = {https://academic.oup.com/mnras/article-pdf/352/1/338/3195813/352-1-338.pdf}
}

@misc{Jarvis2019,
	title        = {{CDS\&E: Two- and Three-point Correlations for Large Data Sets with TreeCorr}},
	author       = {{Jarvis}, Robert M},
	year         = {2019},
	month        = sep,
	pages        = {7610},
	howpublished = {NSF Award Number 1907610. Directorate for Mathematical and Physical Sciences, Division Of Astronomical Sciences. 2019.},
	adsurl       = {https://ui.adsabs.harvard.edu/abs/2019nsf....1907610J}
}

@article{Jeffrey2021,
	title        = {{Likelihood-free inference with neural compression of DES SV weak lensing map statistics}},
	author       = {{Jeffrey}, Niall and {Alsing}, Justin and {Lanusse}, Fran{\c{c}}ois},
	year         = {2021},
	month        = feb,
	volume       = {501},
	number       = {1},
	pages        = {954--969},
	doi          = {10.1093/mnras/staa3594},
	archiveprefix = {arXiv},
	eprint       = {2009.08459},
	primaryclass = {astro-ph.CO},
	adsurl       = {https://ui.adsabs.harvard.edu/abs/2021MNRAS.501..954J}
}

@article{Jeffrey2025,
	title        = {{Dark energy survey year 3 results: likelihood-free, simulation-based wCDM inference with neural compression of weak-lensing map statistics}},
	author       = {{Jeffrey}, N. and {Whiteway}, L. and {Gatti}, M. and {Williamson}, J. and {Alsing}, J. and {Porredon}, A. and {Prat}, J. and {Doux}, C. and {Jain}, B. and {Chang}, C. and {Cheng}, T. -Y. and {Kacprzak}, T. and {Lemos}, P. and {Alarcon}, A. and {Amon}, A. and {Bechtol}, K. and {Becker}, M.~R. and {Bernstein}, G.~M. and {Campos}, A. and {Carnero Rosell}, A. and {Chen}, R. and {Choi}, A. and {DeRose}, J. and {Drlica-Wagner}, A. and {Eckert}, K. and {Everett}, S. and {Fert{\'e}}, A. and {Gruen}, D. and {Gruendl}, R.~A. and {Herner}, K. and {Jarvis}, M. and {McCullough}, J. and {Myles}, J. and {Navarro-Alsina}, A. and {Pandey}, S. and {Raveri}, M. and {Rollins}, R.~P. and {Rykoff}, E.~S. and {S{\'a}nchez}, C. and {Secco}, L.~F. and {Sevilla-Noarbe}, I. and {Sheldon}, E. and {Shin}, T. and {Troxel}, M.~A. and {Tutusaus}, I. and {Varga}, T.~N. and {Yanny}, B. and {Yin}, B. and {Zuntz}, J. and {Aguena}, M. and {Allam}, S.~S. and {Alves}, O. and {Bacon}, D. and {Bocquet}, S. and {Brooks}, D. and {da Costa}, L.~N. and {Davis}, T.~M. and {De Vicente}, J. and {Desai}, S. and {Diehl}, H.~T. and {Ferrero}, I. and {Frieman}, J. and {Garc{\'\i}a-Bellido}, J. and {Gaztanaga}, E. and {Giannini}, G. and {Gutierrez}, G. and {Hinton}, S.~R. and {Hollowood}, D.~L. and {Honscheid}, K. and {Huterer}, D. and {James}, D.~J. and {Lahav}, O. and {Lee}, S. and {Marshall}, J.~L. and {Mena-Fern{\'a}ndez}, J. and {Miquel}, R. and {Pieres}, A. and {Plazas Malag{\'o}n}, A.~A. and {Roodman}, A. and {Sako}, M. and {Sanchez}, E. and {Sanchez Cid}, D. and {Smith}, M. and {Suchyta}, E. and {Swanson}, M.~E.~C. and {Tarle}, G. and {Tucker}, D.~L. and {Weaverdyck}, N. and {Weller}, J. and {Wiseman}, P. and {Yamamoto}, M.},
	year         = {2025},
	month        = jan,
	volume       = {536},
	number       = {2},
	pages        = {1303--1322},
	doi          = {10.1093/mnras/stae2629},
	archiveprefix = {arXiv},
	eprint       = {2403.02314},
	primaryclass = {astro-ph.CO},
	adsurl       = {https://ui.adsabs.harvard.edu/abs/2025MNRAS.536.1303J}
}

@article{Joachimi2011,
	title        = {{Constraints on intrinsic alignment contamination of weak lensing surveys using the MegaZ-LRG sample}},
	author       = {{Joachimi}, B. and {Mandelbaum}, R. and {Abdalla}, F.~B. and {Bridle}, S.~L.},
	year         = {2011},
	month        = mar,
	journal      = {Astronomy and Astrophysics},
	volume       = {527},
	pages        = {A26},
	doi          = {10.1051/0004-6361/201015621},
	eid          = {A26},
	archiveprefix = {arXiv},
	eprint       = {1008.3491},
	primaryclass = {astro-ph.CO},
	adsurl       = {https://ui.adsabs.harvard.edu/abs/2011A&A...527A..26J}
}

@article{Kacprzak2022,
	title        = {{CosmoGridV1: a simulated wCDM theory prediction for map-level cosmological inference}},
	author       = {{Kacprzak}, Tomasz and {Fluri}, Janis and {Schneider}, Aurel and {Refregier}, Alexandre and {Stadel}, Joachim},
	year         = {2022},
	month        = sep,
	journal      = {arXiv e-prints},
	pages        = {arXiv:2209.04662},
	eid          = {arXiv:2209.04662},
	archiveprefix = {arXiv},
	eprint       = {2209.04662},
	primaryclass = {astro-ph.CO}
}

@article{KaiserSquires1993,
	title        = {{Mapping the Dark Matter with Weak Gravitational Lensing}},
	author       = {{Kaiser}, Nick and {Squires}, Gordon},
	year         = {1993},
	month        = feb,
	journal      = {Astrophysical Journal},
	volume       = {404},
	pages        = {441},
	doi          = {10.1086/172297},
	adsurl       = {https://ui.adsabs.harvard.edu/abs/1993ApJ...404..441K}
}

@article{Kids_overview,
	title        = {{The Kilo-Degree Survey}},
	author       = {{de Jong}, Jelte T.~A. and {Verdoes Kleijn}, Gijs A. and {Kuijken}, Konrad H. and {Valentijn}, Edwin A.},
	year         = {2013},
	month        = jan,
	journal      = {Experimental Astronomy},
	volume       = {35},
	number       = {1-2},
	pages        = {25--44},
	doi          = {10.1007/s10686-012-9306-1},
	archiveprefix = {arXiv},
	eprint       = {1206.1254},
	primaryclass = {astro-ph.CO},
	adsurl       = {https://ui.adsabs.harvard.edu/abs/2013ExA....35...25D}
}

@article{Kids_results,
	title        = {{KiDS-Legacy: Cosmological constraints from cosmic shear with the complete Kilo-Degree Survey}},
	author       = {{Wright}, Angus H. and {St{\"o}lzner}, Benjamin and {Asgari}, Marika and {Bilicki}, Maciej and {Giblin}, Benjamin and {Heymans}, Catherine and {Hildebrandt}, Hendrik and {Hoekstra}, Henk and {Joachimi}, Benjamin and {Kuijken}, Konrad and {Li}, Shun-Sheng and {Reischke}, Robert and {von Wietersheim-Kramsta}, Maximilian and {Yoon}, Mijin and {Burger}, Pierre and {Chisari}, Nora Elisa and {de Jong}, Jelte and {Dvornik}, Andrej and {Georgiou}, Christos and {Harnois-D{\'e}raps}, Joachim and {Jalan}, Priyanka and {William}, Anjitha John and {Joudaki}, Shahab and {Lesci}, Giorgio Francesco and {Linke}, Laila and {Loureiro}, Arthur and {Mahony}, Constance and {Maturi}, Matteo and {Miller}, Lance and {Moscardini}, Lauro and {Napolitano}, Nicola R. and {Porth}, Lucas and {Radovich}, Mario and {Schneider}, Peter and {Tr{\"o}ster}, Tilman and {Wittje}, Anna and {Yan}, Ziang and {Zhang}, Yun-Hao},
	year         = {2025},
	month        = mar,
	journal      = {arXiv e-prints},
	pages        = {arXiv:2503.19441},
	doi          = {10.48550/arXiv.2503.19441},
	eid          = {arXiv:2503.19441},
	archiveprefix = {arXiv},
	eprint       = {2503.19441},
	primaryclass = {astro-ph.CO},
	adsurl       = {https://ui.adsabs.harvard.edu/abs/2025arXiv250319441W}
}

@article{Kilbinger2005,
	title        = {{Cosmological parameters from combined second- and third-order aperture mass statistics of cosmic shear}},
	author       = {{Kilbinger}, M. and {Schneider}, P.},
	year         = {2005},
	month        = oct,
	journal      = {Astronomy and Astrophysics},
	volume       = {442},
	number       = {1},
	pages        = {69--83},
	doi          = {10.1051/0004-6361:20053531},
	archiveprefix = {arXiv},
	eprint       = {astro-ph/0505581},
	primaryclass = {astro-ph},
	adsurl       = {https://ui.adsabs.harvard.edu/abs/2005A&A...442...69K}
}

@article{Kilbinger2015,
	title        = {Cosmology with cosmic shear observations: a review},
	author       = {Martin Kilbinger},
	year         = {2015},
	month        = {7},
	journal      = {Reports on Progress in Physics},
	publisher    = {IOP Publishing},
	volume       = {78},
	number       = {8},
	pages        = {086901},
	doi          = {10.1088/0034-4885/78/8/086901},
	url          = {https://dx.doi.org/10.1088/0034-4885/78/8/086901}
}

@article{Lehman2024,
	title        = {{Learning Optimal and Interpretable Summary Statistics of Galaxy Catalogs with SBI}},
	author       = {{Lehman}, Kai and {Krippendorf}, Sven and {Weller}, Jochen and {Dolag}, Klaus},
	year         = {2024},
	month        = nov,
	journal      = {arXiv e-prints},
	pages        = {arXiv:2411.08957},
	doi          = {10.48550/arXiv.2411.08957},
	eid          = {arXiv:2411.08957},
	archiveprefix = {arXiv},
	eprint       = {2411.08957},
	primaryclass = {astro-ph.CO},
	adsurl       = {https://ui.adsabs.harvard.edu/abs/2024arXiv241108957L}
}

@article{Lehman2025,
	title        = {{Cosmological Inference with Cosmic Voids and Neural Network Emulators}},
	author       = {{Lehman}, Kai and {Schuster}, Nico and {Lucie-Smith}, Luisa and {Hamaus}, Nico and {Davies}, Christopher T. and {Dolag}, Klaus},
	year         = {2025},
	month        = feb,
	journal      = {arXiv e-prints},
	pages        = {arXiv:2502.05262},
	doi          = {10.48550/arXiv.2502.05262},
	eid          = {arXiv:2502.05262},
	archiveprefix = {arXiv},
	eprint       = {2502.05262},
	primaryclass = {astro-ph.CO},
	adsurl       = {https://ui.adsabs.harvard.edu/abs/2025arXiv250205262L}
}

@inproceedings{Lemos2023,
	title        = {Sampling-Based Accuracy Testing of Posterior Estimators for General Inference},
	author       = {Lemos, Pablo and Coogan, Adam and Hezaveh, Yashar and Perreault-Levasseur, Laurence},
	year         = {2023},
	month        = {7},
	booktitle    = {Proceedings of the 40th International Conference on Machine Learning},
	publisher    = {PMLR},
	series       = {Proceedings of Machine Learning Research},
	volume       = {202},
	pages        = {19256--19273},
	url          = {https://proceedings.mlr.press/v202/lemos23a.html},
	editor       = {Krause, Andreas and Brunskill, Emma and Cho, Kyunghyun and Engelhardt, Barbara and Sabato, Sivan and Scarlett, Jonathan}
}

@article{Lesgourgues2011,
	title        = {{The Cosmic Linear Anisotropy Solving System (CLASS) I: Overview}},
	author       = {{Lesgourgues}, Julien},
	year         = {2011},
	month        = apr,
	journal      = {arXiv e-prints},
	pages        = {arXiv:1104.2932},
	doi          = {10.48550/arXiv.1104.2932},
	eid          = {arXiv:1104.2932},
	archiveprefix = {arXiv},
	eprint       = {1104.2932},
	primaryclass = {astro-ph.IM},
	adsurl       = {https://ui.adsabs.harvard.edu/abs/2011arXiv1104.2932L}
}

@article{Linke2025,
	title        = {{Euclid and KiDS-1000: Quantifying the impact of source-lens clustering on cosmic shear analyses}},
	author       = {{Linke}, L. and {Unruh}, S. and {Wittje}, A. and {Schrabback}, T. and {Grandis}, S. and {Asgari}, M. and {Dvornik}, A. and {Hildebrandt}, H. and {Hoekstra}, H. and {Joachimi}, B. and {Reischke}, R. and {van den Busch}, J.~L. and {Wright}, A.~H. and {Schneider}, P. and {Aghanim}, N. and {Altieri}, B. and {Amara}, A. and {Andreon}, S. and {Auricchio}, N. and {Baccigalupi}, C. and {Baldi}, M. and {Bardelli}, S. and {Bonino}, D. and {Branchini}, E. and {Brescia}, M. and {Brinchmann}, J. and {Camera}, S. and {Capobianco}, V. and {Carbone}, C. and {Cardone}, V.~F. and {Carretero}, J. and {Casas}, S. and {Castander}, F.~J. and {Castellano}, M. and {Cavuoti}, S. and {Cimatti}, A. and {Congedo}, G. and {Conselice}, C.~J. and {Conversi}, L. and {Copin}, Y. and {Courbin}, F. and {Courtois}, H.~M. and {Da Silva}, A. and {Degaudenzi}, H. and {Dinis}, J. and {Douspis}, M. and {Dubath}, F. and {Dupac}, X. and {Dusini}, S. and {Farina}, M. and {Farrens}, S. and {Ferriol}, S. and {Fosalba}, P. and {Frailis}, M. and {Franceschi}, E. and {Fumana}, M. and {Galeotta}, S. and {Gillis}, B. and {Giocoli}, C. and {Grazian}, A. and {Grupp}, F. and {Guzzo}, L. and {Haugan}, S.~V.~H. and {Holmes}, W. and {Hook}, I. and {Hormuth}, F. and {Hornstrup}, A. and {Hudelot}, P. and {Jahnke}, K. and {Keih{\"a}nen}, E. and {Kermiche}, S. and {Kiessling}, A. and {Kilbinger}, M. and {Kitching}, T. and {Kubik}, B. and {Kuijken}, K. and {K{\"u}mmel}, M. and {Kunz}, M. and {Kurki-Suonio}, H. and {Ligori}, S. and {Lilje}, P.~B. and {Lindholm}, V. and {Lloro}, I. and {Maino}, D. and {Maiorano}, E. and {Mansutti}, O. and {Marggraf}, O. and {Markovic}, K. and {Martinet}, N. and {Marulli}, F. and {Massey}, R. and {McCracken}, H.~J. and {Medinaceli}, E. and {Mei}, S. and {Mellier}, Y. and {Meneghetti}, M. and {Merlin}, E. and {Meylan}, G. and {Moresco}, M. and {Moscardini}, L. and {Munari}, E. and {Nakajima}, R. and {Nichol}, R.~C. and {Niemi}, S. -M. and {Nightingale}, J.~W. and {Padilla}, C. and {Paltani}, S. and {Pasian}, F. and {Pedersen}, K. and {Pettorino}, V. and {Pires}, S. and {Polenta}, G. and {Poncet}, M. and {Popa}, L.~A. and {Raison}, F. and {Rebolo}, R. and {Renzi}, A. and {Rhodes}, J. and {Riccio}, G. and {Romelli}, E. and {Roncarelli}, M. and {Saglia}, R. and {Sakr}, Z. and {Sapone}, D. and {Sartoris}, B. and {Schirmer}, M. and {Secroun}, A. and {Seidel}, G. and {Serrano}, S. and {Sirignano}, C. and {Sirri}, G. and {Stanco}, L. and {Starck}, J. -L. and {Tallada-Cresp{\'\i}}, P. and {Taylor}, A.~N. and {Tereno}, I. and {Toledo-Moreo}, R. and {Torradeflot}, F. and {Tutusaus}, I. and {Valenziano}, L. and {Vassallo}, T. and {Verdoes Kleijn}, G. and {Veropalumbo}, A. and {Wang}, Y. and {Weller}, J. and {Zamorani}, G. and {Zucca}, E. and {Burigana}, C. and {Pezzotta}, A. and {Porciani}, C. and {Scottez}, V. and {Viel}, M. and {Le Brun}, A.~M.~C.},
	year         = {2025},
	month        = jan,
	volume       = {693},
	pages        = {A210},
	doi          = {10.1051/0004-6361/202451494},
	eid          = {A210},
	archiveprefix = {arXiv},
	eprint       = {2407.09810},
	primaryclass = {astro-ph.CO},
	adsurl       = {https://ui.adsabs.harvard.edu/abs/2025A&A...693A.210L}
}

@article{LSST_overview,
	title        = {{LSST: From Science Drivers to Reference Design and Anticipated Data Products}},
	author       = {{Ivezi{\'c}}, {\v{Z}}eljko and {Kahn}, Steven M. and {Tyson}, J. Anthony and {Abel}, Bob and {Acosta}, Emily and {Allsman}, Robyn and {Alonso}, David and {AlSayyad}, Yusra and {Anderson}, Scott F. and {Andrew}, John and {Angel}, James Roger P. and {Angeli}, George Z. and {Ansari}, Reza and {Antilogus}, Pierre and {Araujo}, Constanza and {Armstrong}, Robert and {Arndt}, Kirk T. and {Astier}, Pierre and {Aubourg}, {\'E}ric and {Auza}, Nicole and {Axelrod}, Tim S. and {Bard}, Deborah J. and {Barr}, Jeff D. and {Barrau}, Aurelian and {Bartlett}, James G. and {Bauer}, Amanda E. and {Bauman}, Brian J. and {Baumont}, Sylvain and {Bechtol}, Ellen and {Bechtol}, Keith and {Becker}, Andrew C. and {Becla}, Jacek and {Beldica}, Cristina and {Bellavia}, Steve and {Bianco}, Federica B. and {Biswas}, Rahul and {Blanc}, Guillaume and {Blazek}, Jonathan and {Blandford}, Roger D. and {Bloom}, Josh S. and {Bogart}, Joanne and {Bond}, Tim W. and {Booth}, Michael T. and {Borgland}, Anders W. and {Borne}, Kirk and {Bosch}, James F. and {Boutigny}, Dominique and {Brackett}, Craig A. and {Bradshaw}, Andrew and {Brandt}, William Nielsen and {Brown}, Michael E. and {Bullock}, James S. and {Burchat}, Patricia and {Burke}, David L. and {Cagnoli}, Gianpietro and {Calabrese}, Daniel and {Callahan}, Shawn and {Callen}, Alice L. and {Carlin}, Jeffrey L. and {Carlson}, Erin L. and {Chandrasekharan}, Srinivasan and {Charles-Emerson}, Glenaver and {Chesley}, Steve and {Cheu}, Elliott C. and {Chiang}, Hsin-Fang and {Chiang}, James and {Chirino}, Carol and {Chow}, Derek and {Ciardi}, David R. and {Claver}, Charles F. and {Cohen-Tanugi}, Johann and {Cockrum}, Joseph J. and {Coles}, Rebecca and {Connolly}, Andrew J. and {Cook}, Kem H. and {Cooray}, Asantha and {Covey}, Kevin R. and {Cribbs}, Chris and {Cui}, Wei and {Cutri}, Roc and {Daly}, Philip N. and {Daniel}, Scott F. and {Daruich}, Felipe and {Daubard}, Guillaume and {Daues}, Greg and {Dawson}, William and {Delgado}, Francisco and {Dellapenna}, Alfred and {de Peyster}, Robert and {de Val-Borro}, Miguel and {Digel}, Seth W. and {Doherty}, Peter and {Dubois}, Richard and {Dubois-Felsmann}, Gregory P. and {Durech}, Josef and {Economou}, Frossie and {Eifler}, Tim and {Eracleous}, Michael and {Emmons}, Benjamin L. and {Fausti Neto}, Angelo and {Ferguson}, Henry and {Figueroa}, Enrique and {Fisher-Levine}, Merlin and {Focke}, Warren and {Foss}, Michael D. and {Frank}, James and {Freemon}, Michael D. and {Gangler}, Emmanuel and {Gawiser}, Eric and {Geary}, John C. and {Gee}, Perry and {Geha}, Marla and {Gessner}, Charles J.~B. and {Gibson}, Robert R. and {Gilmore}, D. Kirk and {Glanzman}, Thomas and {Glick}, William and {Goldina}, Tatiana and {Goldstein}, Daniel A. and {Goodenow}, Iain and {Graham}, Melissa L. and {Gressler}, William J. and {Gris}, Philippe and {Guy}, Leanne P. and {Guyonnet}, Augustin and {Haller}, Gunther and {Harris}, Ron and {Hascall}, Patrick A. and {Haupt}, Justine and {Hernandez}, Fabio and {Herrmann}, Sven and {Hileman}, Edward and {Hoblitt}, Joshua and {Hodgson}, John A. and {Hogan}, Craig and {Howard}, James D. and {Huang}, Dajun and {Huffer}, Michael E. and {Ingraham}, Patrick and {Innes}, Walter R. and {Jacoby}, Suzanne H. and {Jain}, Bhuvnesh and {Jammes}, Fabrice and {Jee}, M. James and {Jenness}, Tim and {Jernigan}, Garrett and {Jevremovi{\'c}}, Darko and {Johns}, Kenneth and {Johnson}, Anthony S. and {Johnson}, Margaret W.~G. and {Jones}, R. Lynne and {Juramy-Gilles}, Claire and {Juri{\'c}}, Mario and {Kalirai}, Jason S. and {Kallivayalil}, Nitya J. and {Kalmbach}, Bryce and {Kantor}, Jeffrey P. and {Karst}, Pierre and {Kasliwal}, Mansi M. and {Kelly}, Heather and {Kessler}, Richard and {Kinnison}, Veronica and {Kirkby}, David and {Knox}, Lloyd and {Kotov}, Ivan V. and {Krabbendam}, Victor L. and {Krughoff}, K. Simon and {Kub{\'a}nek}, Petr and {Kuczewski}, John and {Kulkarni}, Shri and {Ku}, John and {Kurita}, Nadine R. and {Lage}, Craig S. and {Lambert}, Ron and {Lange}, Travis and {Langton}, J. Brian and {Le Guillou}, Laurent and {Levine}, Deborah and {Liang}, Ming and {Lim}, Kian-Tat and {Lintott}, Chris J. and {Long}, Kevin E. and {Lopez}, Margaux and {Lotz}, Paul J. and {Lupton}, Robert H. and {Lust}, Nate B. and {MacArthur}, Lauren A. and {Mahabal}, Ashish and {Mandelbaum}, Rachel and {Markiewicz}, Thomas W. and {Marsh}, Darren S. and {Marshall}, Philip J. and {Marshall}, Stuart and {May}, Morgan and {McKercher}, Robert and {McQueen}, Michelle and {Meyers}, Joshua and {Migliore}, Myriam and {Miller}, Michelle and {Mills}, David J.},
	year         = {2019},
	month        = mar,
	volume       = {873},
	number       = {2},
	pages        = {111},
	doi          = {10.3847/1538-4357/ab042c},
	eid          = {111},
	archiveprefix = {arXiv},
	eprint       = {0805.2366},
	primaryclass = {astro-ph},
	adsurl       = {https://ui.adsabs.harvard.edu/abs/2019ApJ...873..111I}
}

@article{Mancini2024,
	title        = {{Field-level cosmological model selection: field-level simulation-based inference for Stage IV cosmic shear can distinguish dynamical dark energy}},
	author       = {{Spurio Mancini}, A. and {Lin}, K. and {McEwen}, J. D.},
	year         = {2024},
	month        = oct,
	journal      = {arXiv e-prints},
	pages        = {arXiv:2410.10616},
	doi          = {10.48550/arXiv.2410.10616},
	eid          = {arXiv:2410.10616},
	archiveprefix = {arXiv},
	eprint       = {2410.10616},
	primaryclass = {astro-ph.CO},
	adsurl       = {https://ui.adsabs.harvard.edu/abs/2024arXiv241010616S}
}

@article{Mead2021,
	title        = {{HMCODE-2020: improved modelling of non-linear cosmological power spectra with baryonic feedback}},
	author       = {{Mead}, A.~J. and {Brieden}, S. and {Tr{\"o}ster}, T. and {Heymans}, C.},
	year         = {2021},
	month        = mar,
	volume       = {502},
	number       = {1},
	pages        = {1401--1422},
	doi          = {10.1093/mnras/stab082},
	archiveprefix = {arXiv},
	eprint       = {2009.01858},
	primaryclass = {astro-ph.CO},
	adsurl       = {https://ui.adsabs.harvard.edu/abs/2021MNRAS.502.1401M}
}

@article{Munshi_2023,
	title        = {Position-dependent correlation function of weak-lensing convergence},
	author       = {Munshi, D. and Jung, G. and Kitching, T. D. and McEwen, J. and Liguori, M. and Namikawa, T. and Heavens, A.},
	year         = {2023},
	month        = feb,
	journal      = {Physical Review D},
	publisher    = {American Physical Society (APS)},
	volume       = {107},
	number       = {4},
	doi          = {10.1103/physrevd.107.043516},
	issn         = {2470-0029},
	url          = {http://dx.doi.org/10.1103/PhysRevD.107.043516}
}

@article{Narayan1996,
	title        = {{Lectures on Gravitational Lensing}},
	author       = {{Narayan}, Ramesh and {Bartelmann}, Matthias},
	year         = {1996},
	month        = jun,
	journal      = {arXiv e-prints},
	pages        = {astro-ph/9606001},
	doi          = {10.48550/arXiv.astro-ph/9606001},
	eid          = {astro-ph/9606001},
	archiveprefix = {arXiv},
	eprint       = {astro-ph/9606001},
	primaryclass = {astro-ph},
	adsurl       = {https://ui.adsabs.harvard.edu/abs/1996astro.ph..6001N}
}

@article{Navarro1997,
	title        = {{A Universal Density Profile from Hierarchical Clustering}},
	author       = {{Navarro}, Julio F. and {Frenk}, Carlos S. and {White}, Simon D.~M.},
	year         = {1997},
	month        = dec,
	journal      = {Astrophysical Journal},
	volume       = {490},
	number       = {2},
	pages        = {493--508},
	doi          = {10.1086/304888},
	archiveprefix = {arXiv},
	eprint       = {astro-ph/9611107},
	primaryclass = {astro-ph},
	adsurl       = {https://ui.adsabs.harvard.edu/abs/1997ApJ...490..493N}
}

@inproceedings{optuna_2019,
	title        = {Optuna: A Next-generation Hyperparameter Optimization Framework},
	author       = {Takuya Akiba and Shotaro Sano and Toshihiko Yanase and Takeru Ohta and Masanori Koyama},
	year         = {2019},
	booktitle    = {Proceedings of the 25th {ACM} {SIGKDD} International Conference on Knowledge Discovery and Data Mining},
	pages        = {2623--2631},
	doi          = {10.1145/3292500.3330701}
}

@article{Papamakarios2017,
	title        = {{Masked Autoregressive Flow for Density Estimation}},
	author       = {{Papamakarios}, George and {Pavlakou}, Theo and {Murray}, Iain},
	year         = {2017},
	month        = may,
	journal      = {arXiv e-prints},
	pages        = {arXiv:1705.07057},
	doi          = {10.48550/arXiv.1705.07057},
	eid          = {arXiv:1705.07057},
	archiveprefix = {arXiv},
	eprint       = {1705.07057},
	primaryclass = {stat.ML},
	adsurl       = {https://ui.adsabs.harvard.edu/abs/2017arXiv170507057P}
}

@article{Porth2024,
	title        = {{A road map to cosmological parameter analysis with third-order shear statistics: III. Efficient estimation of third-order shear correlation functions and an application to the KiDS-1000 data}},
	author       = {{Porth}, Lucas and {Heydenreich}, Sven and {Burger}, Pierre and {Linke}, Laila and {Schneider}, Peter},
	year         = {2024},
	month        = sep,
	volume       = {689},
	pages        = {A227},
	doi          = {10.1051/0004-6361/202347987},
	eid          = {A227},
	archiveprefix = {arXiv},
	eprint       = {2309.08601},
	primaryclass = {astro-ph.CO},
	adsurl       = {https://ui.adsabs.harvard.edu/abs/2024A&A...689A.227P}
}

@article{Potter2017,
	title        = {{PKDGRAV3: beyond trillion particle cosmological simulations for the next era of galaxy surveys}},
	author       = {{Potter}, Douglas and {Stadel}, Joachim and {Teyssier}, Romain},
	year         = {2017},
	month        = may,
	journal      = {Computational Astrophysics and Cosmology},
	volume       = {4},
	number       = {1},
	pages        = {2},
	doi          = {10.1186/s40668-017-0021-1},
	eid          = {2},
	archiveprefix = {arXiv},
	eprint       = {1609.08621},
	primaryclass = {astro-ph.IM},
	adsurl       = {https://ui.adsabs.harvard.edu/abs/2017ComAC...4....2P}
}

@article{Reeves2024,
	title        = {{12 {\texttimes} 2 pt combined probes: pipeline, neutrino mass, and data compression}},
	author       = {{Reeves}, Alexander and {Nicola}, Andrina and {Refregier}, Alexandre and {Kacprzak}, Tomasz and {Machado Poletti Valle}, Luis Fernando},
	year         = {2024},
	month        = jan,
	journal      = {Journal of Cosmology and Astroparticle Physics},
	volume       = {2024},
	number       = {1},
	pages        = {042},
	doi          = {10.1088/1475-7516/2024/01/042},
	eid          = {042},
	archiveprefix = {arXiv},
	eprint       = {2309.03258},
	primaryclass = {astro-ph.CO},
	adsurl       = {https://ui.adsabs.harvard.edu/abs/2024JCAP...01..042R}
}

@article{Schneider1995,
	title        = {{Steps towards nonlinear cluster inversion through gravitational distortions. I. Basic considerations and circular clusters.}},
	author       = {{Schneider}, Peter and {Seitz}, Carolin},
	year         = {1995},
	month        = feb,
	volume       = {294},
	pages        = {411--431},
	doi          = {10.48550/arXiv.astro-ph/9407032},
	archiveprefix = {arXiv},
	eprint       = {astro-ph/9407032},
	primaryclass = {astro-ph},
	adsurl       = {https://ui.adsabs.harvard.edu/abs/1995A&A...294..411S}
}

@article{Schneider2002,
	title        = {{B-modes in cosmic shear from source redshift clustering}},
	author       = {{Schneider}, P. and {van Waerbeke}, L. and {Mellier}, Y.},
	year         = {2002},
	month        = jul,
	journal      = {Astronomy and Astrophysics},
	volume       = {389},
	pages        = {729--741},
	doi          = {10.1051/0004-6361:20020626},
	archiveprefix = {arXiv},
	eprint       = {astro-ph/0112441},
	primaryclass = {astro-ph},
	adsurl       = {https://ui.adsabs.harvard.edu/abs/2002A&A...389..729S}
}

@article{Schneider2005,
	title        = {{The three-point correlation function of cosmic shear. II. Relation to the bispectrum of the projected mass density and generalized third-order aperture measures}},
	author       = {{Schneider}, P. and {Kilbinger}, M. and {Lombardi}, M.},
	year         = {2005},
	month        = feb,
	journal      = {Astronomy and Astrophysics},
	volume       = {431},
	pages        = {9--25},
	doi          = {10.1051/0004-6361:20034217},
	archiveprefix = {arXiv},
	eprint       = {astro-ph/0308328},
	primaryclass = {astro-ph},
	adsurl       = {https://ui.adsabs.harvard.edu/abs/2005A&A...431....9S}
}

@inbook{Schneider2006,
	title        = {Weak Gravitational Lensing},
	author       = {Schneider, P.},
	year         = {2006},
	booktitle    = {Gravitational Lensing: Strong, Weak and Micro},
	publisher    = {Springer Berlin Heidelberg},
	address      = {Berlin, Heidelberg},
	pages        = {269--451},
	doi          = {10.1007/978-3-540-30310-7_3},
	isbn         = {978-3-540-30310-7},
	url          = {https://doi.org/10.1007/978-3-540-30310-7_3}
}

@article{Schneider2015,
	title        = {{A new method to quantify the effects of baryons on the matter power spectrum}},
	author       = {{Schneider}, Aurel and {Teyssier}, Romain},
	year         = {2015},
	month        = dec,
	journal      = {Journal of Cosmology and Astroparticle Physics},
	volume       = {2015},
	number       = {12},
	pages        = {049--049},
	doi          = {10.1088/1475-7516/2015/12/049},
	archiveprefix = {arXiv},
	eprint       = {1510.06034},
	primaryclass = {astro-ph.CO},
	adsurl       = {https://ui.adsabs.harvard.edu/abs/2015JCAP...12..049S}
}

@article{Schneider2019,
	title        = {{Quantifying baryon effects on the matter power spectrum and the weak lensing shear correlation}},
	author       = {{Schneider}, Aurel and {Teyssier}, Romain and {Stadel}, Joachim and {Chisari}, Nora Elisa and {Le Brun}, Amandine M.~C. and {Amara}, Adam and {Refregier}, Alexandre},
	year         = {2019},
	month        = mar,
	volume       = {2019},
	number       = {3},
	pages        = {020},
	doi          = {10.1088/1475-7516/2019/03/020},
	eid          = {020},
	archiveprefix = {arXiv},
	eprint       = {1810.08629},
	primaryclass = {astro-ph.CO},
	adsurl       = {https://ui.adsabs.harvard.edu/abs/2019JCAP...03..020S}
}

@article{SchneiderLombardi2003,
	title        = {{The three-point correlation function of cosmic shear. I. The natural components}},
	author       = {{Schneider}, P. and {Lombardi}, M.},
	year         = {2003},
	month        = jan,
	volume       = {397},
	pages        = {809--818},
	doi          = {10.1051/0004-6361:20021541},
	archiveprefix = {arXiv},
	eprint       = {astro-ph/0207454},
	primaryclass = {astro-ph},
	adsurl       = {https://ui.adsabs.harvard.edu/abs/2003A&A...397..809S}
}

@article{Secco2022a,
	title        = {{Dark Energy Survey Year 3 results: Cosmology from cosmic shear and robustness to modeling uncertainty}},
	author       = {{Secco}, L.~F. and {Samuroff}, S. and {Krause}, E. and {Jain}, B. and {Blazek}, J. and {Raveri}, M. and {Campos}, A. and {Amon}, A. and {Chen}, A. and {Doux}, C. and {Choi}, A. and {Gruen}, D. and {Bernstein}, G.~M. and {Chang}, C. and {DeRose}, J. and {Myles}, J. and {Fert{\'e}}, A. and {Lemos}, P. and {Huterer}, D. and {Prat}, J. and {Troxel}, M.~A. and {MacCrann}, N. and {Liddle}, A.~R. and {Kacprzak}, T. and {Fang}, X. and {S{\'a}nchez}, C. and {Pandey}, S. and {Dodelson}, S. and {Chintalapati}, P. and {Hoffmann}, K. and {Alarcon}, A. and {Alves}, O. and {Andrade-Oliveira}, F. and {Baxter}, E.~J. and {Bechtol}, K. and {Becker}, M.~R. and {Brandao-Souza}, A. and {Camacho}, H. and {Carnero Rosell}, A. and {Carrasco Kind}, M. and {Cawthon}, R. and {Cordero}, J.~P. and {Crocce}, M. and {Davis}, C. and {Di Valentino}, E. and {Drlica-Wagner}, A. and {Eckert}, K. and {Eifler}, T.~F. and {Elidaiana}, M. and {Elsner}, F. and {Elvin-Poole}, J. and {Everett}, S. and {Fosalba}, P. and {Friedrich}, O. and {Gatti}, M. and {Giannini}, G. and {Gruendl}, R.~A. and {Harrison}, I. and {Hartley}, W.~G. and {Herner}, K. and {Huang}, H. and {Huff}, E.~M. and {Jarvis}, M. and {Jeffrey}, N. and {Kuropatkin}, N. and {Leget}, P. -F. and {Muir}, J. and {Mccullough}, J. and {Navarro Alsina}, A. and {Omori}, Y. and {Park}, Y. and {Porredon}, A. and {Rollins}, R. and {Roodman}, A. and {Rosenfeld}, R. and {Ross}, A.~J. and {Rykoff}, E.~S. and {Sanchez}, J. and {Sevilla-Noarbe}, I. and {Sheldon}, E.~S. and {Shin}, T. and {Troja}, A. and {Tutusaus}, I. and {Varga}, T.~N. and {Weaverdyck}, N. and {Wechsler}, R.~H. and {Yanny}, B. and {Yin}, B. and {Zhang}, Y. and {Zuntz}, J. and {Abbott}, T.~M.~C. and {Aguena}, M. and {Allam}, S. and {Annis}, J. and {Bacon}, D. and {Bertin}, E. and {Bhargava}, S. and {Bridle}, S.~L. and {Brooks}, D. and {Buckley-Geer}, E. and {Burke}, D.~L. and {Carretero}, J. and {Costanzi}, M. and {da Costa}, L.~N. and {De Vicente}, J. and {Diehl}, H.~T. and {Dietrich}, J.~P. and {Doel}, P. and {Ferrero}, I. and {Flaugher}, B. and {Frieman}, J. and {Garc{\'\i}a-Bellido}, J. and {Gaztanaga}, E. and {Gerdes}, D.~W. and {Giannantonio}, T. and {Gschwend}, J. and {Gutierrez}, G. and {Hinton}, S.~R. and {Hollowood}, D.~L. and {Honscheid}, K. and {Hoyle}, B. and {James}, D.~J. and {Jeltema}, T. and {Kuehn}, K. and {Lahav}, O. and {Lima}, M. and {Lin}, H. and {Maia}, M.~A.~G. and {Marshall}, J.~L. and {Martini}, P. and {Melchior}, P. and {Menanteau}, F. and {Miquel}, R. and {Mohr}, J.~J. and {Morgan}, R. and {Ogando}, R.~L.~C. and {Palmese}, A. and {Paz-Chinch{\'o}n}, F. and {Petravick}, D. and {Pieres}, A. and {Plazas Malag{\'o}n}, A.~A. and {Rodriguez-Monroy}, M. and {Romer}, A.~K. and {Sanchez}, E. and {Scarpine}, V. and {Schubnell}, M. and {Scolnic}, D. and {Serrano}, S. and {Smith}, M. and {Soares-Santos}, M. and {Suchyta}, E. and {Swanson}, M.~E.~C. and {Tarle}, G. and {Thomas}, D. and {To}, C. and {DES Collaboration}},
	year         = {2022},
	month        = jan,
	volume       = {105},
	number       = {2},
	pages        = {023515},
	doi          = {10.1103/PhysRevD.105.023515},
	eid          = {023515},
	archiveprefix = {arXiv},
	eprint       = {2105.13544},
	primaryclass = {astro-ph.CO},
	adsurl       = {https://ui.adsabs.harvard.edu/abs/2022PhRvD.105b3515S}
}

\end{document}